\documentclass{ws-ijmpa}
\usepackage{url,hyperref,slashed,multirow,cite,verbdef}
\usepackage[utf8]{inputenc}
\hypersetup{
   bookmarks=true,         
   unicode=true,          
   pdftoolbar=true,        
   pdfmenubar=true,        
   pdffitwindow=false,     
   pdfstartview={FitH},    
   pdftitle={RecastingMA5},    
   pdfauthor={E.~Conte and B.~Fuks},     
   pdfsubject={RecastingMA5},   
   pdfcreator={Creator},   
   pdfproducer={Producer}, 
   pdfkeywords={keyword1} {key2} {key3}, 
   pdfnewwindow=true,      
   colorlinks=true,       
   linkcolor=blue,          
   citecolor=magenta,        
   filecolor=magenta,      
   urlcolor=cyan           
}
\newcommand{\MG}{\textsc{MadGraph}~5\_aMC@NLO}
\newcommand{\PY}{\textsc{Pythia}~8}
\newcommand{\MA}{\textsc{MadAnalysis}~5}
\newcommand{\MW}{\textsc{MadWidth}}
\newcommand{\MS}{\textsc{MadSpin}}
\newcommand{\MAnorm}{{MadAnalysis}~5}
\newcommand{\FJ}{\textsc{FastJet}}
\newcommand{\FJnorm}{{FastJet}}
\newcommand{\DEL}{\textsc{Delphes}}
\newcommand{\DELnorm}{{Delphes}}
\newcommand{\ROOT}{\textsc{Root}}
\newcommand{\lhe}{\textsc{Lhe}}
\newcommand{\lhco}{\textsc{Lhco}}
\newcommand{\hepmc}{\textsc{HepMC}}
\newcommand{\stdhep}{\textsc{StdHep}}
\newcommand{\INSP}{\textsc{Inspire}}
\newcommand{\python}{\textsc{Python}}
\newcommand{\spla}{\textsc{SampleAnalyzer}}
\newcommand{\eg}{\textit{e.g.}}
\newcommand{\ie}{\textit{i.e.}}
\newcommand{\etc}{\textit{etc.}}

\def\be{\begin{equation}}
\def\ee{\end{equation}}
\def\bsp#1\esp{\begin{split}#1\end{split}}

\begin{document}
\markboth{E.~Conte and B.~Fuks}
{Confronting new physics theories to LHC data with~\MA}

%
\catchline{}{}{}{}{}
%

\title{Confronting new physics theories to LHC data with~\MA}

\author{Eric Conte}
\address{
Institut Pluridisciplinaire Hubert Curien/D\'epartement Recherches
        Subatomiques, Universit\'e de Strasbourg/CNRS-IN2P3, 23 Rue du Loess,
        F-67037 Strasbourg, France}

\author{Benjamin Fuks}
\address{
  Sorbonne Universit\'e, CNRS, Laboratoire de Physique Th\'eorique et
        Hautes \'Energies, LPTHE, F-75005 Paris, France\\
  Institut Universitaire de France, 103 boulevard Saint-Michel,
    75005 Paris, France}

\maketitle


\begin{abstract}
We provide a comprehensive and pedagogical introduction to the \MA\ framework,
with a particular focus on its usage for reinterpretation studies. To this end,
we first review the main features of the normal mode of the program and how a
detector simulation can be handled. We then detail, step-by-step, how to
implement and validate an existing LHC analysis in the \MA\ framework and how to
use this reimplementation, possibly together with other recast codes available
from the \MA\ Public Analysis Database, for reinterpreting ATLAS and CMS
searches in the context of a new model. Each of these points is illustrated by
concrete examples. Moreover, complete reference cards for the normal and expert
modes of \MA\ are provided in two technical appendices.
\end{abstract}

\tableofcontents

\newpage

\section{Introduction} The Large Hadron Collider (LHC) at CERN has opened a new era in the exploration
of the fundamental laws of Nature, in particular through the delivery of very
high quality results during its first run and (currently on-going) second run.
These results have in particular started to shed light on the electroweak
symmetry breaking mechanism through the discovery of a Higgs boson in 2012 and
the measurement of its properties in the following years. However, despite this
success and having all LHC collaborations carrying out an extensive search
program for new phenomena, no signal for a new particle beyond those of the
Standard Model has been observed. As a consequence, the results of the
experimental searches are interpreted as constraints on
well-defined theoretical contexts, ranging from popular models like the
Minimal Supersymmetric Standard Model to simplified models or effective field
theories. There is however a plethora of motivated new physics theories, that
all come with a large variety of concrete realizations, whose predictions should
be confronted to LHC data. It is therefore crucial to develop a strategy
allowing to exploit the past, present and future results of the LHC in the best
possible manner, so that one could get a full understanding of what physics
beyond the Standard Model could be or could not be.

Many groups have consequently developed public software dedicated to the
reinterpretation of the LHC results. These programs can be classified into two
categories. A first series of tools rely on simplified model results. They aim
to compare predictions for new physics signal cross sections (or event
counts, after including selection efficiency information for a given signature)
with experimentally-derived upper bounds\cite{Kraml:2013mwa,Ambrogi:2017neo,%
Papucci:2014rja}.
Whilst very fast and benefiting from the advantage of reducing generally complex
models to a handful of relevant signatures, this method only allows one to
constrain the part of the signal that maps onto simplified-model signatures. It
moreover suffers from systematic uncertainties related, \eg, to changes in
kinematical distributions under different signal assumptions. A more general
and more precise approach is realized in a second class of tools by means of
Monte Carlo simulations of the new physics signals. By mimicking the
experimental
analysis strategies, predictions for the number of signal event counts are
achieved and next compared with data and the corresponding Standard Model
expectation\cite{Drees:2013wra,%
Dumont:2014tja,Buckley:2010ar,Balazs:2017moi}. Whilst very general and more
precise, these programs generally suffer from being very expensive in terms of
CPU power.

The \MA\ framework\cite{Conte:2012fm,Conte:2014zja} is a platform
for new physics phenomenology. It has been originally developed to
allow for the straightforward design of the analysis of any given collider
signal (together with the associated Standard Model background), thanks to a
user-friendly {\sc Python}-based command line interface and a
developer-friendly {\sc C++} core\cite{Conte:2012fm}. Whilst directly targeting
the LHC in the early days, the program is today additionally used for
prospective studies addressing future colliders. The code has been
extended a few years ago in order to allow for the reinterpretation of the LHC
results~\cite{Conte:2014zja}, so that it lies in the second category of tools
introduced above. In practice, the recasting of the outcome of a
given analysis requires not only the reimplementation, in the \MA\ data format,
of the analysis of interest, but it also implies the validation of the recasted
code through a comparison with officially-provided ATLAS and CMS results. All
validated {\sc C++} reimplementations are released on the \MA\ public analysis
database\cite{Dumont:2014tja}, together with a detailed validation note.
Moreover, the codes are also published on \INSP\ where they are
assigned a Digital Object Identifier (for traceability reasons) and versioned.

The aim of this paper is to ease the reimplementation of an experimental
analysis in the \MA\ framework by providing a step-by-step manual on the tasks
that need to be tackled by the user. In Section~\ref{sec:nut}, we review the
main features of \MA, focusing in particular on its normal mode of running.
Section~\ref{sec:fastsim} is dedicated to the handling of the simulation of the
response of a detector within \MA, which relies on an interface either with the
\FJ\ program\cite{Cacciari:2011ma} or with the \DEL\
package\cite{deFavereau:2013fsa}. The heart of this document consists in the
material provided in Section~\ref{sec:implementation}, where all the steps
necessary for reimplementing and validating an LHC analysis in the {\sc C++}
core of \MA\ are documented in details. An example of using recast codes for
getting constraints on a given physics model is provided in
Section~\ref{sec:recast_example}, and our work is summarized in
Section~\ref{sec:conclusion}. Two reference cards, one for the normal mode and
one for the expert mode of \MA, are finally provided as appendices.

\section{\MA\ in a nutshell} \label{sec:nut}

\subsection{Main features}

The \MA\cite{Conte:2012fm,Conte:2014zja} package allows for the analysis of
Monte-Carlo event files describing one (or more than one) collider process.
These collision events could consist of hard scattering events, parton showered
events, hadronized events or even reconstructed events. They can therefore be
provided in any
of the \lhe\cite{Boos:2001cv,Alwall:2006yp}, \stdhep\cite{stdhep},
\hepmc\cite{Dobbs:2001ck}, or \lhco\cite{lhco} file formats, or as
\ROOT\ files generated by \DEL\cite{deFavereau:2013fsa}. Although those event
file formats are by nature different, \MA\ internally selects an appropriate
reader and accordingly adapts the way in which observables should be evaluated.
Some methods can also be only available for specific classes of events. We refer
to the appendices for more information.
From version v1.6 onwards, \MA\ can moreover be used either in a standalone
fashion, or
from \MG\cite{Alwall:2014hca}. In the latter case, histograms and analyses
can be executed right after event generation, on an event-by-event basis, making
it unnecessary to store huge event files on disk. Furthermore, a choice of
selections and observables of typical interest for the process under
consideration is automatically proposed to the user.

\MA\ has been developed with user-friendliness in mind. For instance,
there is no installation procedure and the building of the core C++ libraries is
automatically achieved behind the scene. Thanks to a \python\ command line
interface, users can rely on an intuitive metalanguage to design a
phenomenological analysis and setup the options of the program in a
straightforward manner. A C++ code is subsequently generated, compiled and
executed on the inputed events. This \python-C++ interplay guarantees an optimal
execution speed through the inclusion of well-tested and optimized methods. The
robustness of the procedure has been
intensively validated by extensive tests on a variety of platforms.

The \MA\ metalanguage is rich, but it cannot handle any possible analysis that
one may dream of. For example, it does not offer predefined keywords
for each observable of the always-growing series of exotic variables used by the
ATLAS and CMS collaborations. To circumvent this issue, the
user can directly implement an analysis in C++, bypassing the \python\
console. This consists of the so-called expert mode of the program in which the
user can benefit from all options and methods already available in \MA\
(readers, writers, the internal data format, observables, services, \etc), and
supplement
them with the few extra more general or non-standard routines that are necessary
for the analysis under consideration. The expert mode is in particular well
suited for the reimplementation of an existing LHC analysis that often relies on
complex object reconstruction criteria and/or complicated observables, going
beyond what could be achieved by using the \MA\ metalanguage only. Present
developments of the metalanguage however target this limitation and aim to allow
for the recasting of the simplest LHC analyses directly in {\sc Python} in a
close future.

In addition, the \MA\ package is interfaced to several
common packages used in the high-energy physics community, like
\FJ\cite{Cacciari:2011ma} or \DEL\cite{deFavereau:2013fsa}. As will be shown in
the next sections, these interfaces allow
the user to simulate the effects of a typical LHC detector on events produced by
a Monte Carlo event generator (assuming that showering and hadronization are
included in
the generation process).

The applications of the program are numerous, although its main usage concerns
phenomenological investigations to assess the reach for the LHC to a
specific theoretical model through a given final-state signature probe.
\MA\ allows the user to define a selection strategy based on criteria
applied on kinematical or geometrical observables. The corresponding cut-flow
charts, exhibiting the selection performance in term of a chosen figure of
merit, are then automatically extracted. \MA\ can also be used in order to
reinterpret LHC experimental results in the context of any theoretical model.
\MA\ is connected to a database of ATLAS and CMS analysis reimplementations in
its internal data format, the so-called \MA\ PAD
(Public Analysis Database). Making use of these reimplementations, the
corresponding ATLAS and CMS selections can be applied on some new physics signal
to test whether it is excluded. This feature is
discussed in details in Section~\ref{sec:recast_example}.

On different footings, \MA\ can additionally be used to validate
Monte Carlo simulations. The \MA\ metalanguage indeed
allows for the production of validation histograms very intuitively, by means of
single-line instructions. This allows one for instance to generate
distributions in given observables or differential jet rate
spectra when multipartonic merging is at stake, and to verify that their
behavior is the expected one.
Moreover, the comparison of different samples
generated, \eg, using different codes or Monte Carlo tunes, is automatic.
Finally, \MA\ can also convert an input event file into an output event file
encoded within a different format, or merge several input event files. For
instance, several
\lhco\ event files can be merged and translated into a single \lhe\ file.

\subsection{Architecture of the package}

\begin{figure}
  \centering
  \includegraphics[scale=0.5]{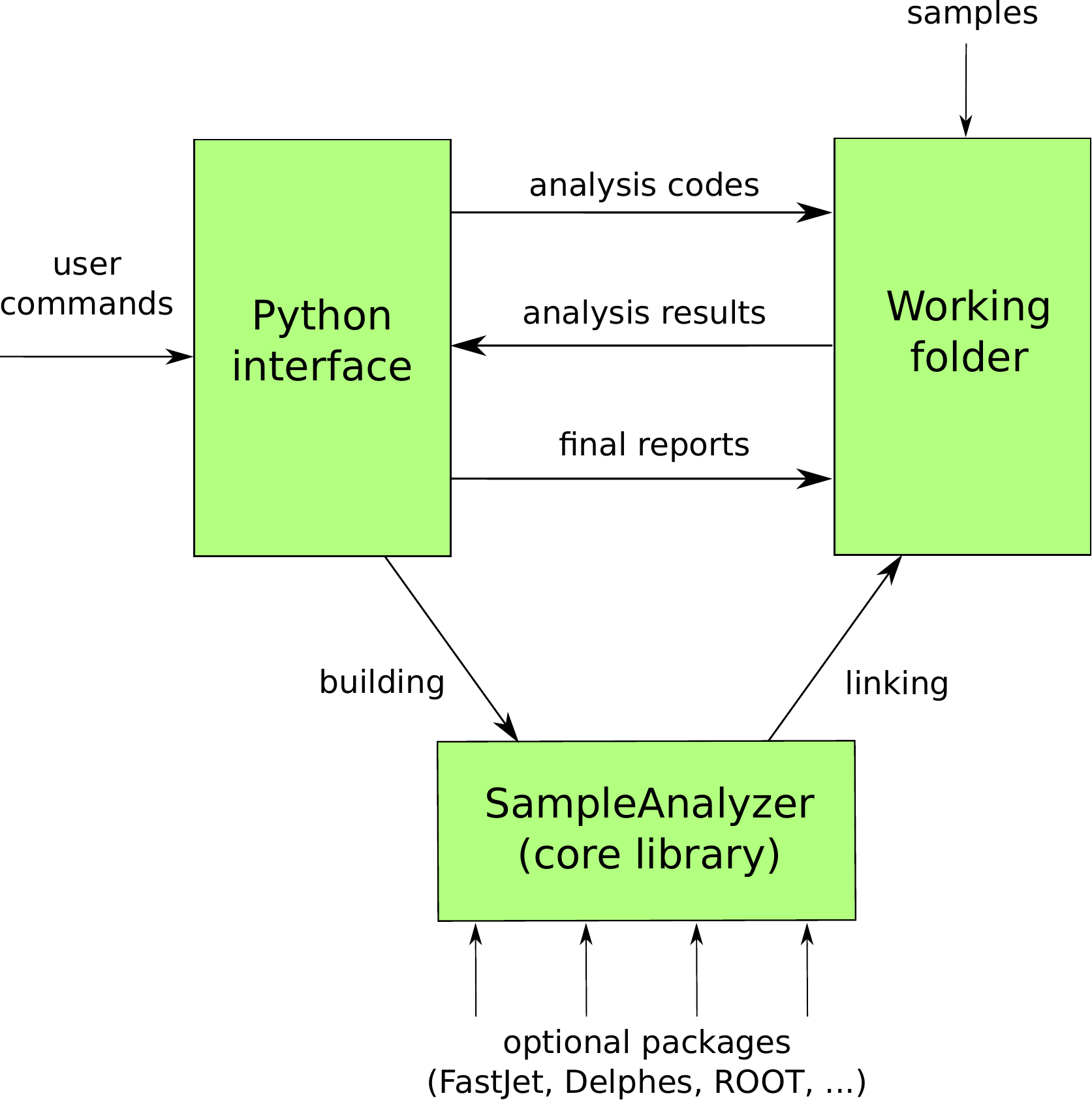}
\caption{Sketch of the architecture of the \MA\ package \label{fig:archi}}
\end{figure}

The architecture of the \MA\ package is summarized in Figure~\ref{fig:archi}. As
mentioned in the previous section, the program is built around a \python\
console which uses
interacted with through an intuitive metalanguage. On the first start of the
program, the \python\ module takes care of building the C++ core library of \MA,
that is named \spla. The latter can be interfaced to several high-energy physics
packages. This is done either automatically, if \MA\ detects the interfaced
programs on the system of the user, or by providing the paths to local
installations by editing the file
\verb+madanalysis/input/installation_options.dat+. Moreover, these programs can
also be installed internally to \MA\ via the \verb+install+ command (see below).

After implementing an analysis through the command line interface, the user
needs
to `submit' his code (through the command \verb+submit+, as explained below).
\MA\ subsequently converts the analysis into a C++ code that is linked, once
compiled, to the \spla\ library and any necessary package
interfaced to \MA\ (like \FJ\ or \DEL). The analysis code is then executed,
and the results of the execution are split into several text files encoded into
the SAF format, the internal format of \spla. These files are read back by the
\python\ interface which finally takes care of generating figures and reports in
the HTML and PDF formats.

In the expert mode of the program, the \python\ interpreter is used to
create a working directory (see Figure~\ref{fig:archi}). It is then up to the
user to implement his/her analysis as a C++ code (the working folder contains a
dummy analysis as an example), compile it (a \verb+makefile+ is present in the
working directory), execute it and handle the outputted
SAF files to get histograms and cutflows.

\subsection{Prerequisites, getting and running the code}

\MA\ is supported on {\sc Linux}, {\sc Unix} and {\sc Mac Os X} systems.
The program is not guaranteed to work with
any version of the {\sc Windows} operating system, even if
the {\sc Cygwin} emulator is used. Three external dependencies are mandatory:
the GNU {\sc G++}
compiler (or {\sc CLang} on {\sc Mac Os X}), a \python\ 2.7 installation (or
newer, but not a \python~3 one), and the GNU
{\sc GMake} package.

Some of the functionalities of \MA\ depend on additional packages. These can be
classified into two categories, the first one addressing data processing and
the second one histogramming and report generation. While not mandatory, the
absence of one or more of these additional packages may imply that some
of the features of \MA\ are disabled on runtime.

In order to be able to read and write compressed event files, the {\sc Gzip}
library must be available, whilst the \FJ\ and \DEL\ functionalities are only
available if these two programs are present on the user system. These three
packages can all be
automatically installed locally, within \MA, by using the \verb+install+
command (see below). Concerning \DEL,
the \ROOT\ framework must have been installed, as \DEL\
relies on \ROOT. Finally, the {\sc SciPy} library should be present in order to
be able to automatically compute limits in the context of LHC analysis
reinterpretations.

Figures can be generated using either \ROOT\ or {\sc MatPlotLib}, if any of
these two package is available. Otherwise, histograms are produced in the SAF
format only, \ie, encoded in a text-based format. PDF reports are additionally
generated if a \LaTeX\ compiler is available.

The \MA\ program can be obtained from the Web,\\
\hspace*{8.5mm} \url{https://launchpad.net/madanalysis5}\\
where a tarball with the latest version of the code can be found. The downloaded
tarball must then be unpacked, by typing in a shell,
\begin{verbatim}
  tar -xzf ma5_v<xxx>.tgz
\end{verbatim}
where \verb+<xxx>+ stands for its version number. Development
branches and future releases of the code are also publicly available, but must
be downloaded via {\sc Bazaar} exclusively. For instance, the current release
can be obtained by typing in a shell
\begin{verbatim}
  bzr branch lp:madanalysis5
\end{verbatim}
assuming that {\sc Bazaar} has been installed on the system. Similar
commands, whose exact form is indicated on the \MA\ webpage, exist for other
versions of \MA.

From the directory in which \MA\ has been downloaded and unpacked, the code can
be launched by typing in a shell
\begin{verbatim}
  ./bin/ma5
\end{verbatim}
This initializes the \MA\ command line interface that starts by checking the
presence of all mandatory packages and which of the optional packages are
available. After a successful initialization step, the prompt \verb+ma5>+ is
shown on the screen and the user can being with typing in any command.

The above \verb+bin/ma5+ command can be supplemented by one or several arguments
that modify the way in which \MA\ works. For instance, the commands
\begin{verbatim}
  ./bin/ma5 -P
  ./bin/ma5 -H
  ./bin/ma5 -R
  ./bin/ma5 -E
\end{verbatim}
initialize the command line interface in the partonic (default), hadronic,
reconstructed and expert mode, respectively. The first three modes are each
dedicated to the analysis of a specific class of events, whilst the last one
allows for the design of an analysis directly in C++ (see
Section~\ref{sec:implementation}). The partonic mode focuses on the analysis of
hard-scattering and parton-level events, the hadronic mode is
dedicated to events including the description of the parton showering and
hadronization, and the reconstructed mode is taking care of
event files where physics objects have been reconstructed. The debug mode of the
program can finally be switched on by starting the program with the \verb+-d+
argument. The screen output consequently includes more detailed information.

\subsection{Basic principles of the normal mode}
In this section, we briefly detail the basic principles behind the normal mode
of running of \MA. For simplicity, we make use of a specific example to
introduce the various concepts instead of listing them sequentially. More
details can be
found in \ref{app:normal}.

The analysis implemented in this section relies on test samples that can be
downloaded by typing, in the \MA\ command line interface,
\begin{verbatim}
  install samples
\end{verbatim}
As a result, a \verb+samples+ directory is created and is used to store four
test samples of Standard Model parton-level events.

We then need to import the Monte Carlo samples to be analyzed, assign them
dataset labels, and tag them as either background or signal events.
By typing in the following commands,
\begin{verbatim}
  import samples/zz* as zz
  import samples/tt* as tt
  set zz.type = signal
  set tt.type = background
\end{verbatim}
we define two datasets, \verb+zz+ and \verb+tt+, the first one corresponding
to diboson events (our signal) and the second one to top-pair production events
(our background). Moreover, the cross section (in pb) associated with each
sample can be modified by the user with the command
\begin{verbatim}
  set tt.xsection = <value-tt>
  set zz.xsection = <value-zz>
\end{verbatim}
which overwrites the cross section values read from the event files, impacting
accordingly the cutflow and histogram normalization.

The second step in our analysis concerns the creation of labels to identify the
physics objects whose properties will be constrained by the analysis strategy.
\MA\ already includes a few predefinitions, like labels for electrons
(\verb+e-+) and
positrons (\verb"e+"), or muons (\verb+mu-+) and antimuons (\verb"mu+"). We
define below extra labels to simultaneously tag, regardless of their electric
charge, electrons and positrons, muons and antimuons, as well as all leptons
from the first two generations,
\begin{verbatim}
  define e  = e+  e-
  define mu = mu+ mu-
  define l  = e   mu
\end{verbatim}

We next declare a few histograms, using different ways in which this could be
achieved,
\begin{verbatim}
  plot MET
  plot PT (mu)
  plot eta(mu) 100 -5 +5 [logY]
\end{verbatim}
With the first line, we request a histogram representing the missing transverse
energy distribution. The range on the $x$-axis and the number of bins are taken
as their default values (that depend on the considered observable). The second
histogram corresponds to the transverse momentum spectrum of all muons and
antimuons present in the events (each muon/antimuon gives rise to a specific
entry in the histogram). The range and binning are again taken as their default
values. Finally, the last line requests a histogram with the pseudorapidity
spectrum of the muons, the number of bins (100) and $x$-axis range
($[-5, 5]$) being this time specified. Moreover, the \verb+logY+ options sets a
log=scale $y$-axis. We refer to
\ref{app:normal} for an extensive descriptions of all available options.

The next part of the analysis concerns the selection itself. It is implemented
as a sequence of criteria that must be satisfied either by an event (that is
then rejected or selected) or by a specific type of objects (that are
then kept in the analysis or put aside). Two complementary
commands, \verb+select+ and \verb+reject+, can be used to this aim. For
example, the following commands allow for the rejection of events that do not
exhibit a large amount of missing transverse energy, for the implementation of a
requirement on the lepton candidates to consider in the analysis and for the
rejection of any surviving event that does not feature at least one final-state
lepton candidate (after accounting for the previous restriction on the leptons),
\begin{verbatim}
  reject MET < 10
  select ABSETA(l) < 5
  select N(l) >= 1
\end{verbatim}
The first line above yields the rejection of events that feature a missing
transverse
energy smaller than 10~GeV, whilst the second line leads to the cleaning of the
event content from all leptons with a pseudorapidity larger than 5, in absolute
value. The last line finally refers to the rejection of any surviving event that
does not feature at least one lepton candidate.

At this stage, we split the analysis into several sub-analyses by introducing
different search (or signal) regions. We begin with the declaration of the
regions by assigning them labels,
\begin{verbatim}
  define_region OneLepton
  define_region SeveralLeptons
\end{verbatim}
This allows for the definition of two regions named  \verb+OneLepton+, that will
be devoted to the analysis of events featuring exactly one lepton, and
\verb+SeveralLeptons+, dedicated to the analysis of events containing strictly
more than one lepton. Selection cuts can then be associated with one or the
other regions as in the self-explanatory following lines of code,
\begin{verbatim}
  select N(l)=1 {OneLepton}
  select N(l)>1 {SeveralLeptons}
  reject PT(l)<25
  reject PT(j)<20 {SeveralLeptons OneLepton}
\end{verbatim}
With the next-to-last line, we introduce a cut that is not explicitly associated
with any region and is is therefore automatically associated with all declared
regions. With the last command, we instead explicitly associate the cut with
two regions.

Global properties of the analysis can be modified by acting on the \verb+main+
object associated with any analysis performed in the \MA\ context. For example,
with
\begin{verbatim}
  set main.lumi = 1000
  set main.graphic_render = matplotlib
\end{verbatim}
we set the integrated luminosity to 1000~fb$^{-1}$, which will impact the
normalization of the histograms and cutflows, and we enforce \MA\ to use
{\sc MatPlotLib} for creating the figures associated with all declared
histograms. The latter command however only works if {\sc MatPlotLib} is
correctly detected by \MA.

In order to execute the analysis, it is enough to cast the \verb+submit+
command,
\begin{verbatim}
  submit
\end{verbatim} 
A working folder is subsequently created, the C++ code corresponding to the
implemented analysis generated, compiled and finally executed. The results are
then loaded back into the \python\ console and the analysis reports are
extracted in the HTML and \LaTeX\ format, the corresponding PDF file being
generated if a \LaTeX\ compiler is found. The HTML report can be opened by
typing, in the \MA\ command line interface, the command
\begin{verbatim}
  open
\end{verbatim}
The report contains the list of commands inputed by the user, a statistical
description of the analyzed samples, the list of declared histograms and cuts,
and the results (figures, selection efficiencies). A cut-flow chart with a
figure of merit (like a signal over background ratio) is also included.

\subsection{Help}

Inline help is available from the \python\ console by typing in, in the command
line interface,
\begin{verbatim}
  help
\end{verbatim}
The list of \MA\ keywords that can be used is subsequently displayed to the
screen. For getting more information about a specific keyword, the user is
invited to type again the \textsl{help} command, followed by the name of the
considered keyword. For instance,
\begin{verbatim}
  help submit
\end{verbatim}
prints out details about the \verb+submit+ command.

For additional questions, we recommend users to visit the \MA\ webpage where
bug reports and questions can be submitted,\\
\hspace*{2.5mm} \url{https://launchpad.net/madanalysis5},\\
and our wiki page,\\
\hspace*{2.5mm}\url{https://madanalysis.irmp.ucl.ac.be}.\\
where detailed step-by-step tutorials are available.

More information on the normal mode can be found in the reference card presented
in \ref{app:normal}, online, as well as in the
original manual of the code\cite{Conte:2012fm}.

\section{Fast detector simulation with \MA} \label{sec:fastsim}
The response of a typical LHC detector can be emulated in \MA\ via interfaces to
\FJ\cite{Cacciari:2011ma} and  \DEL~3\cite{deFavereau:2013fsa}. In addition to
the usual \FJ\ and \DEL\ features, a few new options are available through the
\MA\ interfaces.
For instance, all jet-clustering algorithms implemented in \FJ\ are supplemented
by a parametric simulation of the reconstruction efficiencies, and new \DEL\
modules are available to improve the realism of the detector simulation.

In Section~\ref{sec:fastjet}, we describe how the program reconstruct physics
objects with techniques implemented within \FJ, and Section~\ref{sec:delphes} is
dedicated ot the usage of \DEL\ in the \MA\ framework. A phenomenological
example of usage is presented in Section~\ref{sec:example1}.

\subsection[Simplified detector simulation with \FJ]
{Simplified detector simulation with \FJnorm}
\label{sec:fastjet}

Thanks to its interface to \FJ, \MA\ offers a way to analyze a signal including
the modeling of a perfect detector with an infinite resolution, which
could also be used for undertaking dedicated studies of a
particular detector effect. In this section, we explain how the interface works,
and how to use a jet-clustering algorithm to reconstruct physics objects from
hadronized events (including the modeling of
basic detector effects). As detailed at the end of this subsection, it is
moreover possible to store the reconstructed events in the LHE or LHCO format.
The structure of the format used for the event files generated by \MA, that
slightly differs from the usual standards, is also described.

\subsubsection{Jet clustering algorithms in \MA}

In order to reconstruct the physics objects that could be identified from an
event where hadronization has been simulated, one needs to start \MA\ in the
reconstructed mode,
\begin{verbatim}
  bin/ma5 -R
\end{verbatim}
As jet reconstruction relies on \FJ, the package has first to be found by \MA.
If this
is not the case, a message is printed to the screen and the user is prompted to
type, in the \MA\ command line interface, the command
\begin{verbatim}
  install fastjet
\end{verbatim}
that yields a local installation of \FJ.

\renewcommand{\arraystretch}{1.30}%
\begin{table}
  \tbl{
    Jet algorithms available in \MA, together with their associated keyword.}
  {\begin{tabular}{c|c|c}
     Jet algorithm & Reference & \MA\ keyword\\
   \hline
   The longitudinally invariant $k_T$ algorithm
     & \cite{Catani:1993hr,Ellis:1993tq}
     & \texttt{kt}\\
    The Cambridge/Aachen algorithm
      & \cite{Dokshitzer:1997in,Wobisch:1998wt}
      & \texttt{cambridge}\\
    The anti-$k_T$ algorithm 
      & \cite{Cacciari:2008gp}
      &\texttt{antikt}\\
    The generalized $k_T$ algorithm
      & \cite{Cacciari:2011ma}
      & \texttt{genkt}\\
    The CDF jet clustering algorithm
      & \cite{Abe:1991ui}
      & \texttt{cdfjetclu}\\
    The CDF midpoint algorithm
      & \cite{Blazey:2000qt}
      & \texttt{cdfmidpoint}\\
    The seedless infrared-safe cone (siscone) algorithm
      & \cite{Salam:2007xv}
      & \texttt{siscone}\\
    The grid jet algorithm
      & \cite{Cacciari:2011ma}
      & \texttt{gridjet}\\
   \end{tabular} \label{tab:jets}}
\end{table}

One single jet definition can then be provided, and its configuration is done by
setting the attributes of the \verb+fastsim+ class of the
\verb+main+ \MA\ object. Jet reconstruction is first switched on by typing in
\begin{verbatim}
  set main.fastsim.package = fastjet
\end{verbatim}
This command turns on the usage of \FJ, and the anti-$k_T$ jet
algorithm\cite{Cacciari:2008gp} is invoked by default. Other algorithms can
however be employed by issuing, in the command line interface,
\begin{verbatim}
  set main.fastsim.algorithm = <algo>
\end{verbatim}
where \verb+<algo>+ denotes the keyword of the algorithm of interest (see
Table~\ref{tab:jets}). Several
algorithms available within \FJ\ can be accessed from \MA. This includes both
algorithms based on object recombination like the longitudinally invariant
$k_T$ algorithm\cite{Catani:1993hr,Ellis:1993tq}, the Cambridge/Aachen
algorithm\cite{Dokshitzer:1997in,Wobisch:1998wt}, the anti-$k_T$
algorithm and the generalized $k_T$ algorithm\cite{Cacciari:2011ma}, as well as
cone algorithms like the CDF jet clustering\cite{Abe:1991ui} or
midpoint\cite{Blazey:2000qt} algorithm, the seedless infrared-safe cone
(siscone) algorithm\cite{Salam:2007xv} or the grid jet
algorithm\cite{Cacciari:2011ma}.
Although several cone algorithms are
available within the \MA\ framework, the user should bear in mind that only the
siscone algorithm is infrared-safe.

Jet reconstruction relies on the
combination of protojets, that could be either final-state hadrons or already
combined objects. The combination process follows the
$E$-scheme~\cite{Cacciari:2011ma} in which the
combination of two object always corresponds to the sum of their four-momenta.
The algorithm starts from the inputs selected by \MA. The latter consist of all
visible final-state particles, or equivalently to all final-state particles
whose Particle Data
Group (PDG) identifier\cite{Patrignani:2016xqp} is not included in the
definition of the \verb+invisible+ multiparticle. This multiparticle is
comprised of the PDG codes of all invisible particles. By default, it
includes all Standard Model neutrinos and antineutrinos, as well as the lightest
neutralino and the gravitino that are common invisible particles in
supersymmetric theories. If needed, the user can override the definition of
this \verb+invisible+ multiparticle by typing in
\begin{verbatim}
  define invisible = invisible <new-pdg-code>
\end{verbatim}
In the above example, the \verb+<new-pdg-code>+ value that corresponds to
the PDG identifier of a new invisible exotic particle is added to the default
list of invisible particles.

In a similar way, the user can indicate whether a new
particle is a strongly-interacting particle, which hence participates to the
hadronic activity in the events. The information is provided via the
\verb+hadronic+ multiparticle, whose definition can be superseded by typing in
\begin{verbatim}
  define hadronic = hadronic <new-pdg-code>
\end{verbatim}
The PDG code \verb+<new-pdg-code>+ is hence added to the list of particles that
hadronize.

All jet-clustering algorithms available within the \MA\ framework feature
several options that can be tuned according to the needs of the user. Although
most options are algorithm-dependent, two of them are common to all algorithms.
The user can define the minimum value of the transverse momentum of a
reconstructed object so that it should be returned by the reconstruction process
(the default
threshold value being 5~GeV), and decide whether the algorithm should be
exclusive (default) or inclusive relatively to particle identification. An
inclusive behavior implies that the code includes electrons, muons, taus and
photons originating from hadron decays in the respective electron, muon, tau and
photon collections in addition to consider them as constituents for the
reconstructed objects. An example is given with the following two
commands
\begin{verbatim}
  set main.fastsim.ptmin        = 10
  set main.fastsim.exclusive_id = false
\end{verbatim}
With the first command, we impose that any object whose transverse momentum is
smaller than 10~GeV is ignored by the jet algorithm, the 10~GeV threshold being
passed to the code by setting accordingly the \verb+ptmin+ option
of the \verb+main.fastsim+ object. With the second command, we set the
\verb+exclusive_id+ attribute of the \verb+main.fastsim+ object to false.
This contrasts with the default behavior of the code,
in which leptons and photons originating from hadron decays are solely
considered as constituents of the reconstructed jets and hence do not appear in
the lepton and photon collections.

Concerning the algorithm-dependent options, the user can fix the jet
radius parameter $R$ that enters the definition of the distance measure used by
several algorithms via the \verb+radius+ option of the
\verb+main.fastsim+ object, like in
\begin{verbatim}
  set main.fastsim.radius = 1.0
\end{verbatim}
This option, whose default value is $1.0$, is available for all algorithms but
the grid jet one. The generalized $k_T$ algorithm moreover relies on a distance
measure depending on an additional continuous parameter $p$ that is set to -1
(its default value), 0 and 1 for the anti-$k_T$, Cambridge/Aachen and $k_T$
algorithms, respectively. This parameter can be modified to any real value via
the \verb+p+ attribute of the \verb+main.fastsim+ object, as in
\begin{verbatim}
  set main.fastsim.p = -0.5
\end{verbatim}
where the $p$ parameter is set to $-1/2$.

In addition to the jet radius parameter previously mentioned, the siscone
algorithm depends on the fraction of overlapping momentum above which two
protojets are combined, on the maximum number of passes the algorithm should be
carried out and on a transverse-momentum threshold allowing to remove too soft
reconstructed jets. These parameters can be set up via the \verb+overlap+,
\verb+npassmax+ and \verb+input_ptmin+ attributes of the \verb+main.fastsim+
object, as for instance in
\begin{verbatim}
  set main.fastsim.algorithm   = siscone
  set main.fastsim.radius      = 1.0
  set main.fastsim.overlap     = 0.5
  set main.fastsim.npassmax    = 0
  set main.fastsim.input_ptmin = 0.0
\end{verbatim}
In this example, all parameters are manually fixed to their default values. For
cases in which the \verb+npassmax+ parameter is fixed to zero, the algorithm
stops as soon as a pass does not generate any new stable cone.

The options available for the two CDF reconstruction algorithms are
similar. On top of the radius parameter (\verb+radius+) and the fraction of
overlapping momentum required to combine two protojets (\verb+overlap+), the
user can additionally fix the seed threshold parameter (\verb+seed+) that is
used in the constituent merging procedure.
Moreover, the CDF midpoint algorithm requires the user to fix the cone
area fraction (\verb+cone_areafraction+) that controls the size of the cones
that are searched for within the algorithm, whilst the CDF jet clustering
algorithm
optionally allows for some ratcheting (\verb+iratch+), which implies to retain
the constituents of a combined object from one iteration to the next one. In
practice, the commands
\begin{verbatim}
  set main.fastsim.algorithm    = cdfmidpoint
  set main.fastsim.radius       = 1.0
  set main.fastsim.overlap      = 0.5
  set main.fastsim.seed         = 1.0
  set main.fastsim.areafraction = 1.0
\end{verbatim}
switch on the CDF midpoint algorithm with all parameters manually fixed to their
default values, while the commands
\begin{verbatim}
  set main.fastsim.algorithm = cdfjetclu
  set main.fastsim.radius    = 1.0
  set main.fastsim.overlap   = 0.5
  set main.fastsim.seed      = 0.0
  set main.fastsim.iratch    = 0
\end{verbatim}
switch on the CDF jet clustering algorithm in its default configuration.

Finally, the grid jet algorithm starts by defining a grid in rapidity and
azimuthal angle, and next combines the particles lying in a common grid
cell. The user can fix the maximum allowed value for the rapidity, in absolute
value, as well as the grid spacing via the \verb+ymax+ and \verb+spacing+
attributes of the \verb+main.fastsim+ object. The default values are
respectively 3 and 0.1. The default configuration could equivalently be
obtained by issuing, in the \MA\ command line interface, the commands
\begin{verbatim}
  set main.fastsim.algorithm = gridjet
  set main.fastsim.ymax      = 3
  set main.fastsim.spacing   = 0.1
\end{verbatim}
While algorithms dedicated to large-radius jets are more and more widely used
those days, they are not available in \MA\ yet. This is left for future
developments.

As for any analysis, the clustering is actually performed when the \verb+submit+
command is typed, once the jet algorithm and all its options have been properly
configured. This results in the generation of the corresponding C++ code, its
compilation and its execution on the input event sample(s).

\subsubsection{Detector simulation options}
In its current v1.6 version, \MA\ allows the user to simulate basic detector
effects through efficiencies given as floating-point numbers. Inclusion of
more realistic efficiency functions depending on the transverse momentum and the
pseudorapidity of the particles is currently not possible.

The impact of the detector on the reconstruction
of hadronic taus can be implemented by including a tagging efficiency together
with a mistagging rate of a light jet as a tau lepton. This is achieved in
practice via the self-explanatory commands
\begin{verbatim}
  set main.fastsim.tau_id.efficiency = 0.6
  set main.fastsim.tau_id.misid_ljet = 0.01
\end{verbatim}
where one sets the efficiency of correctly identifying a hadronic tau to 60\%,
the default value being 100\%, and the mistagging rate of a light jet as a
hadronic tau to 1\%, the default value being 0.

The identification of jets originating from the fragmentation of $b$-quarks as
$b$-jets can be included in a similar fashion, the user being allowed to include
a tagging efficiency together with the mistagging rates of charmed and lighter
jets as $b$-jets. Behind the scenes, the tagging procedure works in two steps.
First, it tries to match one of the reconstructed jets to each $B$-hadron in
using the Monte Carlo truth information. The algorithm determines whether there
is a reconstructed jet in a cone of radius $\Delta R$ centered on each
$B$-hadron. If a jet is found, it is considered as $b$-tagged, up to an
efficiency that is specified by the user. This is achieved in practice by
typing, in the \MA\ command line interface,
\begin{verbatim}
  set main.fastsim.bjet_id.matching_dr = 0.2
  set main.fastsim.bjet_id.efficiency  = 0.6
  set main.fastsim.bjet_id.exclusive   = true
\end{verbatim}
The first command sets the size of the cone $\Delta R$ to 0.2 (through the
\verb+bjet_id.matching_dr+ attribute of the \verb+main.fastsim+ object), whilst
the second one fixes the tagging efficiency to 60\% (through the
\verb+bjet_id.efficiency+ attribute of the \verb+main.fastsim+ object). The last
command imposes that only a single $b$-jet can be matched with a given
$B$-hadron, as the \verb+bjet_id.exclusive+ attribute of the \verb+main.fastsim+
object has been set to true. Fixing this attribute to false would have allowed
the code to associate
any number of $b$-tagged jets to a specific $B$-hadron. The misidentification of
charmed and light jets as $b$-jets is implemented in the same way,
\begin{verbatim}
  set main.fastsim.bjet_id.misid_cjet  = 0.1
  set main.fastsim.bjet_id.misid_ljet  = 0.01
\end{verbatim}
the two attributes \verb+bjet_id.misid_cjet+ and \verb+bjet_id.misid_ljet+ of
the \verb+main.fastsim+ object being set to 10\% and 1\% to reflect that in
10\% and 1\% of the cases, a charmed jet and a lighter jet are mistagged as a
$b$-jet, respectively.

With the exception of the calorimetric segmentation, these feature allows to
reproduce the main functionalities of the PGS (Pretty Good Simulation)
program~\cite{pgs}.

\subsubsection{Saving the output in a file}
\label{sec:fastsim_out}

At the end of the reconstruction procedure, the event content is given in terms
of electrons, muons, photons, jets and hadronically-decaying taus. Moreover,
global transverse variables like the missing transverse energy are accessible as
well. \MA\ allows the user to save this information into a file either under a
simplified
Les Houches Events (\lhe) format, or under the LHC Olympics (\lhco) format. The
output file can moreover be optionally compressed in the case where the
{\sc ZLib} library has been detected by \MA. In order to save the reconstructed
output, the user has to type the command
\begin{verbatim}
  set main.outputfile = "output.lhe"
\end{verbatim}
in the \MA\ command line interface. The format of the output file is then
automatically chosen according to the extension of the file, that is
stored in the \verb+<wdir>/Output/<set>/lheEvents0_0+ directory. In this
schematic notation,
\verb+<wdir>+ stands for the working directory created by \MA\ and \verb+<set>+
denotes the label chosen for the dataset.

Files that are generated in the simplified LHE format are compliant with the
standardized LHE format syntax\cite{Boos:2001cv,Alwall:2006yp}. Each event is
encoded as an instance of an XML structure called \verb+event+. Each line of
an \verb+event+ block is then dedicated to the description of a (final-state or
not) object, and the corresponding syntax follows the scheme
\begin{verbatim}
 <ID> <ST> <MTH1> <MTH2> <IC1> <IC2> <PX> <PY> <PZ> <E> <M> <VT> <SP>
\end{verbatim}
Whereas for the case of an initial-state particle, \MA\ strictly follows the LHE
conventions, the latter are extended for final-state reconstructed objects.
The \verb+<ID>+ entry corresponds to a generalized PDG-code connected to
the nature of the object. \MA\ makes use of the 11, 13 and 15 codes for the
reconstructed electrons, muons and hadronic taus, respectively, whilst the
corresponding antiparticles are identified by the -11, -13 and -15 codes.
The PDG codes 22, 21 and 5 refer to a reconstructed photon, non-$b$-tagged jet
and $b$-tagged jet, respectively, and the missing energy is attached to
the code 12.

The \verb+<ST>+ entry is set to either -1, +1 or 3. The -1 value indicates that
an initial-state particle is described, while the +3 value is used for the
intermediate particles originating from the hard process. The +1 value finally
tags a reconstructed final-state object. The \verb+<MTH1>+ and \verb+<MTH2>+
entries denote the line numbers, in a given event record, of the hard-scattering
partons which the reconstructed object is matched with, and the \verb+<PX>+,
\verb+<PY>+, \verb+<PZ>+ and \verb+<E>+ entries stand for the components of the
particle four-momentum $(p_x, p_y, p_z, E)$ whose invariant mass is stored in
the \verb+<M>+ entry. The other options, \verb+IC1+, \verb+IC2+, \verb+VT+ and
\verb+SP+, are set to zero and not used by \MA.

When \MA\ is instructed to output the reconstructed events in the \lhco\
format\cite{lhco}, the output file turns out to be fully compliant with the
\lhco\ syntax (without the need to rely on a generalization of the format). A
given event record is hence comprised of a sequence of lines, the
first line, whose label equals zero, being a tag for the declaration of a new
event. Any other line in the event record corresponds to a final-state
reconstructed object and is encoded following the scheme
\begin{verbatim}
  <nl>  <typ>  <eta>  <phi>  <pt>  <jmas>  <ntrk>  <btag>  <had/em>
\end{verbatim}
In this notation, \verb+<nl>+ corresponds to the line number. It restarts
from zero at each new event that is as stated above identified in this way.
The \verb+<typ>+
entry reflects the nature of the reconstructed object, and is fixed to 0, 1, 2,
3, 4 and 6 for a photon, electron, muon, hadronically-decaying tau, jet and
missing transverse energy, respectively. The four-momentum of the object is
encoded via its pseudorapidity (\verb+<eta>+), azimuthal angle (\verb+<phi>+)
and transverse momentum (\verb+<pt>+) and the invariant mass is additionally
given via the \verb+<jmas>+ entry.

The next three entries have different meanings according to the nature of the
reconstructed objects. In the case of a jet, \verb+<ntrk>+ refers to the number
of tracks constituting the object, whereas for an electron or a muon, it
represents its electric charge. In the case of a hadronically-decaying tau,
\verb+<ntrk>+ stands
for the product of the number of constituents of the tau jet, times its
electric charge (the possible values being thus here $\pm1$ and $\pm 3$). For
the photons and missing energy cases, the \verb+<ntrk>+ variable is irrelevant
and set to zero. The \verb+<bjet>+ element has to be set to 1 or 0 for jets that
are $b$-tagged or not, respectively, whilst it refers to the line number of the
closest jet in cases of muons. It is fixed to zero for all the other objects.

Finally,  the \verb+<had/em>+ entry represents the ratio of the energy deposited
by the object in the hadronic calorimeter to the one deposited in the
electromagnetic calorimeter. It is smaller than 1 for electrons and photons,
larger than 1 for jets and irrelevant in all the other cases (and thus set to
0). This entry is however recycled for muons, for which it contains information
relative to the isolation. It is provided under the \verb+xxx.yyy+ format,
where \verb+xxx+ refers to the scalar sum of the transverse momentum of the
tracks lying in a given cone around the muon, excluding the central track,
and \verb+yyy+ stands for the ratio of the transverse energy present
in the cone to the muon transverse momentum.

\subsection[Realistic detector simulation with \DEL]
{Realistic detector simulation with \DELnorm}
\label{sec:delphes}

While simulating a detector by means of \FJ\ and \MA, as described in
Section~\ref{sec:fastjet}, is simple and efficient, it may not be realistic
enough in some cases. For this reason, \MA\ allows the user to make use of
\DEL\cite{deFavereau:2013fsa}, instead of \FJ, for simulating the response of a
typical collider experiment detector in a more detailed manner. In this section,
we first briefly recall the main features of the \DEL\ package
(Section~\ref{sec:delconcepts}) before detailing the new options that have been
developed in the \MA\ framework (Section~\ref{sec:delma5}). More practical
information on the running of \DEL\ within \MA\ and on the structure of the
output file are provided in the last two subsections (Section~\ref{sec:delrun}
and Section~\ref{sec:delout}).

\subsubsection[Main concepts of the \DEL\ program]{Main concepts of the
 \DELnorm\ program} \label{sec:delconcepts}

The \DEL\ package is dedicated to the simulation of the response of a generic
detector as used in typical high-energy physics collider experiments. It is
hence suitable for simulating in an approximate way not only the ATLAS, CMS and
LHCb detectors of the LHC, but also any detector of the next generation that
could be used in a future collider experiment such as the ILC or the FCC. The
architecture of \DEL\ is modular and the code makes a joint use of different
specialized modules that are each focusing on a specific aspect of the
simulation of a detector. The description of the detector, including the exact
definition of the \DEL\ modules that should be used, is provided by the user via
an input card in which all the modules involved in the simulation under
consideration are enumerated in the sequence following which they should be
called. Moreover, this modular structure easily allows for the implementation of
additional features through the design of new modules, as for instance shown
in Chapter~7 of Ref.~\cite{Fuks:2018yku} where the simulation of displaced
vertices is addressed.

The simulation of the ATLAS and CMS detectors is achieved by using both
parametric and algorithmic methods. First, tracking is simulated by applying
efficiency and smearing functions on the electrically-charged
final-state particles. The final-state objects are next propagated in the
detector electromagnetic field before the calorimetry is simulated. The
ensemble of electromagnetic and hadronic calorimeter energy deposits is derived
by dedicated \DEL\ modules, and one finally ends up with a collection of
tracks (resulting from the tracking simulation) and calorimeter towers
(representing the various energy deposits in the calorimeters). A jet-clustering
algorithm is then applied to cluster the calorimeter towers into jets, which
internally relies on the \FJ\ package. In the aim of improving the
identification of the reconstructed physics objects and the resolution on their
reconstructed four-momenta, \DEL\ additionally calls a particle flow algorithm
that combines both tracker and calorimetric information. As a next
step in the simulation of the detector, lepton and photon isolation is simulated
parametrically, and an algorithm finally takes care of the removal of all
possible overlaps among the reconstructed object. This ensures that a specific
object cannot end up in several collections of reconstructed physics objects, as
this could for instance occur for electrons that can in principle be
reconstructed both as jets and as electrons. The output is stored in a \ROOT\
file.

In addition, \DEL\ offers a way to simulate pileup effects by superimposing to
each event that is read from the input file additional minimum-bias events.
These extra events feature a primary vertex that is different from the one of
the hard-scattering process and the user is required to provide a supplementary
Monte Carlo sample in which they can be found. In a second stage, \DEL\
mimics the action of a pileup removal algorithm. This includes both charged
track subtraction at the tracker level and the removal of
calorimeter deposits stemming from neutral hadrons by means of the jet area
method as implemented in \FJ. The pileup simulation finally includes a modeling
of the loss of performance of object isolation.

\subsubsection[Features of the \MA-\DEL\ interface]
{Features of the \MAnorm-\DELnorm\ interface}
\label{sec:delma5}

The interface of \MA\ with \DEL\ allows the user to access new functionalities
for simulating the response of the detector. These correspond both to new
modules that can be added to the \DEL\ input cards as well as to specific tunes
for tagging and reconstruction efficiencies.

Lepton and photon isolation requirements are often imposed offline, at the
analysis level, instead of at the time of the simulation of the detector (as
it is done by default in \DEL). This would allow, for instance, for the
optimization of the isolation requirements as a function of the analysis under
consideration. Such a task can be performed by tuning the \DEL\ card in the
following way. First, all calls to the isolation modules (the
\verb+PhotonIsolation+, \verb+ElectronIsolation+ and \verb+MuonIsolation+
modules) must be removed, together with the definitions of these modules that
are not necessary anymore. Next, one must turn off the procedure leading to
the removal of objects that are counted twice in different collections as this
only makes sense when isolation is imposed. This is achieved by removing from
the input card all call to the \verb+UniqueObjectFinder+ module, together with
the definition of this module. As a consequence of this last change, two other
standard modules of \DEL, the \verb+ScalarHT+ and \verb+TreeWriter+ modules are
impacted and must be modified accordingly. More precisely, the \verb+ScalarHT+
module, dedicated to the calculation of the $H_T$ variable reflecting the
hadronic activity in the event, has to be removed, and the calculation of
this observable is performed instead automatically by \MA\ at the analysis
level.
On the other hand, the \verb+TreeWriter+ module that is called for saving the
reconstructed event in the output \ROOT\ file has to be modified so that the
collections of non-isolated objects could be stored, instead of the collections
of isolated objects that are not defined anymore. More information on the
output file (and thus on the way to modify the \verb+TreeWriter+ module) is
provided in Section~\ref{sec:delout}.

The \DEL\ output file includes by default the collection
of all (Monte Carlo) hadronic and partonic particles, together with their
properties, prior to the simulation of the detector. This consequently makes the
output \ROOT\ file quite large.
The interface of \MA\ to \DEL\ addresses this issue by enabling a skimming of
the output file so that only Monte Carlo particles that match a reconstructed
object are stored.

On different grounds, there has been a very recent resurgence in the interest
for beyond the Standard Model theories featuring long-lived particles, like
$R$-parity-violated supersymmetric realizations. In these models, event
topologies exhibiting displaced secondary vertices are not uncommon. Related
developments in the \MA\ interface to \DEL\ have started, and the current
version v1.6 of the code can handle displaced leptons.

Displaced leptons are usually associated with tracks pointing towards secondary
vertices that are displaced by a macroscopic distance from the primary vertex.
\MA\ offers the possibility to
include, in the \DEL\ input card, a simulation of the reconstruction of these
displaced tracks via a reconstruction efficiencies. The latter can be
implemented as any efficiency function of the \DEL\ card. They need to be
provided within two new modules, \verb+ElectronTrackingEfficiencyD0+ and
\verb+MuonTrackingEfficiencyD0+, whose names are self-explanatory. These methods
depend on the impact parameter of the displaced vertex $\mathbf{d} = (d_x, d_y,
d_z)$ that can be accessed via the \verb+d0+ (with $d_0 = \sqrt{d_x^2+d_y^2
+d_z^2}$)
and \verb+dz+ ($d_z$) variables as well as on its position $\mathbf{X} = (x, y,
z)$ that can be accessed through the \verb+xd+, \verb+yd+ and \verb+zd+
variables.

In the following, we refer to the \DEL\ version including all these changes as
{\it MA5-tune} of the \DEL\ card, which should not be confused of the old
deprecated MA5tune hacked version of \DEL\ detailed in Section~\ref{sec:pad}.

\subsubsection[Running \DEL\ from \MA]{Running \DELnorm\ from \MAnorm}
\label{sec:delrun}

In this section, we describe how to run \DEL\ from \MA, and as in
Section~\ref{sec:fastjet}, we start from events
where parton showering and hadronization have been simulated. The user who
wishes to use \DEL\ within \MA\ must begin with checking whether both \ROOT\
(that is necessary for the correct running of \DEL) and \DEL\ are available and
detected by \MA. Whereas \ROOT\ must be installed externally, \DEL\ can be
installed from the \MA\ command line interpreter by typing in
\begin{verbatim}
  install delphes
\end{verbatim}

In order to make use of the \MA\ interface to \DEL, the user has to launch the
code in its reconstructed mode,
\begin{verbatim}
  bin/ma5 -R
\end{verbatim}
as any potentially outputted event file would contain reconstructed events. The
simulation of the detector with \DEL\ must then be activated, which is achieved
by typing in the \MA\ console the following command,
\begin{verbatim}
  set main.fastsim.package = delphes
\end{verbatim}
This acts on the \verb+fastsim+ attribute of the \verb+main+ \MA\ object and
configures it for running in the detector simulation mode. Next, the user needs
to select a detector configuration card. This is performed by setting
up the \verb+detector+ attribute of the \verb+main.fastsim+ object accordingly,
\begin{verbatim}
  set main.fastsim.detector = <my-delphes-input-card>
\end{verbatim}
where \verb+<my-delphes-input-card>+ refers to a keyword defining the \DEL\ card
to use. One can pick either the official \DEL\ ATLAS and CMS cards, the
associated keywords being \verb+atlas+ and \verb+cms+, or their corresponding
MA5tune versions (see Section~\ref{sec:delma5}) whose keywords are given by
\verb+cms-ma5tune+ and \verb+atlas-ma5tune+. They correspond to specific LHC
early Run~2 configurations including Run~1 $b$-tagging performances (as their
Run~2 counterparts were not publicly available at the corresponding release
time). At the moment of submission of the analysis, the user has
the option to further edit the chosen card according to his/her needs.

Pileup simulation is by default disabled when \DEL\ is run from \MA. It can
however be included easily. To this aim, the user has to provide the path to a
sample of minimum-bias events, so that such events will be superimposed by \DEL\
to each hard-scattering event that is analyzed. This file must carry the
\verb+.pileup+ extension and satisfy the requirements presented in the \DEL\
documentation. In practice, pileup simulation is activated by tying, in the \MA\
command line interface,
\begin{verbatim}
  set main.fastsim.pileup = <minimum-bias-events.pileup>
\end{verbatim}
The keyword \verb+<minimum-bias-events.pileup>+ refers to the path to the sample
of minimum-bias events that has to be used, and its location is stored in the
\verb+pileup+ attribute of the \verb+main.fastsim+ object.

Finally, \ROOT\ files can be stored or not according to the needs of the user,
which is controled by setting the \verb+output+ attribute of the
\verb+main.fastsim+ object to \verb+true+ or \verb+false+,
\begin{verbatim}
  set main.fastsim.output = <true-or-false>
\end{verbatim}

As mentioned in the previous section, the interface of \MA\ to \DEL\ offers a
way to skim the output file in order to make it lighter. The user can indicate
whether the largest (in terms of storage) physics object collections should be
stored. Tins includes the information on the energy flow (the collection being
labeled as \verb+eflow+), the information on the Monte Carlo truth (the
collection being labeled as \verb+genparticles+), the information on the energy
deposits in the calorimetric towers (the collection being labeled as
\verb+towers+) and the collection of tracks (labeled as \verb+tracks+). In order
to enable the respective storage of these collections, the attributes
\verb+skim_eflow+, \verb+skim_genparticles+, \verb+skim_towers+ and
\verb+skim_tracks+ of the \verb+main.fastsim+ object have to be set to
\verb+true+. The default behavior, that can also be configured by typing in the
command line interface
\begin{verbatim}
   main.fastsim.skim_eflow        = false
   main.fastsim.skim_genparticles = false
   main.fastsim.skim_towers       = false
   main.fastsim.skim_tracks       = false
\end{verbatim}
yields an output \ROOT\ that is not skimmed and contains thus all the
information.

In addition, the user has the opportunity to design an analysis to be performed
on the reconstructed level events, as detailed in Section~\ref{sec:nut}. The
execution of all these tasks is then achieved by typing the \verb+submit+
command. Before starting to analyze the events, \MA\ allows the user to edit the
\DEL\ card, whose syntax is automatically checked prior to the run. In the case
where the card would be not compliant with the \DEL\ requirements, the code
stops and a message is printed to the screen.

Finally, new physics event samples often
include non-standard final-state particles, and it is up to the user to modify
the \DEL\ card in order to indicate how \DEL\ should treat those particles. A
specific attention has to be paid to the \verb+ECal+ and
\verb+Hcal+ instances of the \verb+SimpleCalorimeter+ module and to the
\verb+PdgCodeFilter+ instance of the the \verb+NeutrinoFilter+ module.

\subsubsection{Description of the output file}
\label{sec:delout}
The MA5tune and standard \DEL\ cards yield output files compliant with a
different syntax. While the output of the detector simulation can always be
converted into an LHE or LHCO file (see Section~\ref{sec:fastsim_out}), we focus
in this section on the description of the outputted \ROOT\ file that is stored
in the \verb+<working-directory>/Output/<dataset-label>/RecoEvents0_0+
directory, the \verb+<working-directory>+ folder being the working directory
created by \MA\ on run time and \verb+<dataset-label>+ being the label of the
dataset that has been created.

Many of the collections that are saved in the \ROOT\ output file are common to
both the standard and MA5tune
format. This includes the list of Monte Carlo particles involved in the hard
process and the parton showering (that is stored in the \verb+Particle+ branch),
the collection of tracks (that is stored in the \verb+Track+ branch), the
collection of energy deposits in the calorimetric towers (that is stored in the
\verb+Tower+ branch), the three collections of particle flow objects (that are
stored in the \verb+EFlowTrack+, \verb+EFlowPhoton+ and
\verb+EFlowNeutralHadron+ branches), the collection of jets that would be
reconstructed by a perfect detector (that is stored in the \verb+GenJet+
branch), the missing transverse energy as it would be reconstructed by a perfect
detector (that is stored in the \verb+GenMissingET+ branch) and the actual
missing transverse energy (that is stored in the \verb+MissingET+ branch).

The collections of electrons, muons, photons and jets are in contrast different
in the standard and MA5tune \DEL\ format. Consequently, the names of the
collections in the output file are different. The standard \verb+Electron+,
\verb+Muon+, \verb+Photon+ and \verb+Jet+ branches correspond, in the MA5tune
case, to the \verb+ElectronMA5+, \verb+MuonMA5+, \verb+PhotonMA5+ and
\verb+JetMA5+ branches. This should be accordingly referred to within the
\verb+TreeWriter+ module of the \DEL\ card.

Finally, two extra collections can be stored, namely the pileup contamination
density (stored in the \verb+Rho+ branch) and a potential fat jet collection
(stored in the \verb+FatJet+ branch).

\subsection{Example: monotop phenomenology at the LHC}
\label{sec:example1}

In this section, we work out a simple example in which we study the
phenomenology associated with a specific new physics signal. We intend to pin
down the differences in the signal properties that could be expected when we
consider the simulation of either a perfect detector (object reconstruction by
using a jet algorithm and no detector simulation), a simplified detector (object
reconstruction with a jet algorithm including tagging efficiencies), or a
realistic detector (including the modeling of an LHC detector in \DEL).

\begin{figure}
  \centering \includegraphics[scale=1]{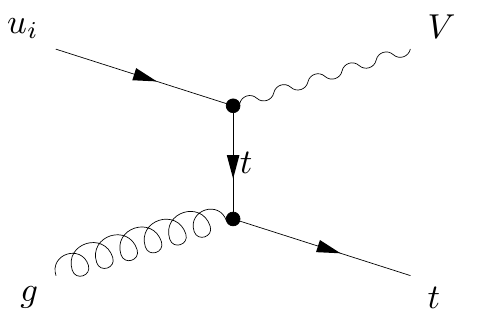}
  \caption{Feynman diagram illustrating the production of a monotop system made
    of an invisibly-decaying vector state $V$ and a top quark.}
  \label{fig:monotop0}
\end{figure}

We here focus on a monotop signal where a single top quark is produced in
association with missing energy originating from a particle decaying into dark
invisible states\cite{Andrea:2011ws},
\be
  p p \to V t \qquad\text{with}\qquad V \to {\rm invisible} \ ,
\ee
such a signal having been
actively searched for by both ATLAS and CMS collaborations\cite{Aad:2014wza,%
Khachatryan:2014uma,CMS:2017gin}. Monotop systems arise in particular when an
invisible boson $V$ is produced via its electrically-neutral flavor-changing
interactions with a top quark and a lighter up-type quark\cite{Andrea:2011ws,%
Davoudiasl:2011fj,Kamenik:2011nb,%
Alvarez:2013jqa,Agram:2013wda,Boucheneb:2014wza}, as illustrated in
Figure~\ref{fig:monotop0}. This boson can be seen as a mediator connecting the
Standard Model sector to a dark sector containing a potential dark matter
candidate.

Following standard conventions\cite{Agram:2013wda}, we describe the new physics
dynamics by the Lagrangian
\be
  \mathcal{L}_{\rm NP} = \mathcal{L}_{\rm kin}
    + V_\mu\ {\bar q_u}\ \gamma^\mu\ a_{\rm FC}\ q_u \,
    + {\cal L}_{\rm dark} + \ldots
\ee
where ${\cal L}_{\rm kin}$ contains kinetic and mass terms for all new
particles, ${\cal L}_{\rm dark}$ includes the interactions driving the $V$-boson
invisible decay and the dots stand for extra neutral flavor-changing
interactions of down-type quarks that are necessary to ensure electroweak gauge
invariance\cite{Boucheneb:2014wza}. The flavor-changing interactions of the $V$
boson with the up-type quarks are
described by the second term of the new physics Lagrangian ${\cal L}_{\rm NP}$,
in which all flavor indices are understood and $q_u$ denotes the up-type quark
field (in the mass basis). The coupling strength $a_{\rm FC}$ hence consists of
a $3\times 3$ matrix in generation space, and we consider a benchmark setup in
which only the up-top component is non zero, $a_{13}=a_{31}=0.1$.

Hard-scattering events at a collision energy of 13~TeV have been simulated with
\MG\cite{Alwall:2014hca}, and the simulation of the parton showering and the
hadronization has been included as implemented in \PY\cite{Sjostrand:2014zea}.
More precisely, we have convolved leading-order matrix elements with the
leading-order set of NNPDF parton densities version 3.0\cite{Ball:2014uwa}, the
latter being accessed via the LHAPDF library\cite{Whalley:2005nh,%
Buckley:2014ana}. We moreover make use of the \MS\
program\cite{Artoisenet:2012st} to enforce leptonic top quark decays, so that
the final-state signature is comprised of one hard lepton, one $b$-tagged jet
and missing energy,
\be
  p p \to t V \to \big(\ell\nu b\big)\ V \quad\equiv\quad
   \ell\ b\ \slashed{E}_T \ ,
\label{eq:signature}\ee
where $V$ is kept undecayed. Its invisible properties will be specified at the
level of the analysis. As above mentioned, we consider three different
ways to implement the simulation of the detector response.

In the case of an ideal detector, that we label by \verb+fastjet+ in the
following, we reconstruct the events by means of the anti-$k_T$ algorithm as
implemented in \FJ\ with a radius parameter set to $R=0.5$. Moreover, we only
retain in our analysis jets with a transverse momentum larger than 20~GeV,
\be
   p_T^{\rm (jet)} > 20~{\rm GeV} \ .
\ee
Following the guidelines of Section~\ref{sec:fastjet}, we instruct \MA\ that the
$V$ boson is invisible by adding its PDG code to the \verb+invisible+ container.
The $b$-tagging efficiency is furthermore assumed to be equal to 100\%, and the
matching of the reconstructed jets with the initial $B$ hadrons is performed by
relying on cones of radius $R=0.5$ centered on the $B$ hadrons. Such
reconstructed jets are often referred to as \textsl{GenJets}.

The second considered option for the detector simulation, denoted by
\verb"fastjet+simulation" in the following, is similar to the ideal detector
case, except that we include non perfect $b$-tagging performances. We mimic the
CSVM tagger of CMS\cite{Chatrchyan:2012jua} and hence include a $b$-tagging
efficiency of 60\% and misidentification rates of 20\% and 1\% for
charm-initiated and lighter jets, respectively.

Finally, for our last option, we consider \DEL\ and make use of the official
configuration card describing the CMS detector. The jet-to-parton matching is
performed like in the previous cases (as \DEL\ internally relies on \FJ),
although the algorithm only uses as inputs the final-state particles whose
transverse momentum $p_T$ and pseudorapidity $\eta$ satisfy
\be
  p_T > 1~{\rm GeV} \qquad\text{and}\qquad
  |\eta| < 2.5 \ .
\ee
We moreover indicate in the \DEL\ configuration file that the
$V$ boson does not hadronize and is invisible. We once again use the CSVM
$b$-tagger of CMS, and we implement the dependence of the efficiency and
mistagging rates on the jet transverse momentum and pseudorapidity on top of the
standard CMS detector description.

\begin{figure}
  \centering
  \includegraphics[width=0.90\columnwidth]{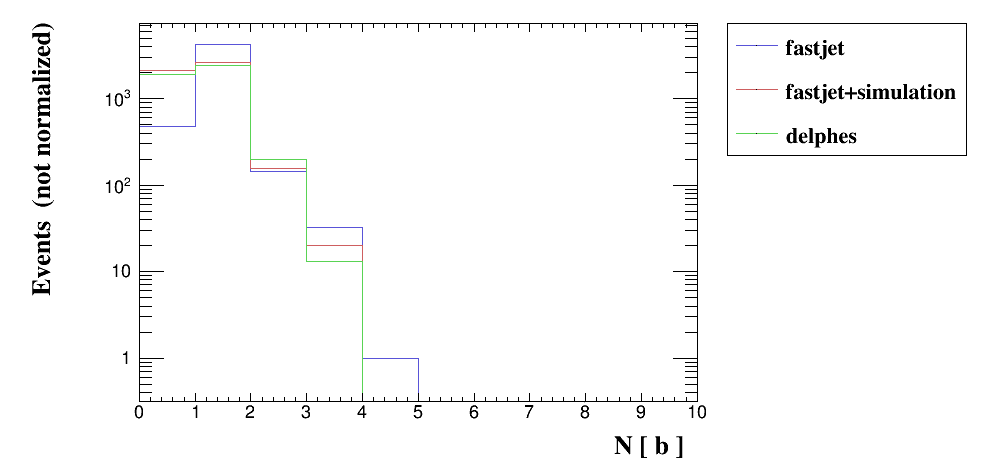}\\[.3cm]
  \includegraphics[width=0.90\columnwidth]{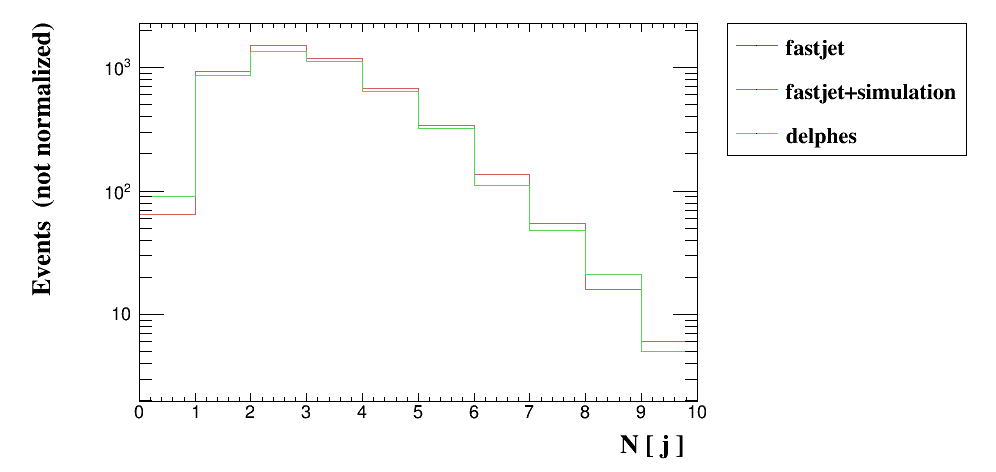}
  \caption{Properties of leptonic monotop events originating from the
    flavor-changing interactions of an invisible vector boson with the up-type
    quarks. We consider three detector configurations, namely an ideal detector
    (blue), a simplified detector (red) and a more realistic choice (green) and
    present the
    distribution in the $b$-tagged jet multiplicity (upper panel) and in the
    light jet multiplicity (lower panel).}
  \label{fig:monotop}
\end{figure}

\begin{figure}
  \centering
  \includegraphics[width=0.90\columnwidth]{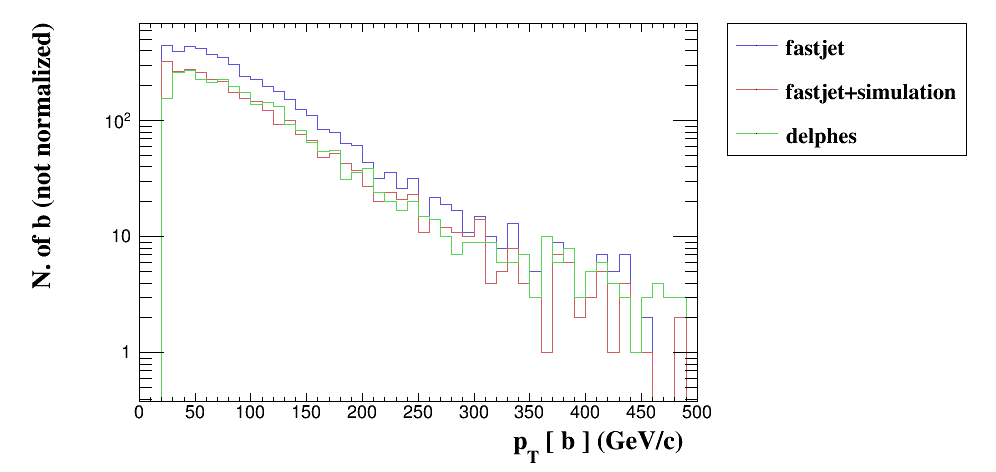}\\[.3cm]
  \includegraphics[width=0.90\columnwidth]{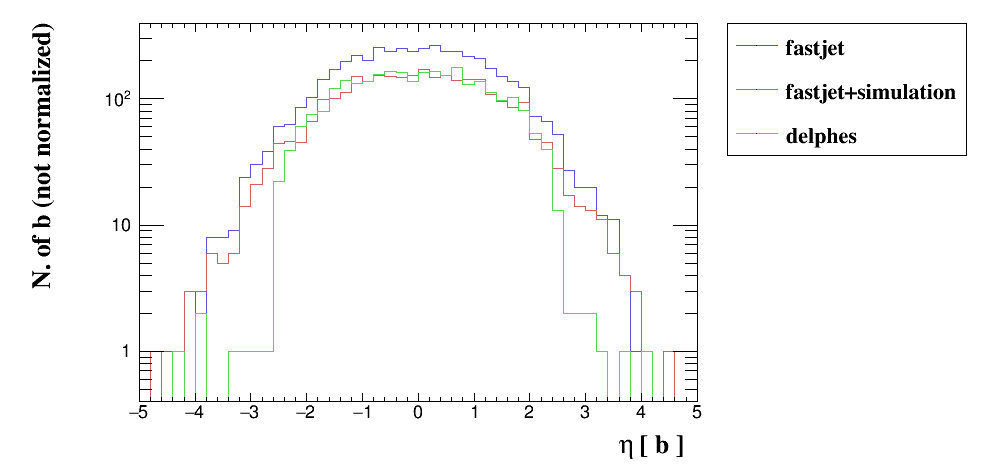}
  \caption{Same as in Figure~\ref{fig:monotop} but for the distribution in the
    transverse momentum of the $b$-tagged jets (upper panel) and in their
    pseudorapidity (lower panel).}
  \label{fig:monotop2}
\end{figure}

For each of the three cases, we store the output of the reconstruction as an
LHCO file, and we further analyze the three resulting files to produce the
comparative distributions shown in Figure~\ref{fig:monotop} and
Figure~\ref{fig:monotop2}. In Figure~\ref{fig:monotop}, we present the
$b$-tagged jet multiplicity (upper panel) and light jet multiplicity (lower
panel). Taking the ideal case as a reference (\verb+fastjet+,
blue curves), we can observe the effect of the non-perfect tagging efficiency in
the simplified (\verb"fastjet+simulation", red curves) and realistic
(\verb+delphes+, green curves) detector simulation cases. While in the ideal
detector setup, only jets issued from the fragmentation of $b$-quarks are
reconstructed (as expected from the signature of Eq.~\eqref{eq:signature}), the
situation changes in the two other configurations. A fraction
of the jets are mistagged and some of the $b$-jets are then
incorrectly identified as light jets. The highest-multiplicity bins turn out to
be less populated, the corresponding events being then largely accounted for in
the $N(b)=1$ and $N(b)=2$ bins. Although the gain mostly compensates the loss
for the $N(b)=2$ bin, the number of 1-$b$-jet events populating the 0-$b$-jet
bin is much larger. We moreover observe that the
simplified detector simulation agree very much with predictions including a more
realistic detector simulation. In Figure~\ref{fig:monotop2}, we investigate the
distribution in the transverse momentum (upper panel) and pseudorapidity (lower
panel) of the $b$-jets. A fairly good agreement is obtained between all three
cases, in particular after accounting for the small tagging efficiency of \DEL\
for large pseudorapidities.

\section{Reimplementing an LHC analysis with \MA} \label{sec:implementation}
  In this section, we describe how to implement a new analysis in the \MA\
framework. In order to make the information useful
for any potential contributor, we provide extensive details by means of a
specific example. Concretely, we focus on a CMS
search for dark matter in the mono-$Z$-boson channel\cite{Sirunyan:2017onm},
and we describe below all the steps that have been necessary to make this search
publicly available within \MA\ and its public analysis database\cite{inspire}.
The recast code is available from \MA\ version v1.6 onwards, and can
as well be directly downloaded from\\
\hspace*{2.5mm}
  \url{http://madanalysis.irmp.ucl.ac.be/wiki/PublicAnalysisDatabase}.


\subsection{Generalities}
The first step towards the implementation of a new analysis in the \MA\
framework consists of selecting an experimental search of interest,
understanding the related article and verifying that the analysis description is
complete. This last task means in particular that the analysis note must include
a clear and unambiguous definition of all the physics objects that have been
used. Object reconstruction is most of the time associated with efficiencies
that are functions of the object transverse momentum and pseudorapidity, and it
is therefore important to check that the related information is publicly
available. In addition, the selection steps defining the analysis signal regions
must be
properly identified, together with the sequence in which they are applied. While
the ordering is not relevant for getting correct final numbers, it is mandatory
for validation purposes, so that the implementation of each intermediate cut of
the analysis could be validated independently from each other. Furthermore,
attention should also be paid to triggers, event cleaning and other similar
issues that cannot generally be well reproduced with standard theory tools
addressing the modeling of a detector response. One must ensure that
these are kept under control, via, \eg, event reweighting.

It is not unlikely that a fraction of the
necessary information is missing, but experimental collaborations in general
provide additional material upon request.

A successful reimplementation of an
experimental analysis requires moreover a rigorous validation
procedure. This is essential but often the most complicated part of the
whole process, and it can usually be achieved only after several
fruitful exchanges with the experimenters involved in the analysis of interest.
The validation of the reimplementation of a specific analysis is
achieved by the comparison of theoretical predictions, for a specific set of
new physics signals, with official experimental numbers. The Standard Model
predictions are less important as official experimental numbers are
available for the backgrounds, so that they can be directly used for physics
purposes. This however assumes that signals only affect signal regions, and do
not contaminate the control regions allowing for a proper background extraction
(which should thus in this case considered as signal regions as well).

For being able to compare theoretical and experimental results for given new
physics signals (or equivalently to validate the reimplementation of an
analysis), one needs to start with well-defined benchmark scenarios
and generate signal events both at the level of the experimental software and
within theory tools. The choice of the benchmark scenarios to consider usually
originates from our experimental colleagues. Although the simulated events are
generally not available, the experimentalists generally provide, for
specific new physics setups, detailed cutflows and intermediate differential
distributions extracted from the Monte Carlo simulations. The theoretical
predictions that must be compared with those results are then also
achieved by using Monte Carlo programs, after making sure that the event
generation process reproduces the experimental simulation chain as much as
possible. This can be achieved, \eg, by sharing the Monte Carlo configuration
files that have been used experimentally. Using the same Monte
Carlo configuration files implies that the only difference in the experimental
and theoretical approach lies at the level of the modeling of the detector
effects.
This introduces genuine differences that must be kept under control, which will
be quantitatively determined by the validation procedure.

Different options exist for unambiguously defining a new physics benchmark
scenario. One could either pass the benchmark definition following a standard
text-based format inspired by the Supersymmetry Les Houches Accord
(SLHA)\cite{Skands:2003cj,Allanach:2008qq}, such a format being commonly used
by all Monte Carlo programs relying on the UFO model
format\cite{Degrande:2011ua}, or one could instead directly provide the
benchmark description via the configuration files of the employed
Monte Carlo program that could then be run immediately.
The precise choice of the new physics scenarios to consider is
irrelevant as we only need to compare numbers for given arbitrary theory
contexts. It is however easier to stick to physics-motivated choices for which
the information necessary for the validation procedure is in general easier to
obtain from the experimental collaborations as it already exists either
privately or publicly. In the best case scenario, all the material is readily
available from {\sc HepData} and the relevant analysis CMS or ATLAS wiki pages,
possibly together with snippets of codes detailing the implementation of
complicated non-standard kinematical variables.

In the next sections, we detail all the steps above-described in the
context of the CMS-EXO-16-010 search for dark matter in the mono-$Z$-boson
channel. We begin in Section~\ref{sec:pad} with instructions on how to install
a local copy of all analyses currently reimplemented in \MA, and how to add a
new blank analysis within this local installation. This blank analysis serves as
a skeleton for the future recasted code. We move on, in Section~\ref{sec:init},
with the description of the way in which an analysis code should be initialized.
We focus on how to
properly reference the reimplementation and why this is important, and on
how to declare signal regions, histograms and selection cuts. The implementation
of the analysis itself, that contains the definition of the physics objects, the
application of the selection cuts and the filling of the histograms, is
detailed in Section~\ref{sec:core}, and the configuration of the \DEL\
card corresponding to the CMS-EXO-16-010 analysis is described in
Section~\ref{sec:delphes_config}. Limit setting requires to use both data and
information on how the Standard Model background populates the different signal
regions of the analysis. This information is provided through an XML file whose
implementation is detailed in Section~\ref{sec:xml-info}.
The validation of the CMS-EXO-16-010 reimplementation in \MA\ is discussed in
Section~\ref{sec:validation}, and we finally explain in Section~\ref{sec:submit}
how to submit a validated recasted code to \INSP\ and how to add the
associated information to the online analysis database of \MA.


\subsection[The physics analysis database of \MA\ and the generation of a
blank analysis]{The physics analysis database of \MAnorm\ and the generation of
a blank analysis}\label{sec:pad}

Before starting implementing any new analysis, it is useful to install locally
all analyses that have been implemented in the \MA\ framework so far. This
allows one to get several
examples of validated codes and hints on the structure of the program.

The database of analyses that have been embedded in \MA\ currently contains (on
August 1$^{\rm st}$, 2018)
8 ATLAS and 10 CMS searches for new physics in LHC Run~I data, as well as 6
ATLAS and 7 CMS analyses from the
Run~II searches. Those analyses can be classified into two sets, according
to the version of \MA\ they are compatible with. The oldest analyses of the
database have been implemented within \MA\ versions prior to v1.2. While they
can be run within the current v1.6 version of the program without any problem,
these analyses rely on the \textit{MA5tune} version of \DEL~3\ for the
simulation of the detector\cite{deFavereau:2013fsa,Dumont:2014tja} (which
contrasts with the \MA-tuned detector cards of the official \DEL\ version
mentioned in Section~\ref{sec:delphes}). This
\DEL-{\it MA5tune} package has been released in 2014 and thus relies on the
\DEL\ version of that time, which is only compatible with \ROOT~5.
It has been tuned so that object isolation
requirements can be implemented at the level of the analysis instead of at the
level of the simulation of the detector by means of standard \MA\
methods\cite{Conte:2014zja}. In addition, a skimming of the \DEL\ output has
been enforced to reduce the size of the output files. From \MA\ version v1.2
onwards, standard releases of \DEL\ are used instead, as these versions support
\ROOT~6 and include all the features that were requiring a tuning at the time
of the \MA\ version v1.1.x series.

The two sets of analyses can be installed locally and separately by typing in
the \MA\ interpreter
\begin{verbatim}
  install PAD
  install PADForMA5tune
\end{verbatim}
The second command is related to the old analyses compliant with the {\it
MA5tune} version of \DEL\ and is automatically ignored by \MA\ when a \ROOT~6
installation is used. As the support for such analyses is planned to be
interrupted in a near future, we recommend, when developing a new recasting
code, to use the latest version of \MA\ and the `PAD' framework, and not the
more ancient `PADForMA5tune' framework.

All analyses to be included in the PAD must currently rely on the expert mode
of \MA. Although up-coming developments could alleviate this limitation, the
capabilities of the \python\ metalanguage is too limited, at least in \MA\
version 1.6, to allow for the full reimplementation of an LHC analysis in
\python. The normal mode of the program can however be used to create a
skeleton for any analysis to be implemented in the expert mode, as described
below, so that it is not necessary to start from scratch.

In order to start implementing a new analysis (in the PAD framework), an empty
analysis has first to be created. This blank analysis code will be updated in a
second step (see the next sections). The
creation of the empty analysis is achieved by running the script
\verb+newAnalyzer.py+
located in the \verb+PAD/Build/SampleAnalyzer+ directory. The only two required
arguments of this script consist of the name of the analysis of interest and the
name of the associated C++ classes that will be designed. Taking them to be
the same and fixed to \verb+cms_exo_16_010+ for the example under consideration,
the creation of the empty analysis is performed by typing in a shell, from the
\verb+PAD/Build/SampleAnalyzer+ directory,
\begin{verbatim}
  python newAnalyzer.py cms_exo_16_010 cms_exo_16_010
\end{verbatim}
The choice of this \verb+cms_exo_16_010+ name follows the standardized naming
scheme employed by the CMS collaboration for the analysis identifiers.

As a result, the C++ header file \verb+analysisList.h+ present in the
\verb+PAD/Build/SampleAnalyzer+ folder is updated, an entry associated
with the new analysis being now present. Moreover, new C++ source and header
files (\verb+cms_exo_16_010.h+ and \verb+cms_exo_16_010.cpp+) containing the
empty analysis have been created and are located
in the \verb+PAD/Build/SampleAnalyzer/User/Analyzer+ folder, together with all
pairs of C++ source and header files related to the other reimplementations
already included in the PAD. These two new files will be modified as
detailed in the below sections.

After the creation of the new analysis files, it is necessary to recompile the
PAD to include the
changes at the level of the executable. To this aim, the environment variables
must be set accordingly in order for \MA\ to properly run in its expert mode.
This can be automatically done by running the setup script available in the
\verb+PAD/Build+ directory. This script can be run either from a
\verb+bash+ shell or from a \verb+tcsh+ shell by typing in one of the commands
\begin{verbatim}
  source setup.sh     source setup.csh
\end{verbatim}
The \verb+PAD/Build+ directory also contains a \verb+Makefile+, so that the
PAD executable can be created via the usual command,
\begin{verbatim}
  make
\end{verbatim}
The obtained executable is named \verb+MadAnalysis5job+ and can be found in the
\verb+PAD/Build+ folder. It originates from the
merging of the \MA\ core libraries, all analysis available within the PAD
and a main program. The libraries contain all the methods embedded in
the {\sc SampleAnalyzer} core of \MA\ for which we refer both to the manual for
their description\cite{Conte:2014zja}, as well as to the reference card of
\ref{app:expert}. The analyses for which a C++ code is available in
the \verb+PAD/Build/SampleAnalyzer/User/Analyzer+ directory are all included
in the executable, and the
source file of the main program is the \verb+PAD/Build/Main/main.cpp+ file. The
latter has been designed to first initialize all analyses, next execute them
over one or more event samples to be specified by the user, and finally store
the results in the \verb+Output+ directory. More information on the format of
the results can be found in Section~\ref{sec:outputruns}.

In the next subsections, we detail how to modify the empty analysis that has
just been created, \ie, the \verb+cms_exo_16_010.cpp+ file, in order to
incorporate the mono-$Z$-boson analysis under consideration. This C++ source
file contains three methods named \verb+Initialize+, \verb+Execute+ and
\verb+Finalize+ that will have to be provided and that are related to the
initialization of the analysis code, its execution, and the generation of
the results, respectively. As the finalization step is automated, nothing has to
be done from the user standpoint so that we focus, in the following subsections,
on the analysis initialization method (Section~\ref{sec:init}) and execution
method (Section~\ref{sec:core}).


\subsection{Analysis initialization}\label{sec:init}

The \verb+Initialize+ method of any analysis contains the declaration of all
signal regions, histograms and selections that are used in the considered
analysis, as well as a header printout that is displayed by \MA\ on run
time.

\subsubsection{Header printout}
Although the implementation of an header printout in a recasted analysis is
not mandatory, this step is important for
traceability reasons. Potential users hence have information on the author of
the recasted code, which kind of analysis has been reimplemented and what are
the relevant publications. For the considered example, we include, at the
beginning of the \verb+Initialize+ method, the following lines,
\begin{verbatim}
 INFO << "<><><><><><><><><><><><><><><><><><><><><><><><>" << endmsg;
 INFO << "<> Analysis: CMS-EXO-16-010, arXiv:1701.02042 <>" << endmsg;
 INFO << "<>           (mono Z-boson)                   <>" << endmsg;
 INFO << "<> Recaster: Benjamin Fuks                    <>" << endmsg;
 INFO << "<> Contact:  fuks@lpthe.jussieu.fr            <>" << endmsg;
 INFO << "<> Based on MadAnalysis 5 v1.6                <>" << endmsg;
 INFO << "<> DOI: 10.7484/INSPIREHEP.DATA.RK53.S39D     <>" << endmsg;
 INFO << "<><><><><><><><><><><><><><><><><><><><><><><><>" << endmsg;
\end{verbatim}
The first lines give information on the recasted analysis, the next ones on the
author of the code and the last three lines refer to the material to be cited if
this analysis is employed for a phenomenology work. In this snippet of code, we
make use of the \verb+INFO+ message service of the {\sc SampleAnalyzer} core.

\subsubsection{Signal region declaration}

The initialization method contains the declaration of all the signal regions of
the analysis. The CMS-EXO-16-010 search under consideration focuses on the
mono-$Z$-boson signature where one leptonically-decaying $Z$-boson is produced
in association with missing energy. The analysis correspondingly contains a pair
of signal regions that only differ by the electronic or muonic nature of the two
leptons issued from the $Z$-boson decay. We arbitrarily name the two regions
\verb+ee+ (for the
electron channel) and \verb+mumu+ (for the muon channel), and they are declared
in the code as
\begin{verbatim}
  Manager()->AddRegionSelection("ee");
  Manager()->AddRegionSelection("mumu");
\end{verbatim}
These C++ commands rely on the \verb+AddRegionSelection+ method of the
\verb+RegionSelectionManager+ class, whose an instance called \verb+Manager()+
is automatically available within any given analysis. Other regions could be
declared by making use of the same syntax, with different region names. The
string
name of a declared region is mandatory as it acts as an identifier both in the
analysis code itself and at the level of the output generated by the execution
of the recasted code. Internally to the code, the region identifiers are used to
associate selection cuts and histograms with the different regions. This
association is mandatory as any specific signal region is defined by an
ensemble of cuts that must be connected to it. Moreover, a histogram
representing the distribution in a particular observable depends on the
previously applied cuts, so that a given distribution could be significantly
different in different regions.

\subsubsection{Selection cut declaration}

The way in which \MA\ treats selection cuts requires to first declare them in
the \verb+Initialize+ method of the analysis and then implement their actual
application in the \verb+Execute+ method of the analysis. Cut declarations
consist in naming the cuts through strings and linking each of them to the
signal region(s) to which they should respectively be applied. This is achieved
with the AddCut method of the RegionSelectionManager class. This method
requires, as a first mandatory argument, a string name that uniquely
identifies the considered
cut. A second optional argument could be specified, and is either
the name of a region or an array of region names in the case where a given cut
is related to several regions. If this second argument is absent, the cut is by
default associated with all declared regions. For instance, the following
commands, that have to be considered in the context of a generic analysis (and
not within the CMS-EXO-16-010 example worked out here),
\begin{verbatim}
  Manager()->AddCut("1lepton");
  Manager()->AddCut("2leptons", "SR1");
  std::string SRs[] = {"SR1", "SR2"};
  Manager()->AddCut("3leptons", SRs);
\end{verbatim}
declare three selection cuts, \verb+Manager()+ being, as indicated in the
previous section, the instance of the \verb+RegionSelectionManager+ class
attached automatically with any analysis embedded in \MA. The first cut is named
\verb+1lepton+ and is
common to all the regions declared in the analysis, whereas the second cut,
named \verb+2leptons+, is only applicable in the case of a region named
\verb+SR1+. The last cut, denoted by
\verb+3leptons+, is in contrast associated with the two regions \verb+SR1+ and
\verb+SR2+.

In the CMS-EXO-16-010 example, the two signal regions are defined by six
preselection and three selection cuts. All cuts are common to both regions,
with the exception of one of the preselection cuts that probes the electronic or
muonic nature of the leptons originating from the $Z$-boson decay.
The mono-$Z$-boson final state signature
steers the preselection where one requires the presence of two isolated leptons
(electrons or muons) of the same flavor and with an opposite electric charge.
Additional constraints are then enforced on the dilepton invariant mass, that
has to be compatible with the $Z$-boson mass, and on the dilepton transverse
momentum. This allows for the rejection of the bulk of the Drell-Yan
background. Moreover, events featuring a third looser lepton or a tau are
vetoed, which consequently reduces the diboson background, as are events
featuring a final-state $b$-tagged jet which are thus compatible with a
top-quark decay. Those preselection cuts are declared in the \verb+Initialize+
method of the \verb+cms_exo_16_010+ class as
\begin{verbatim}
  Manager()->AddCut("2_electrons", "ee");
  Manager()->AddCut("2_muons", "mumu");
  Manager()->AddCut("on-Z");
  Manager()->AddCut("dilepton_pt");
  Manager()->AddCut("3rd_lepton_veto");
  Manager()->AddCut("b_veto");
\end{verbatim}
where we have chosen the string names comprehensibly. With
the exception of the first two cuts that are region-dependent, we have never
specified the second argument when calling the \verb+AddCut+ method as the cuts
apply to all regions.

The final event selection is based on further kinematic requirements. The
missing transverse momentum is enforced to be in a different hemisphere as the
momentum of the reconstructed $Z$-boson, the momentum balance of the event is
constrained to be small and the presence of a minimal amount of missing energy
is imposed. Moreover, the analysis furthermore requires that at most one jet is
present. Such four cuts are declared in the same manner as above,
\begin{verbatim}
  Manager()->AddCut("dphi(met,Z)");
  Manager()->AddCut("momentum_balance");
  Manager()->AddCut("met");
  Manager()->AddCut("at_most_one_jet");
\end{verbatim}
where the cuts identifiers are once again chosen comprehensibly.

\subsubsection{Histogram declaration}

Histograms are declared very similarly to cuts. The \verb+AddHisto+ method of
the \verb+RegionSelectionManager+ class is this time used, its argument being
respectively a name, the number of bins of the histogram, and the upper and
lower bounds defining the range of its $x$-axis. By default, a histogram is
associated with all
declared signal regions but this behavior can be modified by specifying an
additional optional argument that could be either a region identifier, or an
array of region identifiers.

The CMS-EXO-16-010 analysis note includes the missing energy distribution in the
electron and muon channels after the preselection, and after the entire
selection. We consequently declare these four histograms in the
\verb+Initialize+ method as

\begin{verbatim}
  Manager()->AddHisto("MET_preselected_e" ,14,80,1200, "ee");
  Manager()->AddHisto("MET_preselected_mu",14,80,1200, "mumu");
  Manager()->AddHisto("MET_selected_e" ,14,80,1200, "ee");
  Manager()->AddHisto("MET_selected_mu",14,80,1200, "mumu");
\end{verbatim}
As for the cuts, we use self-explanatory identifiers for the different
histograms for comprehensibility purposes, and each histogram is associated with
the relevant signal region (\verb+ee+ or \verb+mumu+). Although the histograms
provided in the analysis note have a variable bin size, the histograms that we
have declared feature bins of a well-defined size. The histograms that will be
generated by \MA\ will then have to be mapped to those of the CMS analysis note
for a comparison purpose. This implies to merge the last seven bins of the \MA\
histograms into two larger bins.


\subsection{Analysis core}
\label{sec:core}

The main part of the reimplementation work consists of the writing of the
analysis core. This task corresponds to the translation, in the \MA\ language,
of the experimental publication of interest.

\subsubsection{Initialization}\label{sec:core-init}

\MA\ internally handles regions and cuts in a way allowing one to avoid having
to test any specific condition several times, which may occur when given cut
conditions are nested and connected to multiple regions. Once all the cuts have
been properly initialized (see Section~\ref{sec:init}), they are applied by
calling, within the \verb+Execute+ function of the analysis class being
implemented, the \verb+ApplyCut+ method of the
\verb+RegionSelectionManager+ class,
\begin{verbatim}
  Manager()->ApplyCut(condition, cut-name)
\end{verbatim}
This method requires two arguments, a boolean object (denoted by
\verb+condition+ in the above snippet of code) and a string (denoted by
\verb+cut-name+ in the above snippet of code). The former indicates whether the
cut condition is true or false, or in other words, whether the currently
analyzed event satisfies the cut. The latter identifies which cut is considered,
and the string name to be used must be one of those previously declared in the
\verb+Initialize+ method of the analysis class (see Section ~\ref{sec:init}).

As a
result of the call to the \verb+ApplyCut+ method, all the signal regions
associated with the considered cut are cycled through. Regions that are
not surviving the preceding cuts are ignored, whereas those that are
still surviving get their cutflow updated according to the cut condition. More
precisely, the event weight is added to the corresponding entry in the cutflow
table provided the cut condition is realized, all cutflow entries being
initialized to zero at the beginning of any run of the code.

Both the surviving nature of each region and the associated cutflows
are internally handled. This requires a proper initialization stage at
the beginning of the treatment of every event, the latter being passed as the
argument \verb+event+ of the \verb+Execute+ method. This implies to
pay attention to the event weights, that can be retrieved via the
\verb+event.mc()->weight()+ method of the event class. A correct handling is
achieved by relying on the \verb+InitializeForNewEvent+ method of the
\verb+RegionSelectionManager+ class and by including, at the beginning of the
\verb+Execute+ method,
\begin{verbatim}
  double myWeight=0.;
  if(Configuration().IsNoEventWeight()) myWeight=1.;
  else if(event.mc()->weight()!=0.) myWeight=event.mc()->weight();
  else
  {
    WARNING << "Found one event with a zero weight. Skipping...\n";
    return false;
  }
  Manager()->InitializeForNewEvent(myWeight);
\end{verbatim}
This tests the presence of a weight in the event file by using the
\verb+Configuration().IsNoEventWeight()+ function, the weight being set to unity
if it cannot be found. If a weight is available, the correct weight value is
instead used. The \verb+InitializeForNewEvent+ method additionally tags
all regions as surviving. The stored value of
the weight is then used internally both for the update of the cutflows and when
histograms are filled. Moreover, in cases where the weight value is vanishing,
the \verb+Execute+ method exits and one can move on with the analysis of the
next event thanks to the last \verb+else+ block.

\subsubsection{Object definitions}
\label{sec:objects}

Once an event is read on run time, it is stored as an instance of the
\verb+EventFormat+ class named \verb+event+ that is passed as an argument of
the \verb+Execute+ method. The information on the objects that have been
reconstructed is stored and available via the \verb+event.rec()+ method, that
returns an instance of the \verb+RecEventFormat+ class. This
object contains five collections that represent the different physics
objects possibly reconstructed in a detector, namely electrons, muons, taus,
photons and jets. These are respectively stored in the \verb+electrons()+,
\verb+muons()+, \verb+taus()+, \verb+photons()+ and \verb+jets()+ vectors.

Not all reconstructed objects are used in a typical experimental
analysis, but only a subset of them whose properties satisfy certain criteria.
For instance, the CMS-EXO-16-010 analysis only relies on electrons,
muons, taus and jets, and more precisely on the actual number of such objects
with definite kinematical properties.
Each selected electron (muon) is hence enforced to have a
transverse momentum $p_T^e$ ($p_T^\mu$) and pseudorapidity $\eta^e$ ($\eta^\mu$)
satisfying
\be\bsp
  p_T^e   > 20~{\rm GeV}\ , \qquad \big|\eta^e\big| < 2.5\ , \\
  p_T^\mu > 20~{\rm GeV}\ , \qquad \big|\eta^\mu\big| < 2.4\ ,
\esp\ee
which also ensures a trigger efficiency compatible with 1. In addition,
electrons lying in the transition region between the electromagnetic
calorimeter barrel and endcap are rejected, as the reconstruction procedure is
not optimal in this region of the detector. This corresponds to reject electron
candidates whose
pseudorapidity satisfies \mbox{$1.44< |\eta^e| < 1.57$}. Moreover, events
featuring extra electrons or muons with a $p_T$ larger than 10~GeV, or isolated
hadronic tau candidates with a $p_T$ larger than 20~GeV, are vetoed.

The above requirements can all be implemented in the \verb+Execute+ method.
Starting with electrons, the analysis requires two types of electrons,
signal electrons that fulfil tighter requirements and that are associated with
the reconstruction of the final-state $Z$-boson, and looser electrons that are
related to the veto of events featuring a third lepton. The corresponding \MA\
code therefore includes the electron definitions,
\begin{verbatim}
  std::vector<const RecLeptonFormat*> SignalLeptons, LooseElectrons;
  for(unsigned int ii=0; ii<event.rec()->electrons().size(); ii++)
  {
    const RecLeptonFormat *myElec = &(event.rec()->electrons()[ii]);
    double eta = std::abs(myElec->eta());
    double pt = myElec->pt();
    double iso_var = PHYSICS->Isol->eflow->sumIsolation(myElec,
        event.rec(),0.4,0.,IsolationEFlow::ALL_COMPONENTS);

    if( (eta>1.44) and (eta<1.57) ) continue;
    if(eta>2.5) continue;

    if(iso_var>0.15*pt) continue;

    if (pt>20)  SignalLeptons.push_back(myElec);
    if (pt>10)  LooseElectrons.push_back(myElec);
  }
  unsigned int ne=SignalLeptons.size();
\end{verbatim}
In the first line, two containers are declared. The first one is dedicated to
the storage all signal leptons (both electrons and muons) whereas the second one
is dedicated to the storage of looser electrons. Loose electrons and muons must
be stored in different containers by virtue of the jet cleaning procedure that
needs to be implemented, as required by the way in which \MA\ works in
conjunction with \DEL\ (see below when jet definitions are discussed).

The above lines of code then include a loop over all event electrons, in which
one first
extracts a few pieces of information for each electron, namely its transverse
momentum (\verb+pt+), its pseudorapidity in absolute value (\verb+eta+) and the
amount of activity in a cone of radius $R=0.4$ centered on the electron
(\verb+iso_var+). The three \verb+if+ statements that follow
guarantee that one ignores all electrons lying outside the detector
acceptance or in the transition region between the barrel and the endcap of the
electromagnetic calorimeter,
and that all considered electrons are isolated. The isolation requirement is
implemented by imposing that the amount of activity in a cone of radius $R=0.4$
centered on the
electron is always lower than 15\% of the electron transverse momentum.

The two lepton containers are finally filled according to the $p_T$ value of the
remaining electron candidates, signal and loose electrons having a $p_T$ larger
than 20 and 10~GeV, respectively. In the way the code has been implemented, any
signal electron is also considered, simultaneously, as a loose electrons. The
selection cuts will be designed in an appropriate manner.
The code ends with the calculation of the number of
signal electrons (\verb+ne+) that is an information that is necessary for the
implementation of the relevant selection cut in Section~\ref{sec:cuts}.

The definition of the muon candidates is similar, the pseudorapidity selection
thresholds being however slightly different,
\begin{verbatim}
  std::vector<const RecLeptonFormat*> LooseMuons;
  for(unsigned int ii=0; ii<event.rec()->muons().size(); ii++)
  {
    const RecLeptonFormat *myMuon = &(event.rec()->muons()[ii]);
    double eta = std::abs(myMuon->eta());
    double pt = myMuon->pt();
    double iso_var = PHYSICS->Isol->eflow->sumIsolation(myMuon,
        event.rec(),0.4,0.,IsolationEFlow::TRACK_COMPONENT);

    if(eta > 2.4) continue;

    if(iso_var>0.20*pt) continue;

    if (pt>20)  SignalLeptons.push_back(myMuon);
    if (pt>10)  LooseMuons.push_back(myMuon);
  }
  unsigned int nmu = SignalLeptons.size() - ne;
  unsigned int nloose = LooseElectrons.size() + LooseMuons.size();
\end{verbatim}
Moreover, following the information available in the CMS article, the
implementation of
the isolation criterion relies this time only on the charged track activity
in a cone of radius $R=0.4$ centered on the muon direction, instead of any type
of surrounding activity as in the electron case. This quantity is
stored in the \verb+iso_var+ variable that is then required to be smaller than
20\% of the muon transverse momentum.

The last two lines
of the above code are dedicated to the calculation of the number of signal muons
(\verb+nmu+), and of the number of loose leptons, two quantities that are
again necessary for the implementation of the selection cuts in
Section~\ref{sec:cuts}.

Our example analysis includes a veto on events
featuring isolated hadronically decaying tau leptons with a transverse momentum
satisfying
\be
  p_T^\tau > 20~{\rm GeV} \ ,
\ee
where tau isolation restricts the activity in a cone of radius $R=0.4$ centered
on the tau candidate to be at most 20\% of the tau $p_T$. The implementation is
similar to what has been presented above, although we only store the number of
tau candidates satisfying the object definition criteria, and not the tau
themselves as they are not necessary for the implementation of a veto. This is
achieved with the following piece of code,
\begin{verbatim}
  unsigned int ntau = 0;
  for(unsigned int ii=0; ii<event.rec()->taus().size(); ii++)
  {
    const RecTauFormat *myTau = &(event.rec()->taus()[ii]);
    double pt = myTau->pt();
    double iso_var = PHYSICS->Isol->eflow->sumIsolation(myTau,
        event.rec(),0.4,0.,IsolationEFlow::ALL_COMPONENTS);

    if(iso_var<0.20*pt and pt>20)  ntau++;
  }
\end{verbatim}
where \verb+ntau+ is the variable that will be used at the level of the cut
implementation in the next section.

We move on with the definition of the jet objects used in the
analysis. This goes along the same lines as for the previous
cases. The signal regions are populated by events featuring at most one jet with
a transverse momentum $p_T^j$ and pseudorapidity $\eta^j$ fulfilling
\be
  p_T^j > 30~{\rm GeV}\ , \qquad \big|\eta^j\big| < 5\ .
\ee
Moreover, events exhibiting the presence of central $b$-tagged jets such that
\be
  p_T^b > 20~{\rm GeV}\ , \qquad \big|\eta^b\big| < 2.5\ ,
\ee
are vetoed. The implementation of the jet definition is thus standard,
\begin{verbatim}
  std::vector<const RecJetFormat*> SignalJets, BtaggedJets;
  for(unsigned int ii=0; ii<event.rec()->jets().size(); ii++)
  {
    const RecJetFormat * myJet = &(event.rec()->jets()[ii]);
    double eta = std::abs(myJet->eta());
    double pt = myJet->pt();

    if(eta > 5.) continue;

    if(pt>30.) SignalJets.push_back(myJet);
    if(pt>20. && eta<2.5 && myJet->btag())
      BtaggedJets.push_back(myJet);
  }
\end{verbatim}
the $b$-tagging information related to a given jet candidate being available via
the standard method \verb+btag()+ of the \MA\ jet class. As a result of the
above code, jet and $b$-tagged jet candidates are respectively stored in the
\verb+SignalJets+ and \verb+BtaggedJets+ containers.

Before evaluating the number of jets and $b$-tagged jets in the event,
an additional treatment must be implemented due to the internal way in which
the \DEL\ detector simulation is embedded within \MA. The reason is that
any electron candidate is also tagged as a potential jet candidate, and a
procedure yielding the removal of this double-counting has to be implemented.
This is also why loose electrons have been stored separately from
loose muons. This double-counting removal is achieved by making use of the
\verb+JetCleaning+ method available within \MA,
\begin{verbatim}
  SignalJets  = PHYSICS->Isol->JetCleaning(SignalJets,
     LooseElectrons, 0.2);
  BtaggedJets = PHYSICS->Isol->JetCleaning(BtaggedJets,
     LooseElectrons, 0.2);
\end{verbatim}
This leads to the removal, from the two jet containers, of any jet candidate
that has already been accounted for as a loose electron. More precisely, the
method compares two collections (for instance, the \verb+SignalJets+ and
\verb+LooseElectrons+ collections in the first call above). If two
objects are found to be separated by an angular distance in the transverse
plane smaller than a given threshold (taken to be $R=0.2$ in the above example),
they are considered as the same object and the corresponding instance of it in
the first container is removed.

As only the number of signal jets and $b$-tagged jets are necessary for the
implementation of the analysis strategy, the jet definition implementation ends
with the calculation of the corresponding quantities that are respectively
stored in the \verb+nj+ and \verb+nb+ variables,
\begin{verbatim}
  unsigned int nb = BtaggedJets.size();
  unsigned int nj = SignalJets.size();
\end{verbatim}

The last objects that are relevant for the implementation of the CMS-EXO-16-010
analysis consist of the missing transverse energy $\slashed{E}_T$ and the
missing transverse momentum $\slashed{\mathbf{p}}_T$. Those quantities are
computed in a standard way in \DEL\ and can be extracted directly within the
analysis code,
\begin{verbatim}
  MALorentzVector pTm = event.rec()->MET().momentum();
  double MET = pTm.Pt();
\end{verbatim}
where the \verb+MALorentzVector+ class, that is similar to the \ROOT\
\verb+TLorentzVector+ class, allows to define four-vectors in the \MA\
framework. The missing transverse momentum is here stored in the
\verb+pTm+ variable, and its norm, the missing transverse energy, in the
\verb+MET+ variable. Those quantities are directly derived from the event
file by \DEL, on the basis of the (inclusive) knowledge of the entire event.

\subsubsection{Cut definitions}\label{sec:cuts}
Once all the objects necessary for the analysis selection strategy are defined,
the next part of the recast code reflects the various selection cuts
themselves. It is important to emphasize that the ordering in which all cuts
have been declared in the analysis \verb+Initialize+ method must match the
ordering in which they are applied in order to get ordered and meaningful
cutflow charts in the output of the program execution.
As in the previous subsection, we focus on the illustrative CMS-EXO-16-010
analysis as this example can easily be adapted to any other analysis to be
considered.

We start by requiring the presence of two isolated charged
lepton of the same flavor and with an opposite electric charge. As indicated in
Section~\ref{sec:core-init}, this is achieved through the \verb+ApplyCut+
method of the \verb+RegionSelectionManager+ class. The cut condition corresponds
to the request that either two signal muons are present (\verb+nmu==2+), so that
the signal region named \verb+mumu+ will be potentially populated later on, or
two signal electrons are present (\verb+ne==2+), so that the signal region named
\verb+ee+ will be potentially populated later. Moreover, the product of the
charges of the two signal leptons must be negative. This is implemented as
\begin{verbatim}
  double charge = 0.;
  if(ne==2 || nmu==2)
    charge=SignalLeptons[0]->charge()*SignalLeptons[1]->charge();
  if(!Manager()->ApplyCut((ne==2 ) && (charge<0.),"2_electrons"))
    return true;
  if(!Manager()->ApplyCut((nmu==2) && (charge<0.),"2_muons"    ))
    return true;
\end{verbatim}
The first \verb+if+ statement refers to the condition that two signal
leptons of the same flavor are needed, \ie, either two electrons or two muons.
In the case where this condition is realized, the product of the charge of the
two leptons is computed and stored in the \verb+charge+ variable (initialized to
zero). The \verb+ApplyCut+ method is finally called twice, once for the electron
channel and once for the muon channel. This allows us to test the presence of
either a positron-electron pair or a muon-antimuon pair and to
update the cutflow accordingly. In each case, the arguments with which the
\verb+ApplyCut+
method is called consist of the string name (or the identifier) of the
cut under consideration (see the cut declaration in the analysis initialization
method \verb+Initialize+) and the respective cut condition that originates from
the merging of the two sub-conditions, \ie, the test on the number of signal
leptons and the one on the product of their electric charges.

The output of the \verb+ApplyCut+ method is a boolean value corresponding to the
result of the
test that at least one of the declared regions is still passing all cuts applied
so far. When it switches from true to false, the event should no longer be
analyzed and one should start analyzing the next event. This explains the
\verb+if+ structure and the \verb+return+ statement dressing the \verb+ApplyCut+
calls. The \verb+return+ statement is hence reached only if the result of
\verb+Manager()->ApplyCut( ... )+ is false, or equivalently if
\verb+!Manager()->ApplyCut( ... )+ is true. This
terminates the analysis of the present event in the case
where all regions (\ie, the two \verb+ee+ and \verb+mumu+ signal regions in the
CMS-EXO-16-010 case) are failing the cuts applied so far, so that the program
moves on with the analysis of the next event.

The second preselection cut (named \verb+on-Z+ at the level of the
implementation of the \verb+Initialize+ method) imposes that the pair of signal
leptons must be compatible with the decay of an on-shell $Z$-boson. The
dilepton invariant mass $m_{\ell\ell}$ is hence imposed to lie in a 20~GeV
mass window centered on the $Z$-boson mass,
\be
  m_{\ell\ell} \in [80, 100]~{\rm GeV}.
\label{eq:mll}\ee
This is implemented by evaluating firstly the four-momentum of the reconstructed
$Z$-boson (stored in the \verb+pZ+ variable),
\begin{verbatim}
  MALorentzVector pZ = SignalLeptons[0]->momentum() +
     SignalLeptons[1]->momentum();
\end{verbatim}
and by secondly requiring that its invariant mass (\ie, $m_{\ell\ell}$)
satisfies Eq.~\eqref{eq:mll},
\begin{verbatim}
  if(!Manager()->ApplyCut(std::abs(pZ.M() - 90.) < 10., "on-Z"))
    return true;
\end{verbatim}
where the string name (\verb+on-Z+) is the identifier of the second cut that has
been declared at the initialization time. This makes use of standard
methods of the \verb+MALorentzVector+ class whose \verb+pZ+ is an instance of.

The third preselection cut further constrains the reconstructed $Z$-boson, such
that its transverse-momentum $p_T^{\ell\ell}$ (or equivalently the transverse
momentum of the dilepton system) satisfies
\be
  p_T^{\ell\ell} > 50~{\rm GeV}.
\ee
The reconstructed $Z$-boson four-momentum being already available in the
analysis code (in the \verb+pZ+ variable defined in the implementation of the
previous selection), this cut is easily implemented as
\begin{verbatim}
  if(!Manager()->ApplyCut(pZ.Pt()>50., "dilepton_pt"))
    return true;
\end{verbatim}
using again standard methods of the \verb+MALorentzVector+ class.

The two final preselection criteria, as declared in the \verb+Initialize+
method, respectively concern the implementation of a veto of events featuring
either a third (looser) lepton or a tau (the \verb+3rd_lepton_veto+ cut), or a
$b$-tagged jet (the \verb+b_veto+ cut). As the number of loose leptons,
identified hadronic taus and $b$-tagged jets have
been computed at the time of the implementation of the object definitions (see
the definition of the \verb+nloose+, \verb+ntau+ and \verb+nb+ variables), these
two cuts are applied by
\begin{verbatim}
  if(!Manager()->ApplyCut((nloose<3) && (ntau==0),"3rd_lepton_veto"))
    return true;
  if(!Manager()->ApplyCut(nb==0,"b_veto"))
    return true;
\end{verbatim}

This concludes the implementation of the preselection requirements.
At this place, two of the
declared histograms (for which validation information is provided in the
CMS-EXO-16-010 analysis note) can be
filled. This is achieved by making use of the \verb+FillHisto+ method of the
\verb+RegionSelectionManager+ class,
\begin{verbatim}
  if(ne==2)
    Manager()->FillHisto("MET_preselected_e",MET);
  if(nmu==2)
    Manager()->FillHisto("MET_preselected_mu",MET);
\end{verbatim}
Here, the first two lines are related to the filling of the histogram assigned
to the electron channel (\ie, when the two signal leptons are electrons) and the
last two lines to the filling of the histogram assigned to the muon channel. In
the \verb+FillHisto+ function, the first argument identifies the histogram
(declared in the \verb+Initialize+ method), and the second one the observable
that is represented. We recall that the latter was calculated at the time of the
object definitions (see Section~\ref{sec:objects}).

The considered mono-$Z$-boson analysis involves four selection cuts related to
the signal region definitions. These cuts make in particular use of the missing
momentum information. They
are straightforwardly implemented by means of standard \MA\ methods,
\begin{verbatim}
 if(!Manager()->ApplyCut(std::abs(pZ.DeltaPhi(pTm))>2.7,"dphi(met,Z)"))
  return true;

 double MomentumBalance = std::abs(MET-pZ.Pt())/pZ.Pt();
 if(!Manager()->ApplyCut(MomentumBalance<0.2,"momentum_balance"))
  return true;

 if(!Manager()->ApplyCut(MET > 80,"met"))
  return true;

 if(!Manager()->ApplyCut(nj <= 1,"at_most_one_jet"))
  return true;
\end{verbatim}
For the first selection, the reconstructed $Z$-boson and the missing momentum
are constrained to lie in different hemisphere, their difference in azimuthal
angle being constrained to be larger than 2.7 radians,
\be
  \big|\Delta\varphi(\slashed{\mathbf{p}}_T,\mathbf{p}^{\ell\ell})\big| > 2.7\ ,
\ee
whereas for the second selection, the momentum balance of the event is required
to be larger than 0.2,
\be
  \frac{\big|\slashed{E}_T - p_T^{\ell\ell}\big|}{p_T^{\ell\ell}} < 0.2 \ .
\ee
These two cuts can easily be implemented from quantities that are already
available within the code, namely the reconstructed $Z$-boson momentum
\verb+pZ+, the missing transverse momentum \verb+pTm+ and the missing transverse
energy \verb+MET+. The last two selections implemented in the above code impose
that the missing transverse energy is larger than 80~GeV,
\be
   \slashed{E}_T > 80~{\rm GeV},
\ee
and that the event features at most one reconstructed jet, this last criterion
being implemented by using the \verb+nj+ variable defined at the time of the
object definitons.

The ordering in which the different selection cuts have been implemented
slightly differs from what has been mentioned in the CMS note. Whilst this does
not change anything for what concerns the final number of events populating each
signal region, the histograms and the intermediate entries in the cutflow
tables may be different. The choice that has been made actually matches the one
underlying the validation material provided by the CMS exotica group. This
consists hence in the only choice that allows us to validate our
reimplementation.

The analysis implementation ends with the filling of the last two histograms,
the missing energy distribution after the final selection for both signal
regions,
\begin{verbatim}
  if(ne==2)
    Manager()->FillHisto("MET_selected_e",MET);
  if(nmu==2)
    Manager()->FillHisto("MET_selected_mu",MET);
\end{verbatim}
the first two lines of code being connected to the electron channel and the last
two to the muon channel. This is similar to what has been implemented for the
filling of the two histograms representing the missing transverse energy
distribution after the preselection cuts.

\subsection{Detector simulation}\label{sec:delphes_config}
As the detector performance may be different from one analysis to another, the
configuration of \DEL\ to be associated with a specific analysis
must be carefully designed.
In order to setup a \DEL\ card including reconstruction properties similar to
those detailed in the CMS analysis note, we start from the default CMS detector
parameterization shipped with the \DEL\ package version 3.4.1 and implement two
modifications.

First, we update each call to
\FJ\cite{Cacciari:2011ma} in order to use, for jet reconstruction, the
anti-$k_T$ algorithm\cite{Cacciari:2008gp} with a radius parameter set to
$R=0.4$ instead of $R=0.5$. Second, we use a more recent parameterization of
the $b$-tagging efficiencies and corresponding mistagging rates, as detailed in
the CMS performance note on the identification of $b$-tagged jets in the second
run of the LHC\cite{CMS:2016kkf}. We implement the loose $b$-jet tagging
efficiency ($\epsilon_b$) that is given, as a function of the jet transverse
momentum, by%
\renewcommand{\arraystretch}{1.35}%
\be
 \epsilon_b(p_T) = \left\{\begin{array}{l l}
   0.707 + 5.6\cdot10^{-3} p_T - 6.27\cdot10^{-5} p_T^2 &\quad
     \multirow{2}{*}{\text{for }$p_T \in [30, 150]$~{\rm GeV}}\\[-.15cm]
         \hspace{5mm} +\ 3.10\cdot10^{-7} p_T^3- 5.63\cdot10^{-10} p_T^4& \\
   0.906 - 6.39\cdot10^{-5} p_T + 4.11\cdot10^{-8} p_T^2 &\quad
     \text{for }p_T \geq 150~{\rm GeV}\\
  \end{array}\right. \ ,
\ee\renewcommand{\arraystretch}{1}%
as well as the corresponding mistagging rates of a charm and lighter jet as a
$b$-jet ($\epsilon_c$ and $\epsilon_\ell$ respectively), that read
\renewcommand{\arraystretch}{1.35}%
\be\bsp
 \epsilon_c(p_T) =& \left\{\begin{array}{l l}
   0.40 + 1.23\cdot10^{-3} p_T - 4.60\cdot10^{-6} p_T^2 &\quad
     \multirow{2}{*}{\text{for }$p_T \in [30, 205]$~{\rm GeV}}\\[-.15cm]
         \hspace{5mm} +\ 5.71\cdot10^{-9} p_T^3& \\
   0.478 + 1.573\cdot10^{-4} p_T &\quad
     \text{for }p_T \geq 205~{\rm GeV}\\
  \end{array}\right. \ ,\\
 \epsilon_\ell(p_T) =& \left\{\begin{array}{l l}
   0.124 - 1.0\cdot10^{-3} p_T + 1.06\cdot10^{-5} p_T^2 &\quad
     \multirow{2}{*}{\text{for }$p_T \in [30, 130]$~{\rm GeV}}\\[-.15cm]
         \hspace{5mm} -\ 3.18\cdot10^{-8} p_T^3 + 3.13\cdot10^{-11} p_T^4& \\
   0.055 + 4.53\cdot10^{-4} p_T - 1.60\cdot10^{-7} p_T^2 &\quad
     \text{for }p_T \geq 130~{\rm GeV}\\
  \end{array}\right. \ .
\esp\ee\renewcommand{\arraystretch}{1}%
All these formulas can be found in the Appendix~A of the CMS performance
note on $b$-tagging for the LHC Run II\cite{CMS:2016kkf}.

Pileup effects are often important, in particular for the LHC Run II where a
large number of proton-proton collisions are expected to occur within a given
bunch crossing or two nearby bunch crossings. In the context of the 2015 data
studied in the CMS-EXO-16-010 analysis, 12 proton-proton interactions
simultaneously occur, in average. In order to check the corresponding effect on
our recasting machinery, we prepare a second \DEL\ card allowing
to superimpose minimum bias events to the signal. To this aim, we start from the
default CMS detector parameterization included in \DEL\ and in which the pileup
modeling is included, and perform all modifications described above. In
addition, the mean number of pileup events is set to 12.

\subsection{Data and Standard Model background information}
\label{sec:xml-info}
The statistical interpretation of the results to be obtained by the recasting of
an LHC analysis in any given theoretical context requires information on the
observations (\ie, how the various signal regions are populated by data) and on
the Standard Model expectation. While we can in principle use the recasting
code to
evaluate the Standard Model contributions, more accurate information can be
directly extracted from the experimental publication. The latter indeed
contains, additionally to the observations, the Standard Model results as
derived when using the full simulation of the LHC detectors. In this section, we
show how this information can be implemented in the \MA\ context, so that the
automatic limit setting procedure shipped with the code can be used. We refer
to Section~\ref{sec:limit-setting} for a practical example where a specific
signal is tested and a limit on the associated cross section is set.

In the previous sections, we described how to write the C++ header and
source files that contain the reimplementation of the analysis under
consideration. In addition, the user has to provide information on the Standard
Model background (and how the different signal regions of the analysis are
populated by data) and on the integrated luminosity of collisions that has been
analyzed. This is done in an XML file whose name is the analysis
name and extension is \verb+info+, and that is stored together with the analysis
C++ files in the \verb+PAD/Build/SampleAnalyzer/User/Analyzer+ directory. In our
example, the C++ header and source files have been named \verb+cms_exo_16_010.h+
and \verb+cms_exo_16_010.cpp+ (see Section~\ref{sec:pad}). Accordingly, the XML
file to be provided will be named \verb+cms_exo_16_010.info+.

The root element of the file is \verb+<analysis>+, and it must be provided
together with a mandatory attribute (\verb+id+) that contains the name of the
analysis. The luminosity (in fb$^{-1}$) is attached to the child element
\verb+<lumi>+, whilst the
information related to the different regions is passed via child elements of
the root named \verb+<region>+. One such element is implemented for each of
the signal regions of the analysis. The structure of the \verb+info+-file thus
reads
\begin{verbatim}
<analysis id="cms_exo_16_010">
  <lumi>2.3</lumi>
  <region type="signal" id="ee">
    ...
  </region>
  <region type="signal" id="mumu">
    ...
  </region>
</analysis>
\end{verbatim}
where the content of the dots will be specified below. The string
\verb+cms_exo_16_010+ has been used as the analysis identifier (set via the
\verb+id+ attribute of the root element), and we have fixed
the luminosity to 2.3~fb$^{-1}$. The analysis containing two signal regions,
two child elements \verb+<region>+ are present with attributes determining their
type (\verb+type+) and name (\verb+id+). The \verb+type+ attribute, whose
value can be either \verb+signal+ or \verb+control+, has been set twice to the
\verb+signal+ value (as we have only implemented the two signal regions of the
analysis), and the identifier attribute \verb+id+ has been set to \verb+ee+ and
\verb+mumu+ for the electron and  muon signal regions, respectively. These names
map those that have been assigned to the signal regions at the time of their
declaration in the \verb+Initialize+ method of the analysis core (see
Section~\ref{sec:init}).

The \verb+<region>+ element contains various child elements dedicated to the
number of observed events populating the region (\verb+<nobs>+), the number of
expected  Standard Model events populating the region (\verb+<nb>+) and the
associated uncertainty given at the level of one standard deviation
(\verb+deltanb>+). Extracting the information from the CMS-EXO-16-10 analysis,
the element related to the electron channel reads
\begin{verbatim}
  <region type="signal" id="ee">
    <nobs>22</nobs>
    <nb>28.9</nb>
    <deltanb>5.53</deltanb>
  </region>
\end{verbatim}
whilst the one corresponding to the muon channel is given by
\begin{verbatim}
  <region type="signal" id="mumu">
    <nobs>44</nobs>
    <nb>45.0</nb>
    <deltanb>7.15</deltanb>
  </region>
\end{verbatim}

\subsection{Validation}
\label{sec:validation}
The results presented in the CMS-EXO-16-010 experimental analysis are
interpreted in the framework of several models. This includes a simplified
tree-level ultraviolet-complete model for dark matter where a fermionic dark
matter particle couples to the Standard Model sector through interactions
with a spin-1 mediator\cite{Abercrombie:2015wmb}, an effective dark matter
model featuring higher-dimensional four-point interactions of a pair of dark
matter particles with two of the Standard Model electroweak gauge
bosons\cite{Abercrombie:2015wmb}, and an effective model describing
unparticle dynamics\cite{Georgi:2007ek,Georgi:2007si}.

The validation material provided by the CMS collaboration (upon
our request) concerns only
the first of these three models. In this context, the Standard Model
is extended by a dark matter particle $\chi$ that is assumed to be a Dirac
fermion of mass $m_\chi$ whose interactions with the Standard Model quarks
are mediated by a spin-1 mediator $Z'$ of mass $m_{Z'}$. The interactions of the
$Z'$ boson are moreover imposed to be of a purely vector nature, the
corresponding Lagrangian reading thus
\be\bsp
  {\cal L} = &\ {\cal L}_{\rm SM} - \frac14 Z'_{\mu\nu} Z'^{\mu\nu} + \frac12
    m_{Z'}^2 Z^{\prime\mu}Z'_\mu + i \bar\chi \slashed{\partial}\chi
    - m \bar\chi\chi \\ &\quad
    - Z'_\mu \bigg[ \sum_{q=u,d} g_q \bar q \gamma^\mu q +
        g_\chi \bar\chi\gamma^\mu\chi \bigg] \ .
\esp\ee
The ${\cal L}_{\rm SM}$ Lagrangian represents the Standard Model Lagrangian,
$Z'_{\mu\nu}$ denotes the field strength tensor of the $Z'$ boson and flavor
indices are moreover understood for what concern the new physics interactions of
the quarks. The strength of the interaction of the $Z'$ mediator with the dark
matter particle is denoted by $g_\chi$, whilst the couplings to the Standard
Model quarks are universal and read $g_u$ and $g_d$ for up-type and down-type
quarks, respectively. Although interactions with charged leptons and neutrinos
are in principle possible, they are ignored for simplicity.

\begin{figure}
  \centering
  \includegraphics[width=0.35\columnwidth]{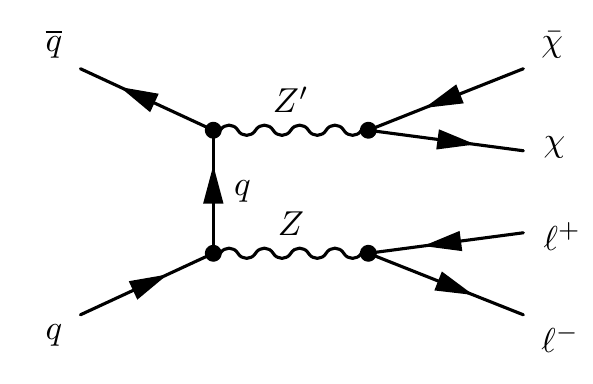}
  \caption{Representative Feynman diagrams for the production of a pair of dark
    matter particles $\chi$ with a leptonically-decaying $Z$-boson.}
  \label{fig:monoZ_diag}
\end{figure}

In this theoretical framework, a mono-$Z$-boson final state could emerge from
quark-antiquark annihilation. This is illustrated by the representative Feynman
diagram shown in Figure~\ref{fig:monoZ_diag} in which the $Z$-boson decay into a
leptonic final state is included.

The validation material provided by the CMS collaboration contains five cutflow
charts related to benchmark setups in which the dark matter mass and the new
physics couplings have been fixed to
\be
  m_\chi = 50~{\rm GeV} \qquad\text{and}\qquad
  g_\chi = g_u = g_d = 1 \ ,
\ee
and that differ by the choice of the mediator mass,
\be
  m_{Z'} \in \big[10, 200, 500, 1000, 5000\big]~{\rm GeV}.
\ee
For each scenario, CMS has provided official cutflow charts, as
well as information on the configuration of the Monte Carlo event generators
that have been used to generate the dark matter signal. This material is
available on the \MA\ Public Analysis Database,\\
\hspace*{2.5mm}
  \url{http://madanalysis.irmp.ucl.ac.be/wiki/PublicAnalysisDatabase},\\
and includes Monte Carlo setup files. This prevents us from introducing any bias
at the level of the generation of the signal events, which could impact, in an
uncontroled way, the comparison of the \MA\ results with the CMS results. In
addition, our validation procedure includes a comparison of the histograms
that have been implemented in the \MA\ code with those shown in the CMS note for
the same dark matter model.

The other theoretical contexts in which CMS has
interpreted the results have not been addressed in our validation procedure,
as precise information on event generation has not been provided. The performed
comparisons however make us confident about the reasonable level of accuracy
reached by our reimplementation.

\renewcommand{\arraystretch}{1.20}%
\begin{table}
  \tbl{
    Comparison of results obtained with our \MA~reimplementation (MA5) and those
    provided by the CMS collaboration (CMS) for a dark matter benchmark scenario
    where the dark matter mass has been set to 50~GeV and the mediator mass to
    10~GeV. All vector couplings of the mediator have been fixed to 1. The
    results are expressed in
    terms of selection efficiencies as defined in Eq.~\eqref{eq:effic}, and the
    relative difference between the CMS and the \MA\ efficiencies,
    $\delta_i^{\rm rel}$, stems from Eq.~\eqref{eq:delma5cms}.}
  {\begin{tabular}{c|c||c|c|c||c|c|c}
   \multicolumn{2}{c||}{\multirow{2}{*}{Selection step}}&
     \multicolumn{3}{c||}{Electron channel}&\multicolumn{3}{c}{Muon channel}\\
   \cline{3-8}
   \multicolumn{2}{c||}{}&
     $\epsilon_i^{\rm CMS}$ & $\epsilon_i^{\rm MA5}$ & $\delta_i^{\rm rel}$ &
     $\epsilon_i^{\rm CMS}$ & $\epsilon_i^{\rm MA5}$ & $\delta_i^{\rm rel}$ \\
   \hline
   1&Two leptons                      &   -   &   -   &  -
                                      &   -   &   -   &  -   \\
   2&$m_{\ell\ell} \in [80, 100]$~GeV & 0.929 & 0.933 & 0.4\%
                                      & 0.919 & 0.933 & 1.5\%\\
   3&$p_T^{\ell\ell} > 50$~GeV        & 0.647 & 0.648 & 0.2\%
                                      & 0.644 & 0.638 & 1.0\%\\
   4&Third lepton and tau veto        & 0.963 & 0.992 & 3.0\%
                                      & 0.961 & 0.991 & 3.2\%\\
   5&$b$-jet veto                     & 0.983 & 0.933 & 5.1\%
                                      & 0.984 & 0.931 & 5.4\%\\
   6&$\big|\Delta\varphi(\slashed{\mathbf{p}}_T,\mathbf{p}^{\ell\ell})\big|>2.7$
                                      & 0.694 & 0.761 & 9.6\%
                                      & 0.705 & 0.762 & 8.0\%\\
   7&$\big|\slashed{E}_T - p_T^{\ell\ell}\big|/p_T^{\ell\ell} < 0.2$
                                      & 0.623 & 0.715 &15.0\%
                                      & 0.623 & 0.704 &13.0\%\\
   8&$\slashed{E}_T > 80~{\rm GeV}$   & 0.744 & 0.691 & 7.2\%
                                      & 0.724 & 0.677 & 6.4\%\\
   9&At most one jet                  & 0.973 & 0.979 & 0.7\%
                                      & 0.972 & 0.980 & 0.8\%\\
  \end{tabular} \label{tab:m10}}\vspace*{4mm}

  \tbl{Same as in Table~\ref{tab:m10} but for a mediator mass of 200~GeV.}
  {\begin{tabular}{c|c||c|c|c||c|c|c}
   \multicolumn{2}{c||}{\multirow{2}{*}{Selection step}}&
     \multicolumn{3}{c||}{Electron channel}&\multicolumn{3}{c}{Muon channel}\\
   \cline{3-8}
   \multicolumn{2}{c||}{}&
     $\epsilon_i^{\rm CMS}$ & $\epsilon_i^{\rm MA5}$ & $\delta_i^{\rm rel}$ &
     $\epsilon_i^{\rm CMS}$ & $\epsilon_i^{\rm MA5}$ & $\delta_i^{\rm rel}$ \\
   \hline
   1&Two leptons                      &   -   &   -   &  -
                                      &   -   &   -   &  -   \\
   2&$m_{\ell\ell} \in [80, 100]$~GeV & 0.929 & 0.932 & 0.4\%
                                      & 0.920 & 0.929 & 1.0\%\\
   3&$p_T^{\ell\ell} > 50$~GeV        & 0.676 & 0.683 & 1.0\%
                                      & 0.658 & 0.674 & 2.5\%\\
   4&Third lepton and tau veto        & 0.959 & 0.991 & 3.4\%
                                      & 0.959 & 0.992 & 3.5\%\\
   5&$b$-jet veto                     & 0.983 & 0.931 & 5.4\%
                                      & 0.984 & 0.930 & 5.5\%\\
   6&$\big|\Delta\varphi(\slashed{\mathbf{p}}_T,\mathbf{p}^{\ell\ell})\big|>2.7$
                                      & 0.703 & 0.771 & 9.7\%
                                      & 0.708 & 0.765 & 8.1\%\\
   7&$\big|\slashed{E}_T - p_T^{\ell\ell}\big|/p_T^{\ell\ell} < 0.2$
                                      & 0.625 & 0.724 &16.0\%
                                      & 0.641 & 0.706 &10.0\%\\
   8&$\slashed{E}_T > 80~{\rm GeV}$   & 0.735 & 0.714 & 2.8\%
                                      & 0.745 & 0.711 & 4.6\%\\
   9&At most one jet                  & 0.979 & 0.977 & 0.4\%
                                      & 0.982 & 0.978 & 0.2\%\\
  \end{tabular} \label{tab:m200}}\vspace{4mm}

  \tbl{Same as in Table~\ref{tab:m10} but for a mediator mass of 500~GeV.}
  {\begin{tabular}{c|c||c|c|c||c|c|c}
   \multicolumn{2}{c||}{\multirow{2}{*}{Selection step}}&
     \multicolumn{3}{c||}{Electron channel}&\multicolumn{3}{c}{Muon channel}\\
   \cline{3-8}
   \multicolumn{2}{c||}{}&
     $\epsilon_i^{\rm CMS}$ & $\epsilon_i^{\rm MA5}$ & $\delta_i^{\rm rel}$ &
     $\epsilon_i^{\rm CMS}$ & $\epsilon_i^{\rm MA5}$ & $\delta_i^{\rm rel}$ \\
   \hline
   1&Two leptons                      &   -   &   -   &  -
                                      &   -   &   -   &  -   \\
   2&$m_{\ell\ell} \in [80, 100]$~GeV & 0.929 & 0.931 & 0.2\%
                                      & 0.922 & 0.930 & 0.9\%\\
   3&$p_T^{\ell\ell} > 50$~GeV        & 0.783 & 0.775 & 1.0\%
                                      & 0.770 & 0.765 & 0.6\%\\
   4&Third lepton and tau veto        & 0.953 & 0.990 & 3.9\%
                                      & 0.952 & 0.990 & 4.0\%\\
   5&$b$-jet veto                     & 0.980 & 0.918 & 6.3\%
                                      & 0.982 & 0.918 & 6.5\%\\
   6&$\big|\Delta\varphi(\slashed{\mathbf{p}}_T,\mathbf{p}^{\ell\ell})\big|>2.7$
                                      & 0.719 & 0.770 & 7.2\%
                                      & 0.718 & 0.767 & 6.8\%\\
   7&$\big|\slashed{E}_T - p_T^{\ell\ell}\big|/p_T^{\ell\ell} < 0.2$
                                      & 0.672 & 0.726 & 8.0\%
                                      & 0.662 & 0.718 & 8.4\%\\
   8&$\slashed{E}_T > 80~{\rm GeV}$   & 0.860 & 0.819 & 4.7\%
                                      & 0.854 & 0.809 & 5.3\%\\
   9&At most one jet                  & 0.954 & 0.966 & 1.3\%
                                      & 0.956 & 0.972 & 1.7\%\\
  \end{tabular} \label{tab:m500}}\vspace*{4mm}

\end{table}

Samples of 300.000 simulated dark matter events (including electronic or
muonic $Z$-boson decays) have been generated using
\MG\cite{Alwall:2014hca} for the simulation of the hard scattering process and
\PY\cite{Sjostrand:2014zea} for the simulation of the hadronic environment
(parton showering and hadronization), using the Monash
tune\cite{Skands:2014pea}. The hard scattering matrix element associated with
the signal process
\be
  p p \to \big(Z' \to \bar \chi \chi\big) \big(Z \to \ell^+ \ell^-\big)
\ee
is evaluated at the leading-order accuracy and convolved with the
leading-order set of NNPDF parton densities version 3.0\cite{Ball:2014uwa}, the
latter being accessed via the LHAPDF library\cite{Whalley:2005nh,%
Buckley:2014ana}. The
renormalization and factorization scales are moreover set to the geometric mean
of the transverse mass of all final-state particles, and the width of the
mediator is calculated automatically by means of the \MW\
program\cite{Alwall:2014bza}. In order to evaluate the number of signal events
surviving each cut and the number of signal events populating both signal
regions, we use the standard recasting methods implemented within \MA\ and
detailed in Section~\ref{sec:outputruns}. We begin with a recasting procedure
ignoring any pileup effect (see Section~\ref{sec:delphes_config}).

\begin{table}
  \tbl{Same as in Table~\ref{tab:m10} but for a mediator mass of 1000~GeV.}
  {\begin{tabular}{c|c||c|c|c||c|c|c}
   \multicolumn{2}{c||}{\multirow{2}{*}{Selection step}}&
     \multicolumn{3}{c||}{Electron channel}&\multicolumn{3}{c}{Muon channel}\\
   \cline{3-8}
   \multicolumn{2}{c||}{}&
     $\epsilon_i^{\rm CMS}$ & $\epsilon_i^{\rm MA5}$ & $\delta_i^{\rm rel}$ &
     $\epsilon_i^{\rm CMS}$ & $\epsilon_i^{\rm MA5}$ & $\delta_i^{\rm rel}$ \\
   \hline
   1&Two leptons                      &   -   &   -   &  -
                                      &   -   &   -   &  -   \\
   2&$m_{\ell\ell} \in [80, 100]$~GeV & 0.928 & 0.931 & 0.4\%
                                      & 0.921 & 0.927 & 0.7\%\\
   3&$p_T^{\ell\ell} > 50$~GeV        & 0.835 & 0.822 & 1.6\%
                                      & 0.825 & 0.807 & 2.2\%\\
   4&Third lepton and tau veto        & 0.948 & 0.988 & 4.2\%
                                      & 0.949 & 0.990 & 4.3\%\\
   5&$b$-jet veto                     & 0.977 & 0.904 & 7.5\%
                                      & 0.979 & 0.903 & 7.7\%\\
   6&$\big|\Delta\varphi(\slashed{\mathbf{p}}_T,\mathbf{p}^{\ell\ell})\big|>2.7$
                                      & 0.705 & 0.766 & 8.6\%
                                      & 0.695 & 0.759 & 9.1\%\\
   7&$\big|\slashed{E}_T - p_T^{\ell\ell}\big|/p_T^{\ell\ell} < 0.2$
                                      & 0.678 & 0.725 & 6.9\%
                                      & 0.668 & 0.708 & 5.9\%\\
   8&$\slashed{E}_T > 80~{\rm GeV}$   & 0.915 & 0.870 & 4.9\%
                                      & 0.902 & 0.863 & 4.3\%\\
   9&At most one jet                  & 0.936 & 0.960 & 2.5\%
                                      & 0.943 & 0.961 & 1.9\%\\
  \end{tabular} \label{tab:m1000}}\vspace{4mm}

  \tbl{Same as in Table~\ref{tab:m10} but for a mediator mass of 5000~GeV.}
  {\begin{tabular}{c|c||c|c|c||c|c|c}
   \multicolumn{2}{c||}{\multirow{2}{*}{Selection step}}&
     \multicolumn{3}{c||}{Electron channel}&\multicolumn{3}{c}{Muon channel}\\
   \cline{3-8}
   \multicolumn{2}{c||}{}&
     $\epsilon_i^{\rm CMS}$ & $\epsilon_i^{\rm MA5}$ & $\delta_i^{\rm rel}$ &
     $\epsilon_i^{\rm CMS}$ & $\epsilon_i^{\rm MA5}$ & $\delta_i^{\rm rel}$ \\
   \hline
   1&Two leptons                      &   -   &   -   &  -
                                      &   -   &   -   &  -   \\
   2&$m_{\ell\ell} \in [80, 100]$~GeV & 0.928 & 0.931 & 0.3\%
                                      & 0.921 & 0.928 & 0.7\%\\
   3&$p_T^{\ell\ell} > 50$~GeV        & 0.841 & 0.839 & 0.2\%
                                      & 0.832 & 0.827 & 0.6\%\\
   4&Third lepton and tau veto        & 0.947 & 0.988 & 4.3\%
                                      & 0.945 & 0.988 & 4.6\%\\
   5&$b$-jet veto                     & 0.977 & 0.893 & 8.6\%
                                      & 0.978 & 0.894 & 8.6\%\\
   6&$\big|\Delta\varphi(\slashed{\mathbf{p}}_T,\mathbf{p}^{\ell\ell})\big|>2.7$
                                      & 0.708 & 0.760 & 7.3\%
                                      & 0.698 & 0.754 & 8.0\%\\
   7&$\big|\slashed{E}_T - p_T^{\ell\ell}\big|/p_T^{\ell\ell} < 0.2$
                                      & 0.687 & 0.720 & 4.9\%
                                      & 0.684 & 0.703 & 2.7\%\\
   8&$\slashed{E}_T > 80~{\rm GeV}$   & 0.923 & 0.889 & 3.7\%
                                      & 0.908 & 0.879 & 3.2\%\\
   9&At most one jet                  & 0.932 & 0.953 & 2.2\%
                                      & 0.935 & 0.954 & 2.1\%\\
  \end{tabular} \label{tab:m5000}}\vspace*{4mm}
\end{table}\renewcommand{\arraystretch}{1}%

The comparison of the results obtained with \MA\ to the official numbers
provided by the CMS collaboration is shown in Table~\ref{tab:m10},
Table~\ref{tab:m200}, Table~\ref{tab:m500}, Table~\ref{tab:m1000} and
Table~\ref{tab:m5000} for a benchmark scenario in which the mediator mass has
been fixed to 10~GeV, 200~GeV, 500~GeV, 1000~GeV and 5000~GeV, respectively.
For each cut, we present the selection efficiency defined by
\be
  \epsilon_i = \frac{n_i}{n_{i-1}} \ ,
\label{eq:effic}\ee
where $n_{i-1}$ and $n_i$ correspond to the event number before and after the
considered cut, respectively. The relative difference between the \MA\ recasting
and the CMS official results is normalized to the CMS result, and thus
estimated, for each cut, by
\be
   \delta_i^{\rm rel} =
     \bigg|1 - \frac{\epsilon_i^{\rm MA5}}{\epsilon_i^{\rm CMS}}\bigg| \ .
\label{eq:delma5cms}\ee
Results for the firsts selection, \ie, the requirement of the presence of two
electrons (in the electron channel) or muons (in the muon channel) are not shown
as they have not been provided by the CMS collaboration.

At each step of the validation, the \MA\ predictions and the official CMS
results have been found to agree at a level of about 10\% or below, with the
exception of the selection on the momentum balance of the event for scenarios
where the mediator mass is small (10~GeV or 200~GeV). In this case, the CMS
results are only described at the level of 10\%--15\% by \MA. This discrepancy
can be traced back to the fast simulation of the detector by \DEL\ for which a
proper description of the missing energy is harder to achieve. The latter indeed
depends on all the other objects in the events and is thus sensitive to all the
aspect of the detector simulation.

\renewcommand{\arraystretch}{1.30}%
\begin{table}
  \tbl{
    Comparison of the selection efficiencies obtained with our \MA\
    reimplementation (MA5) to those provided by the CMS collaboration (CMS) for
    a dark matter benchmark scenario where the dark matter mass has been set to
    50~GeV and the mediator mass $m_{Z'}$ is varying. All vector couplings of
    the mediator have been fixed to 1. The results are expressed in terms of the
    total selection efficiency as defined in Eq.~\eqref{eq:tot_effic}, and the
    relative difference $\delta^{\rm rel}$ between the CMS and the \MA\
    efficiencies is computed as in Eq.~\eqref{eq:tot_delma5cms}.}
  {\begin{tabular}{c||c|c|c||c|c|c}
   \multirow{2}{*}{Scenario}&
     \multicolumn{3}{c||}{Electron channel}&\multicolumn{3}{c}{Muon channel}\\
   \cline{2-7}
   & $\epsilon^{\rm CMS}$ & $\epsilon^{\rm MA5}$ & $\delta^{\rm rel}$ &
     $\epsilon^{\rm CMS}$ & $\epsilon^{\rm MA5}$ & $\delta^{\rm rel}$ \\
   \hline
   $m_{Z'} = 10$~GeV                  & 0.178 & 0.219 &16.0\%
                                      & 0.173 & 0.200 &13.0\%\\
   $m_{Z'} = 200$~GeV                 & 0.186 & 0.230 &23.0\%
                                      & 0.189 & 0.220 &15.0\%\\
   $m_{Z'} = 500$~GeV                 & 0.269 & 0.291 & 8.0\%
                                      & 0.258 & 0.280 & 8.6\%\\
   $m_{Z'} = 1000$~GeV                & 0.294 & 0.320 & 7.9\%
                                      & 0.279 & 0.300 & 6.8\%\\
   $m_{Z'} = 5000$~GeV                & 0.302 & 0.320 & 5.7\%
                                      & 0.287 & 0.300 & 4.8\%\\
  \end{tabular} \label{tab:efficiencies}}
\end{table}\renewcommand{\arraystretch}{1}%

In Table~\ref{tab:efficiencies}, we calculate the total efficiency of the
analysis for both signal regions and confront the results obtained in the \MA\
framework with the official ones provided by the CMS collaboration. The
efficiencies are defined as
\be
  \epsilon = \frac{n_9}{n_1} \ ,
\label{eq:tot_effic}\ee
where $n_1$ and $n_9$ are the number of events that are selected after the first
cut (\ie, the selection on the number of signal leptons) and the number of
events surviving all cuts, respectively. The difference between \MA\ and CMS is
evaluated again relatively to the CMS results,
\be
   \delta^{\rm rel} =
     \bigg|1 - \frac{\epsilon^{\rm MA5}}{\epsilon^{\rm CMS}}\bigg| \ .
\label{eq:tot_delma5cms}\ee
An agreement of the level of about 10\%--20\% has been found, resulting from the
cumulative effect of all the cuts, the agreement being once again better for the
heavy mediator case.

\begin{figure}
  \centering
  \includegraphics[width=0.495\columnwidth]{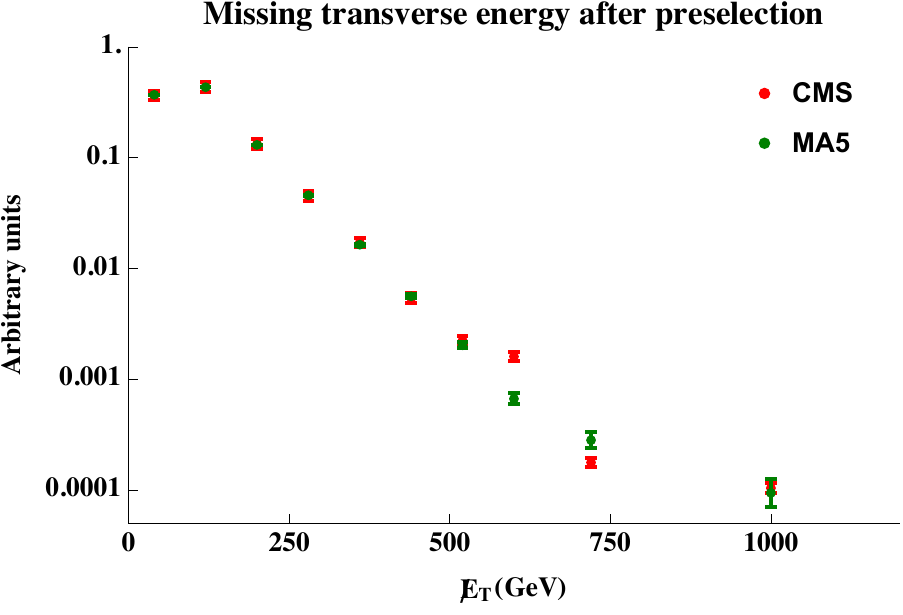}
  \includegraphics[width=0.495\columnwidth]{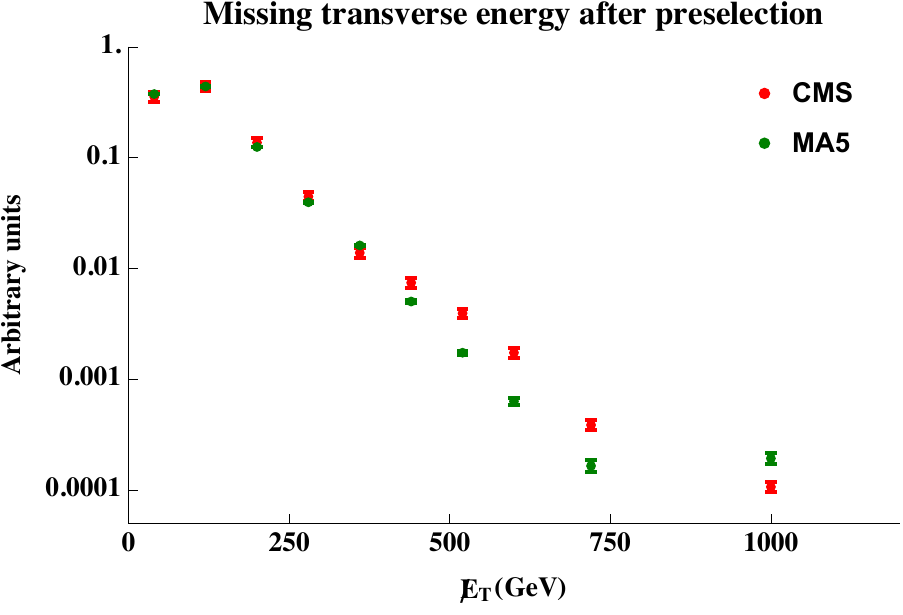}\\[3mm]
  \includegraphics[width=0.495\columnwidth]{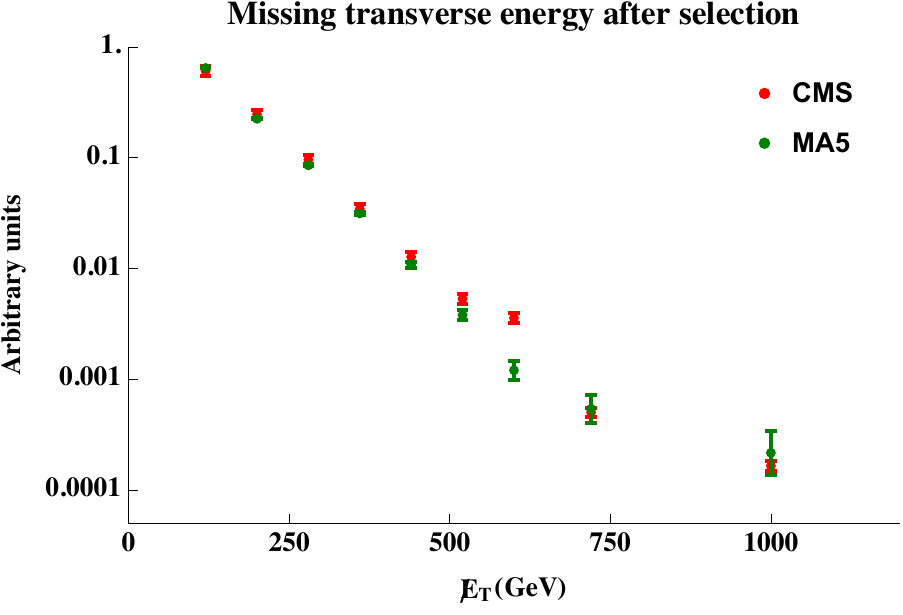}
  \includegraphics[width=0.495\columnwidth]{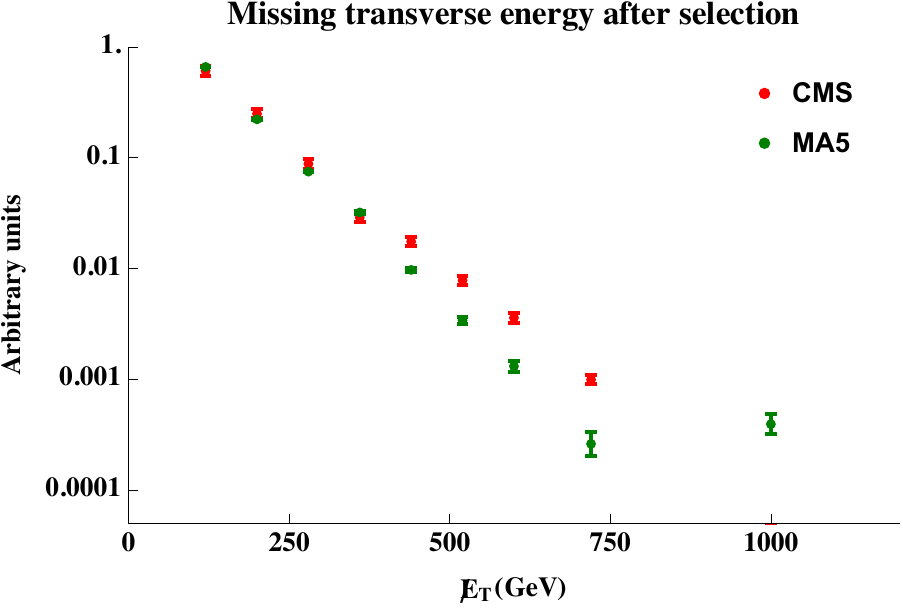}
  \caption{Missing transverse energy spectrum in the electron channel (left)
    and in the muon channel (right) after all preselection cuts including the
    requirement on the number of jets (upper panel) and after all cuts (lower
    panel).
    We compare the official CMS results (red) including a 10\% Monte Carlo
    uncertainty with the results obtained with \MA\ (green), the
    statistical uncertainties being included for the latter.}
  \label{fig:histos}
\end{figure}

In Figure~\ref{fig:histos}, we compare normalized missing transverse
energy distributions at different level of the analysis. Results for the
electron and muon channels are presented in the left and right panel of the
figure, respectively, and are given for the benchmark scenario
in which the mediator mass has been set to 200~GeV.

In the upper panel of the figure, all the preselection cuts (\ie, the first
five cuts) have been applied,
together with the requirement that at most one jet is present in the selected
events (\ie, the last of all cuts). This change in the cut ordering is necessary
to map what has been done by CMS to produce the validation material. In the
lower panel of the figure, we show the same distributions but after imposing all
selection cuts. We observe a fair agreement between the \MA\ predictions and the
official numbers, the shapes of the distributions qualitatively matching well.
The peaking bins are indeed in accordance with the official results and the
differences for the higher-missing energy bins are of at most 10\%-15\%.
This level of accuracy is similar to what has been found for the cutflow in
Table~\ref{tab:m200}. We also observe that the largest difference of 23\% and
15\% obtained for the total efficiencies in the electron and muon signal
region, respectively, is due to the impact of the tail of the missing energy
distributions.

\renewcommand{\arraystretch}{1.25}%
\begin{table}
  \tbl{Same as in Table~\ref{tab:m10} but when the pileup modeling is
    included in the \MA\ recasting.}
  {\begin{tabular}{c|c||c|c|c||c|c|c}
   \multicolumn{2}{c||}{\multirow{2}{*}{Selection step}}&
     \multicolumn{3}{c||}{Electron channel}&\multicolumn{3}{c}{Muon channel}\\
   \cline{3-8}
   \multicolumn{2}{c||}{}&
     $\epsilon_i^{\rm CMS}$ & $\epsilon_i^{\rm MA5}$ & $\delta_i^{\rm rel}$ &
     $\epsilon_i^{\rm CMS}$ & $\epsilon_i^{\rm MA5}$ & $\delta_i^{\rm rel}$ \\
   \hline
   1&Two leptons                      &   -   &   -   &  -
                                      &   -   &   -   &  -   \\
   2&$m_{\ell\ell} \in [80, 100]$~GeV & 0.929 & 0.931 & 0.2\%
                                      & 0.919 & 0.929 & 1.1\%\\
   3&$p_T^{\ell\ell} > 50$~GeV        & 0.647 & 0.651 & 0.7\%
                                      & 0.644 & 0.637 & 1.1\%\\
   4&Third lepton and tau veto        & 0.963 & 0.989 & 2.7\%
                                      & 0.961 & 0.989 & 2.9\%\\
   5&$b$-jet veto                     & 0.983 & 0.933 & 5.1\%
                                      & 0.984 & 0.935 & 5.0\%\\
   6&$\big|\Delta\varphi(\slashed{\mathbf{p}}_T,\mathbf{p}^{\ell\ell})\big|>2.7$
                                      & 0.694 & 0.700 & 0.9\%
                                      & 0.705 & 0.699 & 0.9\%\\
   7&$\big|\slashed{E}_T - p_T^{\ell\ell}\big|/p_T^{\ell\ell} < 0.2$
                                      & 0.623 & 0.585 & 6.1\%
                                      & 0.623 & 0.577 & 7.3\%\\
   8&$\slashed{E}_T > 80~{\rm GeV}$   & 0.744 & 0.760 & 2.2\%
                                      & 0.724 & 0.744 & 2.8\%\\
   9&At most one jet                  & 0.973 & 0.965 & 0.7\%
                                      & 0.972 & 0.973 & 0.1\%\\
  \end{tabular} \label{tab:m10_pileup}}\vspace*{4mm}

  \tbl{Same as in Table~\ref{tab:m200} but when the pileup modeling is
    included in the \MA\ recasting.}
  {\begin{tabular}{c|c||c|c|c||c|c|c}
   \multicolumn{2}{c||}{\multirow{2}{*}{Selection step}}&
     \multicolumn{3}{c||}{Electron channel}&\multicolumn{3}{c}{Muon channel}\\
   \cline{3-8}
   \multicolumn{2}{c||}{}&
     $\epsilon_i^{\rm CMS}$ & $\epsilon_i^{\rm MA5}$ & $\delta_i^{\rm rel}$ &
     $\epsilon_i^{\rm CMS}$ & $\epsilon_i^{\rm MA5}$ & $\delta_i^{\rm rel}$ \\
   \hline
   1&Two leptons                      &   -   &   -   &  -
                                      &   -   &   -   &  -   \\
   2&$m_{\ell\ell} \in [80, 100]$~GeV & 0.929 & 0.930 & 0.1\%
                                      & 0.920 & 0.927 & 0.7\%\\
   3&$p_T^{\ell\ell} > 50$~GeV        & 0.676 & 0.680 & 0.6\%
                                      & 0.658 & 0.675 & 2.7\%\\
   4&Third lepton and tau veto        & 0.959 & 0.987 & 3.0\%
                                      & 0.959 & 0.989 & 3.1\%\\
   5&$b$-jet veto                     & 0.983 & 0.934 & 5.0\%
                                      & 0.984 & 0.930 & 5.5\%\\
   6&$\big|\Delta\varphi(\slashed{\mathbf{p}}_T,\mathbf{p}^{\ell\ell})\big|>2.7$
                                      & 0.703 & 0.705 & 0.3\%
                                      & 0.708 & 0.707 & 0.1\%\\
   7&$\big|\slashed{E}_T - p_T^{\ell\ell}\big|/p_T^{\ell\ell} < 0.2$
                                      & 0.625 & 0.599 & 4.1\%
                                      & 0.641 & 0.585 & 8.7\%\\
   8&$\slashed{E}_T > 80~{\rm GeV}$   & 0.735 & 0.781 & 6.4\%
                                      & 0.745 & 0.777 & 4.2\%\\
   9&At most one jet                  & 0.979 & 0.963 & 1.1\%
                                      & 0.982 & 0.962 & 1.7\%\\
  \end{tabular} \label{tab:m200_pileup}}\vspace{4mm}

  \tbl{Same as in Table~\ref{tab:m500} but when the pileup modeling is
    included in the \MA\ recasting.}
  {\begin{tabular}{c|c||c|c|c||c|c|c}
   \multicolumn{2}{c||}{\multirow{2}{*}{Selection step}}&
     \multicolumn{3}{c||}{Electron channel}&\multicolumn{3}{c}{Muon channel}\\
   \cline{3-8}
   \multicolumn{2}{c||}{}&
     $\epsilon_i^{\rm CMS}$ & $\epsilon_i^{\rm MA5}$ & $\delta_i^{\rm rel}$ &
     $\epsilon_i^{\rm CMS}$ & $\epsilon_i^{\rm MA5}$ & $\delta_i^{\rm rel}$ \\
   \hline
   1&Two leptons                      &   -   &   -   &  -
                                      &   -   &   -   &  -   \\
   2&$m_{\ell\ell} \in [80, 100]$~GeV & 0.929 & 0.929 & 0.0\%
                                      & 0.922 & 0.927 & 0.5\%\\
   3&$p_T^{\ell\ell} > 50$~GeV        & 0.783 & 0.777 & 0.7\%
                                      & 0.770 & 0.765 & 0.7\%\\
   4&Third lepton and tau veto        & 0.953 & 0.987 & 3.6\%
                                      & 0.952 & 0.987 & 3.7\%\\
   5&$b$-jet veto                     & 0.980 & 0.919 & 6.2\%
                                      & 0.982 & 0.922 & 6.1\%\\
   6&$\big|\Delta\varphi(\slashed{\mathbf{p}}_T,\mathbf{p}^{\ell\ell})\big|>2.7$
                                      & 0.719 & 0.726 & 1.1\%
                                      & 0.718 & 0.726 & 1.0\%\\
   7&$\big|\slashed{E}_T - p_T^{\ell\ell}\big|/p_T^{\ell\ell} < 0.2$
                                      & 0.672 & 0.634 & 5.7\%
                                      & 0.662 & 0.622 & 6.2\%\\
   8&$\slashed{E}_T > 80~{\rm GeV}$   & 0.860 & 0.872 & 1.4\%
                                      & 0.854 & 0.861 & 0.8\%\\
   9&At most one jet                  & 0.954 & 0.952 & 0.2\%
                                      & 0.956 & 0.954 & 0.2\%\\
  \end{tabular} \label{tab:m500_pileup}}\vspace*{4mm}

\end{table}

\begin{table}
  \tbl{Same as in Table~\ref{tab:m1000} but when the pileup modeling is
    included in the \MA\ recasting.}
  {\begin{tabular}{c|c||c|c|c||c|c|c}
   \multicolumn{2}{c||}{\multirow{2}{*}{Selection step}}&
     \multicolumn{3}{c||}{Electron channel}&\multicolumn{3}{c}{Muon channel}\\
   \cline{3-8}
   \multicolumn{2}{c||}{}&
     $\epsilon_i^{\rm CMS}$ & $\epsilon_i^{\rm MA5}$ & $\delta_i^{\rm rel}$ &
     $\epsilon_i^{\rm CMS}$ & $\epsilon_i^{\rm MA5}$ & $\delta_i^{\rm rel}$ \\
   \hline
   1&Two leptons                      &   -   &   -   &  -
                                      &   -   &   -   &  -   \\
   2&$m_{\ell\ell} \in [80, 100]$~GeV & 0.928 & 0.929 & 0.2\%
                                      & 0.921 & 0.925 & 0.4\%\\
   3&$p_T^{\ell\ell} > 50$~GeV        & 0.835 & 0.824 & 1.4\%
                                      & 0.825 & 0.810 & 1.8\%\\
   4&Third lepton and tau veto        & 0.948 & 0.985 & 3.9\%
                                      & 0.949 & 0.986 & 4.0\%\\
   5&$b$-jet veto                     & 0.977 & 0.908 & 7.1\%
                                      & 0.979 & 0.910 & 7.1\%\\
   6&$\big|\Delta\varphi(\slashed{\mathbf{p}}_T,\mathbf{p}^{\ell\ell})\big|>2.7$
                                      & 0.705 & 0.733 & 4.0\%
                                      & 0.695 & 0.730 & 4.9\%\\
   7&$\big|\slashed{E}_T - p_T^{\ell\ell}\big|/p_T^{\ell\ell} < 0.2$
                                      & 0.678 & 0.648 & 4.4\%
                                      & 0.668 & 0.631 & 5.5\%\\
   8&$\slashed{E}_T > 80~{\rm GeV}$   & 0.915 & 0.910 & 0.6\%
                                      & 0.902 & 0.899 & 0.4\%\\
   9&At most one jet                  & 0.936 & 0.940 & 0.4\%
                                      & 0.943 & 0.941 & 0.2\%\\
  \end{tabular} \label{tab:m1000_pileup}}\vspace{4mm}

  \tbl{Same as in Table~\ref{tab:m5000} but when the pileup modeling is
    included in the \MA\ recasting.}
  {\begin{tabular}{c|c||c|c|c||c|c|c}
   \multicolumn{2}{c||}{\multirow{2}{*}{Selection step}}&
     \multicolumn{3}{c||}{Electron channel}&\multicolumn{3}{c}{Muon channel}\\
   \cline{3-8}
   \multicolumn{2}{c||}{}&
     $\epsilon_i^{\rm CMS}$ & $\epsilon_i^{\rm MA5}$ & $\delta_i^{\rm rel}$ &
     $\epsilon_i^{\rm CMS}$ & $\epsilon_i^{\rm MA5}$ & $\delta_i^{\rm rel}$ \\
   \hline
   1&Two leptons                      &   -   &   -   &  -
                                      &   -   &   -   &  -   \\
   2&$m_{\ell\ell} \in [80, 100]$~GeV & 0.928 & 0.931 & 0.1\%
                                      & 0.921 & 0.923 & 0.3\%\\
   3&$p_T^{\ell\ell} > 50$~GeV        & 0.841 & 0.844 & 0.4\%
                                      & 0.832 & 0.833 & 0.2\%\\
   4&Third lepton and tau veto        & 0.947 & 0.985 & 4.0\%
                                      & 0.945 & 0.986 & 4.3\%\\
   5&$b$-jet veto                     & 0.977 & 0.895 & 8.4\%
                                      & 0.978 & 0.897 & 8.3\%\\
   6&$\big|\Delta\varphi(\slashed{\mathbf{p}}_T,\mathbf{p}^{\ell\ell})\big|>2.7$
                                      & 0.708 & 0.739 & 4.4\%
                                      & 0.698 & 0.728 & 4.3\%\\
   7&$\big|\slashed{E}_T - p_T^{\ell\ell}\big|/p_T^{\ell\ell} < 0.2$
                                      & 0.687 & 0.661 & 3.8\%
                                      & 0.684 & 0.644 & 5.9\%\\
   8&$\slashed{E}_T > 80~{\rm GeV}$   & 0.923 & 0.928 & 0.6\%
                                      & 0.908 & 0.018 & 1.1\%\\
   9&At most one jet                  & 0.932 & 0.926 & 0.7\%
                                      & 0.935 & 0.930 & 0.5\%\\
  \end{tabular} \label{tab:m5000_pileup}}\vspace*{4mm}
\end{table}\renewcommand{\arraystretch}{1}%

\renewcommand{\arraystretch}{1.30}%
\begin{table}
  \tbl{
    Same as in Table~\ref{tab:efficiencies} when the pileup modeling is included
    in the \MA\ recasting.}
  {\begin{tabular}{c||c|c|c||c|c|c}
   \multirow{2}{*}{Scenario}&
     \multicolumn{3}{c||}{Electron channel}&\multicolumn{3}{c}{Muon channel}\\
   \cline{2-7}
   & $\epsilon^{\rm CMS}$ & $\epsilon^{\rm MA5}$ & $\delta^{\rm rel}$ &
     $\epsilon^{\rm CMS}$ & $\epsilon^{\rm MA5}$ & $\delta^{\rm rel}$ \\
   \hline
   $m_{Z'} = 10$~GeV                  & 0.178 & 0.170 & 5.5\%
                                      & 0.173 & 0.160 & 7.6\%\\
   $m_{Z'} = 200$~GeV                 & 0.186 & 0.186 & 0.3\%
                                      & 0.189 & 0.180 & 5.8\%\\
   $m_{Z'} = 500$~GeV                 & 0.269 & 0.250 & 6.9\%
                                      & 0.258 & 0.240 & 7.3\%\\
   $m_{Z'} = 1000$~GeV                & 0.294 & 0.280 & 5.4\%
                                      & 0.279 & 0.260 & 6.1\%\\
   $m_{Z'} = 5000$~GeV                & 0.302 & 0.290 & 4.3\%
                                      & 0.287 & 0.270 & 5.1\%\\
  \end{tabular} \label{tab:eff_pileup}}
\end{table}\renewcommand{\arraystretch}{1}%

\begin{figure}
  \centering
  \includegraphics[width=0.495\columnwidth]{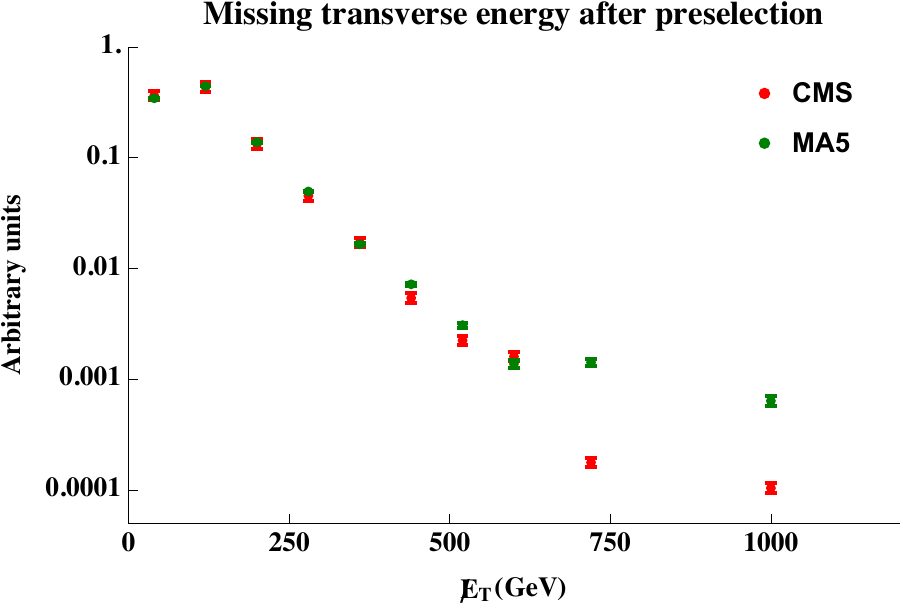}
  \includegraphics[width=0.495\columnwidth]{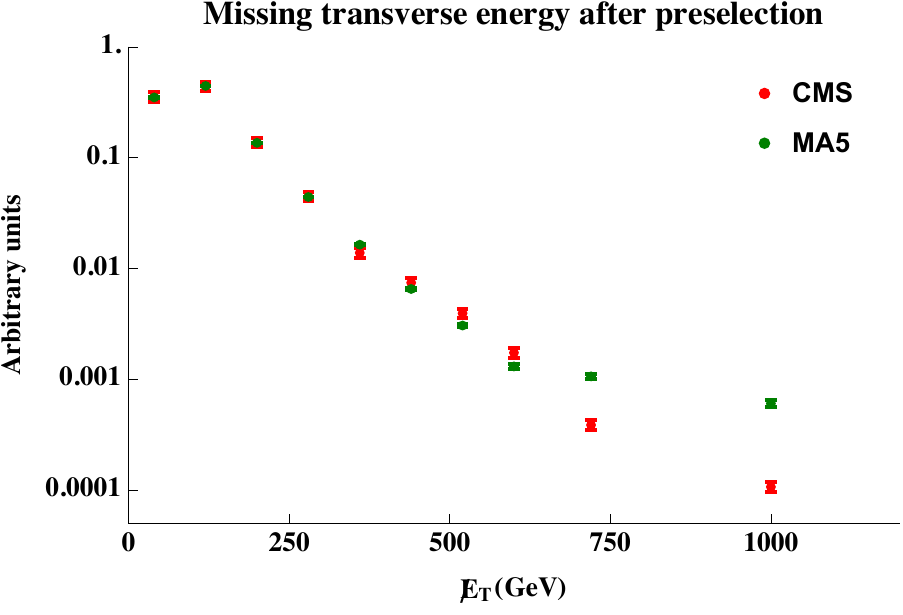}\\[3mm]
  \includegraphics[width=0.495\columnwidth]{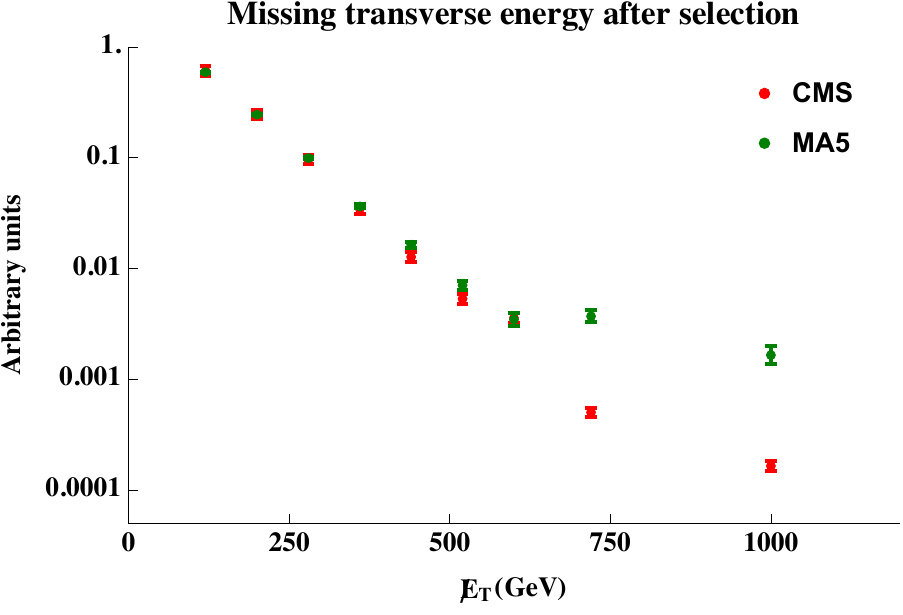}
  \includegraphics[width=0.495\columnwidth]{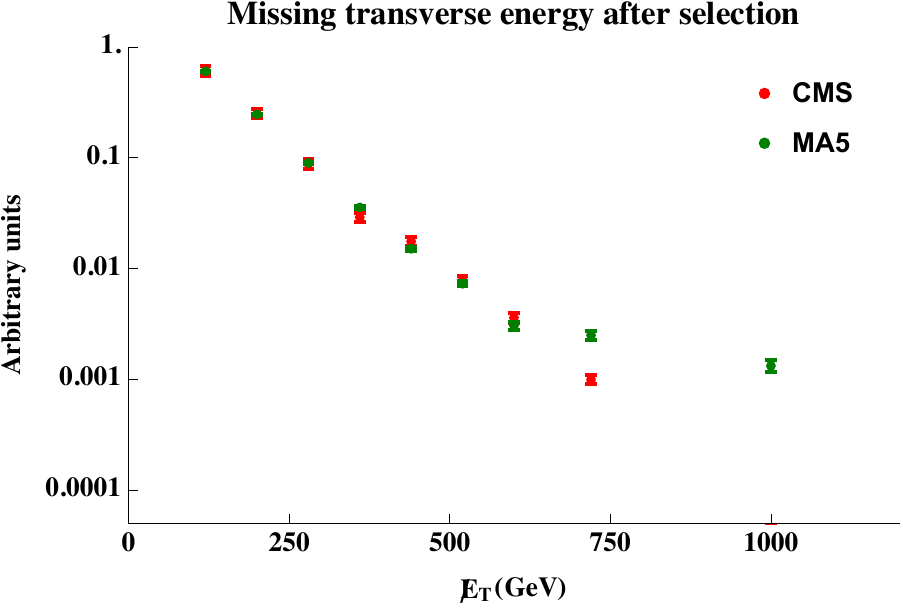}
  \caption{Same as in Figure~\ref{fig:histos} but when the pileup modeling is
    included in the \MA\ recasting.}
  \label{fig:hist_pileup}
\end{figure}

Altogether, the differences between the \MA\ and CMS results seem to point
towards a mismodeling of the missing energy.
The latter is known to be very sensitive to pileup effects, so that we perform
the exercise a second time, using instead a \DEL\ cards including the modeling
of the pileup (see Section~\ref{sec:delphes_config}). The results are shown in
Tables~\ref{tab:m10_pileup}--\ref{tab:m5000_pileup} for the cutflows
associated with the different
signals under consideration, as well as in Table~\ref{tab:eff_pileup}
for the total signal efficiency. Distributions in the missing energy for the two
signal regions are given in Figure~\ref{fig:hist_pileup}. We observe a much
better agreement at all levels.

This leads us to the conclusion that the implementation of the CMS-EXO-16-010
analysis within the \MA\ framework can be considered as validated. We have found
a level of agreement with the experimental results that is of  5--10\%, so
that any prediction that would be made with the \MA\ framework could be seen as
reasonably accurate. The \DEL\ parameterization including the pileup effects is
the one that is recommended to be used.

\subsection[Submission to the Public Analysis Database of \MA]{Submission to the
Public Analysis Database of \MAnorm} \label{sec:submit}

Once the recasted analysis is validated, it has to be submitted to the Public
Analysis Database of \MA\ so that it can be merged with the current release of
the program. This task is achieved in two steps.

First, a tarball containing the
three analysis files, namely the C++ header and source files and the information
file (in our case, the \verb+cms_exo_16_010.h+, \verb+cms_exo_16_010.cpp+ and
\verb+cms_exo_16_010.info+ files) should be submitted to \INSP\ via a web-based
platform located at\\
\hspace*{2.5mm}
  \url{https://inspirehep.net/help/data-submission}.

The submission platform contains a form in which the user is required to provide
a title for the analysis
to be submitted, a short description of what the code does, and a tarball with
the files. We recommend the user to choose a title including a couple of
keywords, the experimental identifier for the analysis and the integrated
luminosity that has been analyzed. The description should in particular detail
where more information on the reimplementation can be found, in particular
concerning the validation and the experimental inputs. In our
case, the adopted title is
\begin{verbatim}
  MadAnalysis5 implementation of the mono-Z analysis of CMS with
  2.3 fb-1 of data (CMS-EXO-16-010)
\end{verbatim}
and the description reads
\begin{verbatim}
  This consists of the reimplementation, in the MadAnalysis 5
  framework, of a CMS search for dark matter when it is produced in
  association with a leptonically decaying Z-boson. 2.3 fb-1 of LHC
  proton-proton collisions at 13 TeV has been analyzed.

  Information on how to use this code and a detailed validation
  summary are available on the Public Analysis Database of
  MadAnalysis 5
  (http://madanalysis.irmp.ucl.ac.be/wiki/PhysicsAnalysisDatabase).

  The CMS analysis is documented on the collaboration wiki
  http://cern.ch/cms-results/public-results/publications/EXO-16-010.
\end{verbatim}

Moreover, the web-based \INSP\ form contains a box that should be ticked for the
creation of a Digital Object Identifier (DOI). This ensures that a DOI will
be assigned by \INSP\ to the reimplemented code, so that the latter becomes
uniquely identifiable, searchable and citable. Thanks to the \INSP\ versioning
system, potential changes in the code can moreover be reliably traced.
The DOI that has been assigned to the CMS-EXO-16-010 \MA\ reimplementation is
given by
\begin{verbatim}
  10.7484/INSPIREHEP.DATA.RK53.S39D
\end{verbatim}

In a second step, the code is merged with the current release of \MA. To this
end, the user is requested to contact the \MA\ authors
(\verb+ma5team@iphc.cnrs.fr+) and provide the \INSP\ reference together with the
DOI, the \DEL\ configuration file and a detailed note
allowing any potential reader to reproduce the validation procedure. This note
may optionally be linked to files (like Monte Carlo event generation
configuration cards or SLHA benchmark definitions). Subsequently,
an entry will be (manually) added to the online public analysis database
of \MA\ and the \verb+install PAD+ command detailed in Section~\ref{sec:pad}
will be updated so that the PAD installation process would also download the
files related to the newly validated analysis from \INSP. Updates of the
reimplemented code are automatically handled as long as
they are submitted to \INSP.

\section{Using \MA~for reinterpreting LHC results} \label{sec:recast_example}
\subsection[Recasting LHC analyses with \MA]{Recasting LHC analyses with
\MAnorm}\label{sec:outputruns}

Reinterpreting the results of an analysis in a different theoretical context has
been made straightforward and fairly easy in the normal mode of
\MA\cite{Conte:2012fm}. The only task left to the user is to provide signal
events and the list of analyses
included in the database for which a reinterpretation is desired.

Practically, we assume that a signal Monte Carlo event sample, that we denote in
the following by \verb+myevents.hepmc.gz+, has been generated. This event
generation process has to include both the simulation of the hard-scattering
process, the one of
the parton showering and the hadronization. \MA\ has to be started
in the reconstructed-level mode, which is achieved by typing in a shell, from
the root directory in which the \MA\ tarball has been unpacked,
\begin{verbatim}
  ./bin/ma5 -R
\end{verbatim}
This subsequently starts the \MA\ command line interface in which the user can
access all the functionalities of the program.

First, the user is requested to switch the recasting mode on, which is achieved
by typing in the command
\begin{verbatim}
  set main.recast = on
\end{verbatim}
He/she then needs to import his/her event sample and link it to a dataset label,
\begin{verbatim}
  import <path-to-events.hepmc.gz>
\end{verbatim}
Subsequently, the event sample \verb+myevents.hepmc.gz+ is attached to the
default dataset \verb+defaultset+, as no explicit label as been provided at the
time of calling the \verb+import+ command. In the case where the user would like
to assign another label to the dataset, the latter command has to be issued
slightly differently,
\begin{verbatim}
  import <path-to-events.hepmc.gz> as <my-label>
\end{verbatim}
where \verb+<my-label>+ is replaced by the desired label.

Several event files could also be imported and either collected under a unique
dataset (by using the same \verb+my-label>+ label at each relevant occurence of
the import command), or split through a set of different datasets (being thus
tagged by different dataset labels). The analysis is performed on all defined
datasets, that can thus possibly refer to several event files each, and each
event of each imported file is hence analyzed.

The user is allowed to enter two optional commands that impact how the code
behaves. They consist of
\begin{verbatim}
  set main.recast.store_root = <value>
  set main.recast.card_path = <path-to-a-recasting-card>
\end{verbatim}
The former command tells the code whether to store the \ROOT\ file outputted by
\DEL, the possible values being \verb+True+ or \verb+False+ (the default
behavior), whereas the latter command indicates to use a predefined recasting
card by providing its path. In the case where no recasting card is provided,
\MA\ checks which analyses are included in the PAD and accordingly
creates a recasting card. On run time, the user is asked whether he/she
wants to modify the card.

Such a recasting card contains a sequence of lines, where one given line maps
one of the available analyses. The syntax for a line of the card follows
\begin{verbatim}
 <analysis-tag>   <type>   <switch>  <delphes-card>   #   <comment>
\end{verbatim}
The analysis tag corresponds to the file name of the C++ code containing the
relevant analysis. In the case of the CMS-EXO-16-010 example worked out in
Section~\ref{sec:implementation}, this would read \verb+cms_exo_16_010+. The
second tag is related to the version of the PAD that has been used for the
analysis implementation. For analyses implemented in the framework of the more
ancient `PADForMA5tune' PAD version, the keyword \verb+v1.1+ should be included,
whilst for the (recommended and) current version, it is enough to set this tag
to \verb+v1.2+. This reflects the fact that the version v1.2 of \MA\ has been
the first \MA\ version compatible with the current PAD format. The
\verb+<switch>+ tag controls whether the analysis has to be recasted,
its value being either \verb+on+ or \verb+off+. Furthermore, information on the
\DEL\ card to use for the simulation of the detector is provided via the
\verb+<delphes-card>+ tag, that must refer to the \DEL\ card filename (including
the \verb+tcl+ extension). We recall that all cards are
located in the \verb+PAD/Input/Cards+ directory. Finally, the last tag contains
a comment generally providing an information on the analysis.

In order to reinterpret the CMS-EXO-16-010 analysis of the
previous section, we have set the switches of all entries of the default
recasting card to \verb+off+, with the exception of the one related to the
considered analysis that is kept to the \verb+on+ value. The corresponding line
in the recasting card is given by
\begin{verbatim}
 cms_exo_16_010 v1.2 on delphes_card_cms_exo_16_010.tcl # CMS13-MonoZ
\end{verbatim}

\MA\ has then all the information necessary for performing the recasting tasks.
The run is initiated by submitting the recasting job and typing in the command
line interface
\begin{verbatim}
  submit
\end{verbatim}
The code starts by simulating the response of the detector using all referenced
\DEL\ cards for which at leat one analysis has been switched on, and it next
computes, analysis by analysis, how the different signal regions are populated
by the
analyzed signal events. From these results and the information on the Standard
Model background and on the data provided in the different \verb+info+ files,
exclusion limits are computed using the CL$_s$
prescription\cite{Read:2002hq}. More information on the output and the way in
which the calculation of the exclusion limits is done is given in
Section~\ref{sec:reca_output} and Section~\ref{sec:limit-setting} respectively.

\subsection[Output of the automatic recasting procedure of \MA]{Output of the
  automatic recasting procedure of \MAnorm} \label{sec:reca_output}
Once the \MA\ reinterpretation run detailed in the previous section has been
performed, the output is stored in a folder named \verb+ANALYSIS_X+, where
\verb+X+ is a number iteratively incremented at each run of \MA. The name of the
output folder can also be chosen by the user, by
providing an argument when issuing the \verb+submit+ command (see the end of
Section~\ref{sec:outputruns}), as in
\begin{verbatim}
  submit <name-of-the-output-folder>
\end{verbatim}

This output folder contains one file, named \verb+history.ma5+, as well as two
subfolders named \verb+Input+ and \verb+Output+. The file consists of a history
of the commands that have been typed in the \MA\ command line interface prior to
the \verb+submit+ command, and the \verb+Input+ directory contains the recasting
card (\verb+recasting_card.dat+) that has been used by the code (see
Section~\ref{sec:outputruns}), together with a text file, whose extension is
\verb+list+, that contains the path to all the event samples that have been
analyzed (the path to the \verb+myevents.hepmc.gz+ file in our practical
example).

The \verb+Output+ directory is the folder in which the results are collected. It
contains a single text file, \verb+CLs_output_summary.dat+, as well as at least
five
subfolders. Four of these subfolders (\verb+DVI+, \verb+Histos+, \verb+HTML+ and
\verb+PDF+) are there for the purpose of having a single output structure for
all the possible usages of \MA, and are not relevant for the recasting purpose.
The other folders, that are named according to the labels which the analyzed
datasets have been assigned to (see
Section~\ref{sec:outputruns}), contain the reinterpretation results. Each
dataset is hence related to a specific folder. A fraction
of the information is moreover summarized in the \verb+CLs_output_summary.dat+
file. This file is comprised of a series of lines, each line being each
associated with a given dataset and a given signal region of one of the
analyses that have been reinterpreted. The format of a line follows
\begin{verbatim}
<set>  <analysis>  <SR>  <exp>  <obs> || <eff>  <stat>  <syst>  <tot>
\end{verbatim}
The first element of the line consists of the name of one of the datasets that
have been analyzed. The \verb+<analysis>+ tag then refers to the name of one of
the analyses that have been reinterpreted, and \verb+<SR>+ to one of the
implemented signal regions of this analysis. \verb+<exp>+ and \verb+<obs>+ are
the values of the cross section that would be excluded at the 95\% confidence
level when Standard Model predictions (expected exclusion) and data (observed
exclusion) are used for the number of events populating the different signal
regions, respectively. These cross sections are
computed by using the results of the sole single signal region \verb+<SR>+ of
the analysis \verb+<analysis>+, and correspond to the cross sections that should
be associated with the analyzed event sample(s) to yield its exclusion at the
95\% confidence level. We provide more information in
Section~\ref{sec:limit-setting} on the actual way in which those numbers are
derived.

A specific cross section value can be attached to the dataset under
consideration, and can be inputted by using standard \MA\ commands. It is then
required to type, in the \MA\ interpreter,
\begin{verbatim}
  set <dataset-label>.xsection = <value>
\end{verbatim}
before the submission command. \verb+<dataset-label>+ refers to the label of the
dataset under consideration and \verb+<value>+ to the value of the associated
cross section.
In this case, an extra number is included in the outputted summary file, before
the double vertical line, and corresponds to the confidence level at which the
analyzed sample is excluded, given the provided cross section. 

On the right hand-side of the double vertical line, the four tags \verb+<eff>+,
\verb+<stat>+, \verb+<syst>+ and \verb+<tot>+ respectively refer to the
selection efficiency associated with the \verb+<SR>+ region of the
\verb+<analysis>+ analysis, and the associated statistical, systematical and
total uncertainties. The systematics are for the moment always set to zero, the
corresponding location in the output file being reserved for future
developments.

As above-mentioned, the \verb+Output+ folder contains a series of subfolders
whose name is the label of one of the defined datasets. In our example, no
dataset label has been assigned and a single event file has been imported, so
that there is a single of such a `dataset' folder. It is named
\verb+defaultset+.

A dataset folder contains
two files, a subfolder named \verb+RecoEvents+ and a set of extra subfolders
whose name is one of the recasted analyses. As in our example a single analysis
has been reinterpreted, only one of such `analysis' subfolders is present, and
it is named
\verb+cms_exo_16_010+. We start by discussing the information stored in the two
available files included in the dataset folder. One of them is named
\verb+CLs_output.dat+ and contains the same information as provided in
the \verb+CLs_output_summary.dat+ file, restricted to the relevant dataset and
encoded following the same syntax. The second file carries the dataset name and
the \verb+saf+ extension, so that its name is \verb+defaultset.saf+ in our
example. This file contains global information on the dataset and includes
the associated cross section (if available), the number of events, weight
information and the path to the various event files which have been attached to
the dataset. Detailed information on each
file then follows (number of events, cross section and weight information). The
format of the SAF file follows an XML-like syntax. In addition to a header and a
footer (\verb+SAFheader+ and \verb+SAFfooter+), the information is organized as
follows. The global information on the
sample is provided via the \verb+<SampleGlobalInfo>+ XML element,
\begin{verbatim}
 <SampleGlobalInfo>
    # xsection  xsec_error  nevents  sum_wgt+  sum_wgt-
    0.0e+00     0.0e+00     0        0.0e+00   0.00e+00
 </SampleGlobalInfo>
\end{verbatim}
where the numerical values have been fixed to zero for the sake of the
illustration. The meaning of each value is self-explanatory. The paths to the
different event files attached to the dataset are stored in the
\verb+<FileInfo>+ element, which reads in our example
\begin{verbatim}
 <FileInfo>
 "<reco>/RecoEvents_v1x2_delphes_card_cms_exo_16_010.root" # file 1/1
 </FileInfo>
\end{verbatim}
where \verb+<reco>+ stands for the absolute path to the \verb+RecoEvents+ folder
above-mentioned. The detailed information on the different files is provided
similarly to the global information on the dataset, but uses the
\verb+<SampleDetailedInfo>+ XML root element instead.

The \verb+RecoEvents+ folder is the one used to store all \ROOT\ files
outputted by \DEL, one file being connected to one given \DEL\ parameterization.
As sketched above (see the \verb+<FileInfo>+ snippet), the filename contains a
reference to the corresponding \DEL\ card to facilitate its identification. In
the case where the \MA\ run does not requires to store the \ROOT\ files
generated by \DEL, the \verb+RecoEvents+ folder is kept empty. We now move on
with a description of the analysis folders containing the
cutflows and the histograms.

Each of these folders contain one SAF file
with a reference to all the signal regions implemented in the considered
analysis. The syntax used to encode this SAF file is again XML-like. In
addition to a header and a footer, the file contains a \verb+<RegionSelection>+
element with the name of the regions defined in the corresponding analysis. In
our example, the folder name is \verb+cms_exo_10_016+, as already mentioned, and
the name of the SAF file is \verb+cms_exo_10_016.saf+. The region
information is then encoded as
\begin{verbatim}
  <RegionSelection>
      "ee"
      "mumu"
  </RegionSelection>
\end{verbatim}
one line being used for each signal region.
The analysis folder additionally contains two subfolders \verb+Cutflows+ and
\verb+Histograms+. The former contains one SAF file for each signal
region (thus two SAF files named \verb+ee.saf+ and \verb+mumu.saf+ in our
example), and the latter a single file named \verb+histos.saf+. Each cutflow
SAF file is once again encoded following an XML-like syntax. It contains two
elements, in addition to a header and footer, denoted by \verb+<InitialCounter>+
and \verb+<Counter>+. The first of these two tags is used once, for the initial
number of events prior to any cut. The second of these tags is used multiple
times, one occurence being associated with each cut. Taking again the
CMS-EXO-16-010 example, the head of the file associated to the \verb+ee+ region
could be given by
\begin{verbatim}
  <InitialCounter>
  "Initial number of events"      #
  300000          0               # nentries
  1.339665e+00    0.000000e+00    # sum of weights
  5.985929e-06    0.000000e+00    # sum of weights^2
  </InitialCounter>

  <Counter>
  "2_electrons"                   # 1st cut
  40832           0               # nentries
  1.819489e-01    0.000000e+00    # sum of weights
  8.111897e-07    0.000000e+00    # sum of weights^2
  </Counter>
\end{verbatim}
The different lines are respectively related to the number
of events surviving the cut, the weighted number of events surviving the cut and
the number of events, weighted by the square of the event weight (which allows
to calculate the associated variance). The first column is related to the
positive-weighted
events where the second column contains the information connected to the
negative-weighted events, if any.

The histogram file \verb+histos.saf+ is also following an XML-like structure and
contains \verb+<Histo>+ elements. Each instance corresponds to one of the
declared histograms of the analysis. The histogram information is spread over
three subtags, named \verb+<Description>+, \verb+<Statistics>+ and
\verb+<Data>+, so that one entry of the \verb+histos.saf+ file schematically
read
\begin{verbatim}
  <Histo>
    <Description>
        ...
    </Description>
    <Statistics>
        ...
    </Statistics>
    <Data>
        ...
    </Data>
  </Histo>
\end{verbatim}
The description part contains general information about the histogram, such as
the histogram name, the number of bins, the minimum and maximum value on the
$x$-axis and the regions which the histogram is attached to. Taking the example
of the first histogram declared in Section~\ref{sec:init}, the content of the
description field is
\begin{verbatim}
  <Description>
    "MET_preselected_e"
    # nbins   xmin           xmax
      15      0.000000e+00   1.200000e+03
    # Defined regions
      ee    # Region nr. 1
  </Description>
\end{verbatim}
This indicates that there is a histogram named \verb+MET_preselected_e+ that
contains 15 bins ranging from 0 to 1200~GeV. Moreover, the histogram is only
relevant for the region \verb+ee+. The statistics part includes typical
information like the number of events populating the histogram, the
number of entries and various weighted sums. Continuing with the above example,
we have a statistics information block
\begin{verbatim}
  <Statistics>
    21609          0              # nevents
    9.672496e-02   0.000000e+00   # sum of event-weights over events
    21609          0              # nentries
    9.672496e-02   0.000000e+00   # sum of event-weights over entries
    4.331157e-07   0.000000e+00   # sum weights^2
    1.130326e+01   0.000000e+00   # sum value*weight
    1.870627e+03   0.000000e+00   # sum value^2*weight
  </Statistics>
\end{verbatim}
which is again self-explanatory. Like for the cuts, the first columnm is related
to the positive-weighted events and the second one to the negative-weighted
events. The values for each bin is then stored within the \verb+<Data>+ XML
element,
\begin{verbatim}
  <Data>
      0.000000e+00   0.000000e+00    # underflow
      3.561772e-02   0.000000e+00    # bin 1 / 15
      4.170943e-02   0.000000e+00    # bin 2 / 15
      .
      .
      .
      0.000000e+00   0.000000e+00    # bin 14 / 15
      0.000000e+00   0.000000e+00    # bin 15 / 15
      0.000000e+00   0.000000e+00    # overflow
  </Data>
\end{verbatim}
where the dots stand for all bin values omitted for brevity. Underflow and
overflow bins are also included, and the two columns again refer to the
positive-weighted and negative-weighted events.

\subsection{Getting an exclusion limit}
\label{sec:limit-setting}
In the previous section, we have mentioned that when running \MA, the limits on
a new physics signal are given in two or three different ways. In order to fix
the notations, we assume that an event sample has been generated in a
specific new physics context, and that the associated cross section is denoted
by $\sigma$. A first manner to quote limits is to associate with $\sigma$ the
exclusion level of the signal, using the CL$_s$ prescription.
Two other ways correspond to the computation of the minimum value
that the $\sigma$ cross section should be in order to lead to a signal excluded
at the 95\% confidence level, firstly when accounting for the observed number of
events and secondly by using as the observed number of events the expected
number of events. These excluded cross sections are respectively denoted by
$\sigma_{95}^{\rm exp}$ and $\sigma_{95}^{\rm obs}$. This additionally allows
\MA\ to identify the best signal region that can exclude the signal, \ie, the
signal
region that is statistically the most significant. Whereas \MA\ calculations of
the exclusions are performed for each signal region of each reinterpreted
analysis independently, we consider, in the following, a given signal region of
a given analysis.

The observed number of events $n_{\rm obs}$ populating the signal region under
consideration and the associated expected number of Standard Model background
events $n_{\rm b}$ are extracted from the analysis \verb+info+ file
(\verb+cms_exo_16_010.info+ in our example), together with the uncertainty
$\Delta n_{\rm b}$ on $n_{\rm b}$. As a result of the run of \MA, the number of
signal events $n_{\rm s}$ that are expected can be evaluated as
\be
  n_{\rm s} = \varepsilon_{\rm s}\ {\cal L}\ \sigma\ .
\ee
The ratio of the last to the first entry of the cutflow table associated with
the considered signal region allows for the evaluation of the signal efficiency
$\varepsilon_{\rm s}$, whereas the information on the integrated luminosity
${\cal L}$ is read from the \verb+info+ file and the cross section value
$\sigma$ is the provided one.

A large number of toy experiments ($N$) is then generated. For each of
those, the code computes the background and signal-plus-background probabilities
$p_{\rm b}$ and $p_{{\rm b} + {\rm s}}$ and derive a CL$_s$ number from these.

More precisely, for each toy
experiment, the expected number of background events $N_{\rm b}$ is randomly
chosen assuming that it follows a Gaussian distribution with a mean $n_{\rm b}$
and a width $\Delta n_{\rm b}$, the corresponding probability density
$f(N_{\rm b} \ | \ n_{\rm b}, \Delta n_{\rm b})$ being thus given by
\be
  f\big(N_{\rm b} \ \big| \ n_{\rm b}, \Delta n_{\rm b}\big) =
     \frac{1}{\sqrt{2 \pi \big(\Delta n_{\rm b}\big)^2}} \exp\Bigg[
        -\frac{\big(N_{\rm b}-n_{\rm b}\big)^2}{2 \big(\Delta n_{\rm b}\big)^2}
   \Bigg] \ .
\ee
Ignoring negative $N_{\rm b}$ values, the actual number of background events
$\hat N_{\rm b}$ is randomly generated after assuming that it follows a
Poisson distribution with parameter $N_{\rm b}$, the corresponding probability
mass function $f(\hat N_{\rm b} \ \big| \ N_{\rm b})$ reading
\be\label{eq:poisson}
  f\big(\hat N_{\rm b} \ \big| \ N_{\rm b} \big) =
   \frac{\big(N_{\rm b}\big)^{\hat N_{\rm b}}\  e^{-N_{\rm b}}}
     {\hat N_{\rm b}!} \ .
\ee
Knowing that $n_{\rm obs}$ events have been observed, we define the background
probability $p_{\rm b}$ as the probability for the background to fluctuate as
low as $n_{\rm obs}$, which is also given by the fraction of toy experiments for
which
\be
  \hat N_{\rm b} \leq n_{\rm obs} \ .
\ee

In a second step, we evaluate the signal-plus-background probability
$p_{{\rm b} + {\rm s}}$. We start
by randomly generating the actual number of signal-plus-background events
$\hat N_{\rm b}  + \hat N_{\rm s}$ in assuming that it follows a Poisson
distribution with parameter $n_{\rm s} + N_{\rm b}$. The corresponding
probability mass function $f(\hat N_{\rm b} + \hat N_{\rm s} \ \big| \
N_{\rm b} + n_{\rm s})$ is in this case similar to Eq.~\eqref{eq:poisson},
\be
  f\big(\hat N_{\rm b}  + \hat N_{\rm s}\ \big| \ N_{\rm b} + n_{\rm s}\big) =
   \frac{\big(N_{\rm b} + n_s\big)^{\hat N_{\rm b} + \hat N_{\rm s}}\
      e^{-(N_{\rm b} + n_{\rm s})}}
     {\big(\hat N_{\rm b} + \hat N_{\rm s}\big)!} \ .
\ee
after retaining only experiments for which $N_{\rm b} + n_{\rm s}$ is strictly
positive. The probability $p_{{\rm b} + {\rm s}}$ is then defined as the
probability for the signal-plus-background to fluctuate as low as $n_{\rm obs}$.
Once again, we evaluate this probability by determining the fraction of toy
experiments for which
\be
  \hat N_{\rm b}  + \hat N_{\rm s} \leq n_{\rm obs} \ .
\ee

The CL$_s$ value is then derived as
\be
  {\rm CL}_s = {\rm max}\Big(0, 1-\frac{p_{{\rm b}+{\rm s}}}{p_{\rm b}}\Big) \ .
\ee
The corresponding observed cross section excluded at the 95\% confidence level
$\sigma_{95}^{\rm obs}$ is calculated as above, but after setting $\sigma$ free
and determining the minimum value for which we get ${\rm CL}_s = 95 \%$. The
expected cross section excluded at the 95\% confidence level
$\sigma_{95}^{\rm exp}$ is evaluated in the same way, but when one assumes that
$n_{\rm obs} = n_{\rm b}$.

By default, all calculations are made with a number of toy experiments set to
\be
  N = 100000 \ .
\ee
This value can be modified by typing, in the \MA\ command line interface,
\begin{verbatim}
  set main.recast.CLs_numofexps = <value>
\end{verbatim}
where \verb+<value>+ is the desired integer number.

\subsection{Manual execution of the recasted analyses}
In the previous sections, we have shown how to reinterpret the results of an
experimental analysis starting from a signal event file at the hadronic level,
\ie, after that hadronization has been simulated. It may however be sometimes
desirable to start the process from a \ROOT\ file, after the simulation of the
detector response, either using a different detector parameterization or
because the \ROOT\ file has already been generated (so that CPU time could
be spared). In this case, there is a way to run the PAD manually, without
relying on the pythonic mode of \MA.

First, an input file has to be created in the \verb+PAD/Inputs+ directory. This
file includes the paths to all event files to be analyzed, these event files
being in the \DEL\ output format.
Second, the user must edit the \verb+PAD/Build/Main/main.cpp+ file
in order to switch on and off the analyses that must be reinterpreted. This is
achieved by commenting out the relevant \verb+Execute+ and
\verb+Initialize+ functions in the file. For instance, switching off the run of
the CMS-EXO-16-036 analysis is done by commenting out its initialization,
\begin{verbatim}
  //  std::map<std::string, std::string> prmcms_sus_16_036;
  //  AnalyzerBase* analyzer_cms_sus_16_036=
  //    manager.InitializeAnalyzer("cms_sus_16_036",
  //     "cms_sus_16_036.saf",prmcms_sus_16_036);
  //  if (analyzer_cms_sus_16_036==0) return 1;
\end{verbatim}
and execution,
\begin{verbatim}
  // if (!analyzer_cms_sus_16_036->Execute(mySample,myEvent))
  //   continue;
\end{verbatim}

 After compiling the
\verb+MadAnalysis5job+ executable as mentioned in Section~\ref{sec:pad} (\ie, by
typing \verb+make+ in a shell from the \verb+PAD/Build+ directory), the code can
be executed on the event sample(s) to analyze. This is done by typing in a
shell, from the \verb+PAD/Build+ directory,
\begin{verbatim}
   ./MadAnalysis5job  ../Inputs/<input>
\end{verbatim}
where \verb+<input>+ is the name of the input file. The output is generated and
stored in the \verb+PAD/Output/<input>+ directory, when the name \verb+<input>+
refers to the input file that has been used. The structure of the output
directory is similar to the one described in Section~\ref{sec:reca_output}, with
the exception of the files with the CL$_s$ information that are this time
absent. The corresponding exclusions are indeed not evaluated, and it is up
to the user to derive them from the cutflow SAF files and the analysis
\verb+info+ files.

\subsection{Example: constraining top-philic dark matter with mono-$Z$ probes}

In this section, we illustrate with a physics case the capabilities of \MA\ for
recasting. We make use of the CMS-EXO-16-010 analysis that we have picked as an
example, and reinterpret its results to constrain a simplified dark matter model
where dark matter couples dominantly via a scalar mediator to top quarks. This
class of models is commonly called top-philic dark matter models and is usually
constrained by a variety of collider searches\cite{Arina:2016cqj}, including
monojet\cite{Feng:2005gj,Bai:2010hh}, mono-$Z$-boson\cite{Bell:2012rg},
mono-Higgs\cite{Petrov:2013nia} and monolepton\cite{Bai:2012xg} probes.
Monophoton searches\cite{Birkedal:2004xn,Fox:2011fx} are here
irrelevant as photon emission is forbidden at the leading-order by virtue of
charge conjugation invariance.

We focus in this section on the evaluation of the sensitivity of
mono-$Z$ probes to a specific class of dark matter models when 2.3~fb$^{-1}$ of
LHC data is analyzed. The simplified model that we consider is built by
complementing the Standard Model with a dark matter particle $\chi$ of mass
$m_\chi$ that is assumed to be a Dirac fermion, and a scalar mediator $\varphi$
of mass $m_\varphi$. The interactions of the two particles are described by the
new physics Lagrangian
\be
  {\cal L}_{\rm NP} =
    \partial_\mu \varphi^\dag \partial^\mu \varphi - m_\varphi^2 \varphi^\dag
      \varphi
    + i \bar\chi \slashed{\partial}\chi - m_\chi \bar\chi\chi
    - g_t \frac{y_t}{\sqrt{2}} \bar t t \varphi
    - g_\chi \bar \chi\chi \varphi \ ,
\ee
where the new physics coupling strengths are denoted by $g_t$ and $g_\chi$
($y_t$
being the top quark Yukawa coupling). In this model description, the form of the
interactions of the mediator to the Standard Model particles is inspired by the
minimal flavor-violation paradigm\cite{DAmbrosio:2002vsn,Buras:2000dm} so that
they are proportional to the Standard Model fermion masses. Although this model
is neither ultraviolet-complete nor gauge invariant, its Lagrangian consists of
a good example for describing in a simplified way a top-philic dark matter
scenario. The model has four free parameters,
\be
  \Big\{ g_t, g_\chi, m_\chi, m_\varphi \Big\} \ ,
\ee
and for the sake of the example, we fix both couplings to
\be
  g_\chi = g_t = 4 \ ,
\label{eq:coup}\ee
and keep the two new physics masses free.

\begin{figure}
  \centering
  \includegraphics[width=0.35\columnwidth]{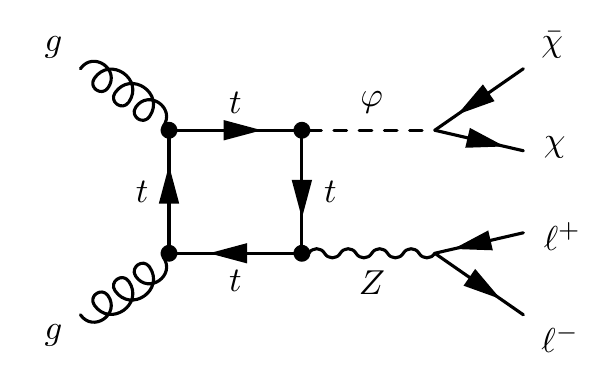}
  \caption{Representative leading-order Feynman diagram for the production of a
    pair of dark matter particles $\chi$, originating from the decay of a
    (possibly off-shell) top-philic scalar mediator $\varphi$, in association
    with a (possibly off-shell) leptonically-decaying $Z$-boson.}
  \label{fig:LImonoz}
\end{figure}
\begin{figure}
  \centering
  \includegraphics[width=0.850\columnwidth]{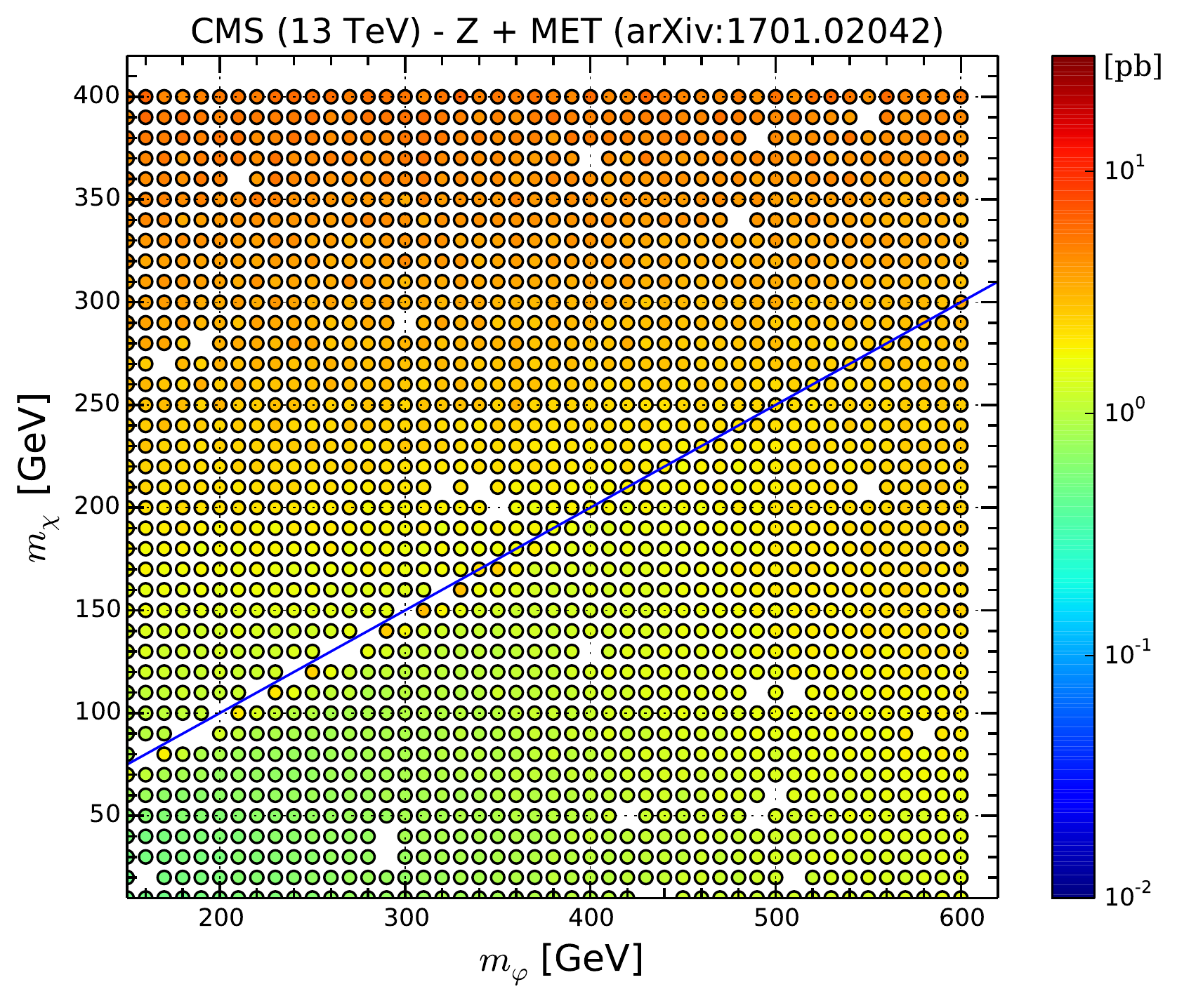}
  \caption{Constraints derived from the mono-$Z$-boson CMS-EXO-16-010 analysis
    of 3.2~fb$^{-1}$ of LHC data at a collision energy of 13~TeV on the
    top-philic dark matter model under consideration. The results are presented
    in the $(m_\varphi, m_\chi)$ plane under the form of a limit on the
    production
    cross section of the $pp\to \ell^+\ell^- \varphi$ process. We show, for
    different choices of new physics masses, the cross section value that is
    excluded at the 95\% confidence level. We moreover recall that both new
    physics couplings are set to $g_\chi = g_t = 4$. The blue line represents the
    $m_\varphi = 2 m_\chi$ regime where dark matter can be resonantly produced
    from the mediator decay.}
  \label{fig:constraints_monoz}
\end{figure}

In the model under consideration, a mono-$Z$-boson signal can originate from the
loop-induced process,
\be
  p p \to \ell^+\ell^- \bar\chi\chi\ ,
\label{eq:monoz2}\ee
for which a representative Feynman diagram is shown in Figure~\ref{fig:LImonoz}.
The corresponding diagrams involve the exchange of both a mediator and a
$Z$-boson, and they both can be either on-shell or off-shell. The compatibility
of the dilepton system with the $Z$-boson is however further imposed at the
analysis level. Hard-scattering events are generated with the help of the loop
module of \MG\cite{Alwall:2014hca,Mattelaer:2015haa}, and the hard process
matrix element is convolved with the leading-order set of
NNPDF parton densities version 3.0\cite{Ball:2014uwa}. The unphysical scales are
all set to the geometric mean of the transverse mass of all final-state
particles, and the width of the scalar mediator is calculated with
\MW\cite{Alwall:2014bza}. The simulation of the parton showering and of the
hadronization has been performed with \PY\cite{Sjostrand:2014zea}, and we
simulate the detector response with \DEL\cite{deFavereau:2013fsa} using the
detector parameterization designed in Section~\ref{sec:delphes_config}. The
signal selection and limit calculations are performed with the help of
\MA\cite{Conte:2014zja,Conte:2012fm}, as described in
Section~\ref{sec:implementation} and in the beginning of this section.

In Figure~\ref{fig:constraints_monoz}, we represent the constraints that are
imposed by the CMS-EXO-16-010 analysis on the considered top-philic dark matter
model. The results are represented in the $(m_\varphi, m_\chi)$ mass plane, and
as the cross section which the process of Eq.~\eqref{eq:monoz2} should have to
induce a mono-$Z$-boson signal excluded at the 95\% confidence level. Two
regimes are considered and distinguished by the $m_\varphi = 2 m_\chi$ blue
line. In the low dark-matter mass region where the dark-matter mass $m_\chi$ is
at least twice smaller than the mediator mass $m_\varphi$ (\ie, below the blue
line), the final state can be produced through the production of an on-shell
mediator that then decay into a pair of dark matter particles. For scenarios in
which the mediator cannot decay into a top-antitop pair, one observes stronger
constraints when the $\chi\bar\chi$ pair can be resonantly produced. On
different lines, in
contrast, for heavier dark matter particles where the mediator has to be
produced off-shell, the kinematics of the process changes on top of being
phase-space suppressed, so that larger cross sections can only be constrained.
In a regime in which the mediator can decay into a pair of top-antitop quarks
(regardless of the dark matter mass), it becomes broad
by virtue of the adopted strong couplings. In this case, cross sections of the
order of the pb can only be excluded, regardless of the mass configuration.

Nevertheless, even under the hypothesis of the strong
new physics couplings given by Eq.~\eqref{eq:coup}, the corresponding signal
cross sections lie at at least one or two orders of magnitude below the
exclusion thresholds, so that no mono-$Z$-boson constraints can actually be
extracted.

\section{Conclusion} \label{sec:conclusion}
Particle physics is living an exciting time, with the Run 2 of the LHC ongoing
until the end of 2018 and its Run 3 being expected to start in 2021. The results
to be collected promise a great understanding of the Standard Model and
beyond, provided they can be exploited in the best possible way. This
requires the development of a strategy allowing for the comparison of
the experimental results with theoretical predictions in the context of various
models of new physics. Many different groups have consequently released public
frameworks allowing for such a reinterpretation of the LHC results in general
theoretical setups. Those frameworks rely either on simplified model results so
that bounds on new physics can be extracted from a comparison of predicted
signal rates with experimentally-derived upper bounds, or on the
simulation of LHC collisions so that the experimental selection strategy can be
mimicked to obtain the new physics event counts.

In this work, we have focused on the \MA\ platform dedicated to particle physics
phenomenology. In a nutshell, \MA\ contains a user-friendly {\sc Python} command
line
interface that allows users to design cut-and-count collider analyses via a
small set of commands, as well as a C++ core which can be used for implementing
the most complex LHC analyses directly in C++ instead of through the possibly
too-limited {\sc Python} metalanguage.  All those features allow \MA\
to be used for reinterpretation studies, the \python\ interface being a
mean to create an analysis template to be modified, in the expert mode, in order
to recast an actual LHC analysis. From simulated LHC collisions at the Monte
Carlo truth level (\ie, after parton showering and hadronization), \MA\ moreover
takes care of adding the simulation of the response of the
detector, evaluates how these simulated events populate the search regions of
all LHC searches that are reimplemented in its Public Analysis
Database, and finally derives confidence level exclusions on
the signal.

The goal of this document is to complement the \MA\ manuals of
Refs.~\cite{Conte:2012fm,Conte:2014zja} and illustrate two important features
with practical examples. First, we presented how the normal mode of the program
could be used to simulate the response of a typical high-energy physics detector
when event reconstruction was at stake. We emphasized the different ways in
which this could be handled and performed their comparison on the example of
monotop production at the LHC. Next, we reviewed how the expert mode of \MA\ was
suitable for the reimplementation of existing LHC searches for new physics and
for subsequently constraining a given beyond the Standard Model framework. We
comprehensively
addressed all the steps necessary to tackle these tasks, which included the
implementation of the analysis in the \MA\ framework itself, the validation of
the recast code and its public release on the \MA\ Public Analysis Database.
This last step ensures that a DOI is assigned to the analysis, which makes the
recast code uniquely identifiable, citable and searchable. As a practical
example, we considered the CMS-EXO-16-010 search for dark matter in the
mono-$Z$-boson channel, and presented how to use the code to constrain one of
the simplified models investigated in traditional LHC searches for dark matter
with the results of CMS-EXO-16-010.

The outcome of the LHC experiments includes not only data itself, but also all
the analyses with the associated results. The best exploitation of the LHC
legacy however requires that the high-energy physics community can easily reuse
those results for further phenomenological studies. All recent developments
undertaken within the \MA\ framework aim to tackle this issue. We hence offer a
scheme, extensively detailed in the present document, allowing one not
only to reimplement an existing LHC analysis but also to reuse it in a
straightforward manner to constrain any given model of new physics. In this way,
we hope that the \MA\ project can contribute significantly to the community
efforts to unravel the electroweak symmetry breaking mechanism and the
potential discovery of new phenomena.

\section*{Acknowledgements}
We thank both ATLAS and CMS for providing plentiful information on their
searches to make them recastable by people outside the collaborations, this
effort being crucial for the legacy of the LHC. This indeed allows the whole
high-energy physics community to exploit the LHC experimental results
for new physics studies in the best possible manner.

We are grateful to our CMS experimental colleagues from the exotica working
group, and in particular to Andreas Albert, Olivier Buchm\"uller and Viatcheslav
Valuev, for their help in the validation of the reimplementation of the
CMS-EXO-16-010 analysis. We also heartfully thank Guillaume Chalons, Sabine
Kraml and Dipan Sengupta for valuable comments on the manuscript and all lively
discussions on LHC recasting of the last 5 years.

BF has been supported in part by French state funds managed by the Agence
Nationale de la Recherche (ANR), in the context of the LABEX ILP
(ANR-11-IDEX-0004-02, ANR-10-LABX-63).

\newpage
\appendix
\section{Reference card for the normal mode}
\label{app:normal}
\subsection*{Starting the \MAnorm\ interpreter}
The \MA\ command line interface can be started by typing in a shell
\begin{verbatim} bin/ma5 [options] [script] \end{verbatim}
where the potential options (\verb+[options]+) are given in the table below.
\verbdef{\vv}{-v}
\renewcommand{\arraystretch}{1.2}%
\begin{center}\begin{tabular}{l p{2.5cm} p{8.1cm}}
\hline
Short & {\centering{Full}} & Description\\
\hline
\verb?-P? & \verb?--partonlevel? & Parton-level mode.\\
\verb?-H? & \verb?--hadronlevel? & Hadron-level mode.\\
\verb?-R? & \verb?--recolevel?   & Reconstructed-level mode.\\
\verb?-E? & \verb?--expert?      & Expert mode.\\
\multirow{2}{*}{\vv} & \verb?--version?
          & \multirow{2}{*}{Displays the current \MA\ version number.}\\
          &  \verb?--release?    & \\
\verb?-b? & \verb?--build?       & Builds of the \spla\ library.\\
\verb?-f? & \verb?--forced?      & Skips \MA\ confirmation messages.\\
\verb?-s? & \verb?--script?      & Executes a script containing all analysis
    commands and exits the program. The file containing the script has to be
    provided as \verb+[script]+. Removing the \verb+-s+
    option prevents \MA\ from exiting.\\
\verb?-h? & \verb?--help?        & Print all the available commands.\\
\verb?-i? & \verb?--installcard? & Generates a file with information on the
     \MA\ dependencies, \verb?installation_card.dat?, that is located in the
     \verb+madanalysis/input+ folder.\\
\verb?-d? & \verb?--debug?       & Debug mode.\\
\hline
\end{tabular}
\end{center}
In case of installation issues, the usage of the dependencies can be tuned by
editing the \verb?madanalysis/input/installation_options.dat? file. More
information can be found in the original \MA\ manual~\cite{Conte:2012fm}.

\subsection*{Installation of optional packages}
\MA\ is interfaced to several high-energy physics packages and can be linked to
a variety of external plugins. Any of those can be installed by typing in the
\MA\ interpreter,
\begin{verbatim} install [package]\end{verbatim}
where the different choices for \verb+[package]+ are given in the table below.
\renewcommand{\arraystretch}{1.2}%
\begin{center}\begin{tabular}{l p{9.2cm}}
\hline
Package & Description\\
\hline
\verb?PAD?            & The \MA\ Public Analysis Database of recasted LHC
   analyses relying on \DEL~3 for the simulation of the detector effects.\\
\verb?PADforMA5tune?  & The \MA\ Public Analysis Database of recasted LHC
   analyses relying on the old tuned version of \DEL~3 for the simulation of the
   detector effects.\\
\verb?delphes?        & The current release of \DEL~3.\\
\verb?delphesMA5tune? & The old tuned version of \DEL~3.\\
\verb?fastjet?        & The \FJ\ and {\sc FastJetContrib} packages.\\
\verb?samples?        & Some test Monte Carlo samples.\\
\verb?zlib?           & The {\sc ZLib} library allowing to handle gzipped
                        compressed files.\\
\hline
\end{tabular}
\end{center}

\subsection*{List of commands available from the \MAnorm\ interpreter}
A small set of commands, to be typed from the \MA\ interpreter, are related to
console actions.
\renewcommand{\arraystretch}{1.2}%
\begin{center}\begin{tabular}{l p{8.8cm}}
\hline
Command & Description\\
\hline
\verb?quit?           & \multirow{2}{*}{Exits \MA.}\\
\verb?EOF?            & \\
\verb?help?           & Displays the list of available commands.\\
\verb?help [command]? & Displays details about a specific command.\\
\verb?history?        & Displays the history of all commands that have been
                        typed by the user.\\
\verb?reset?          & Clears the \MA\ memory as when the program is started.\\
\verb?restart?        & Restarts \MA.\\
\verb?shell [command]?& \multirow{2}{*}{Executes a {\sc Unix} command from the
   interpreter.}\\
\verb?![command]?     & \\
\verb?#[text]?        & Indicates a comment.\\
\hline
\end{tabular}
\end{center}
\verbdef{\verbtexta}{define [(multi)particle label] = [ID1] [ID2] ...}
\verbdef{\verbtextb}{define_region [label1] [label2] ...}
\verbdef{\verbtextc}|plot [obs]([part1] [part2] ...) nbins min max [ [opts] ] { [regs] }|
\verbdef{\verbtextda}{reject ([part]) [criterion]}
\verbdef{\verbtextdb}{select ([part]) [criterion]}
\verbdef{\verbtexte}{[criterion]}
\verbdef{\verbtextf}{swap main.selection[n1] main.selection[n2]}
In order to design an analysis in the \MA\ framework, the user can rely on the
following commands.
\renewcommand{\arraystretch}{1.2}%
\begin{center}\begin{tabular}{p{3.40cm} p{8.75cm}}
\hline
Command & Description\\
\hline
\multicolumn{2}{l}{\verbtexta}\\
   & Creates a new (multi)particle label attached to the provided PDG codes
      \verb+[ID1]+, \verb+[ID2]+, $\ldots$\\
\multicolumn{2}{l}{\verbtextb}\\
   & Creates one or more signal regions named \verb+[label1]+, \verb+[label2]+,
   ...\\
\verb?import [label]?        & Imports a sample or a UFO model.\\
\verb?open [folder]?         & Opens a report from the folder \verb?[folder]?.
   If the folder is unspecified, the last created report is open.\\
\multicolumn{2}{l}{\verbtextc}\\
   & Defines a histogram with the distribution in the observable \verb?[obs]?.
    Its computation may require to combine the momenta of several objects
    \verb?[part1]?, \verb+[part2]+, $\ldots$ Any other parameter is optional.
    The \verb?nbins?,
    \verb?min? and \verb?max? quantities respectively correspond to the number
    of bins, and the upper and lower bound of the $x$-axis of the histogram. The
    options \verb?[opts]? (see below) allow the user to tune the
    display of the histogram, and \verb+[regs]+ indicates to which signal region
    one needs to attach this histogram.\\
\verb?reject [criterion]?
\verb?select [criterion]? & Defines a selection cut that leads
  to the rejection (or selection) of an event if the \verbtexte condition is
  satisfied.\\
\multicolumn{2}{l}{\verbtextda}\\
\multicolumn{2}{l}{\verbtextdb}\\
  & Defines a selection cut that leads to the rejection (or selection) of an
    object candidate if the \verb?[criterion]? condition is satisfied.\\
\verb?remove [object]? & Deletes an object (a (multi)particle label, a region, a
 histogram or a cut).\\
\verb?resubmit? & Adjusts the last generated C++ code relatively to the commands
  issued after the last \verb?submit? command, and executes it.\\
\verb?set [obj] = [val]? & Sets an attribute of a specific object to a given
  value.\\
\verb?submit [folder]? & Generates, compiles and executes the C++ code
  corresponding to the current analysis either in a folder named
  \verb?[folder]?, if specified, or in an arbitrary folder otherwise.\\
\multicolumn{2}{l}{\verbtextf}\\ & Swaps the analysis steps
   number \verb?[n1]? and \verb?[n2]?.\\
\hline
\end{tabular}
\end{center}
Throughout the analysis, information can be printed to the screen by means of
the following commands.
\renewcommand{\arraystretch}{1.2}%
\begin{center}\begin{tabular}{p{4.00cm} p{7.70cm}}
\hline
Command & Description\\
\hline
\verb+display [object]+       & Displays the properties of a specific object.\\
\verb+display_datasets+       & Displays the list of all defined datasets.\\
\verb+display_multiparticles+ &
   Displays the list of all defined multiparticle labels.\\
\verb+display_particles+      &
   Displays the list of all defined particle labels.\\
\verb+display_regions+        &
   Displays the list of all defined signal regions.\\
\hline
\end{tabular}
\end{center}

\subsection*{Properties of the main object}
The \verb+main+ object of \MA\ allows to setup varied options (see the table
below) to impact the code on run time. They can be modified and
displayed by using the \verb+set+ and \verb+display+ commands introduced above.
\renewcommand{\arraystretch}{1.2}%
\begin{center}\begin{tabular}{l p{7.50cm}}
\hline
Command & Description\\
\hline
\verb?main.currentdir?     & Folder containing \MA.\\
\verb?main.fastsim.package?& The package used for the fast-simulation of the
  detector response. The available choices are \verb?fastjet?, \verb?delphes?,
  \verb?delphesMA5tune? and \verb?none?.\\
\verb?main.fom.formula?    & Formula to be used to calculate the figure of merit
  in the cutflow charts. Denoting by $S$ and $B$ the number of signal and
  background events, the available choices are 1 ($S/B$), 2 ($S/\sqrt{B}$), 3
  ($S/\sqrt{B}$), 4 ($S/\sqrt{S+B}$) and 5 ($S/\sqrt{S+B+x_B^2}$). For this last
  formula, the $x_B$ parameter is specified via \verb?main.fom.x?.\\
\verb?main.graphic_render? & Package to use for figure generation. The
  available choices are \verb?root?, \verb?matplotlib? and \verb?none?.\\
\verb?main.isolation.algorithm? & Algorithm to be used for particle isolation.
  The available choices are \verb?cone? (no activity in a cone of radius
  specified by \verb+main.isolation.radius+) and \verb+sumpt+ (the scalar sum of
  the transverse momenta of all particles lying in a given cone around the
  candidate must be smaller than \verb?main.isolation.sumPT? and the ratio of
  the transverse energy of all particles in this cone to the transverse momentum
  of the candidate must be smaller than
  \verb?main.isolation.ET_PT?).\\
\end{tabular}
\begin{tabular}{l p{8.4cm}}
\verb?main.lumi? & Integrated luminosity, in $\textrm{fb}^{-1}$, to use for
  histogram and cutflow normalization.\\
\verb?main.normalize? & Way in which histograms have to be normalized. The
  available choices are \verb?none? (each event counts for 1), \verb?lumi?
  (normalization to the integrated luminosity without taking into account the
  event weights), \verb?lumi_weight? (as \verb?lumi? but with the
  event weights).\\
\verb?main.outputfile? & Name of the output file to write events onto.\\
\verb?main.recast? & Switching the recasting mode \verb?on? and \verb?off?.\\
\verb?main.stacking_method? & Way in which the contributions of the different
  datasets to a histogram are displayed. The available choices are
  \verb?normalize2one? (the integral of each contribution equals 1),
  \verb?stack? (each contribution is stacked) and \verb?superimpose? (each
  contribution is superimposed).\\
\hline
\end{tabular}
\end{center}

\subsection*{Observables to be used for histograms and cuts}
As shown in the tables above, the definition of a histogram or of a cut
condition necessitates to provide an observable that could depend on the momentum
of one or more particles or objects. We list in the tables below all observables
supported by \MA, and begin with those that do not depend on the momenta of any
object. They are therefore called without any argument.
\renewcommand{\arraystretch}{1.2}%
\begin{center}\begin{tabular}{l p{10.2cm}}
\hline
Symbol& Description\\
\hline
\verb?ALPHA_QCD? & Value of the QCD coupling constant.\\
\verb?ALPHA_QED? & Value of the electromagnetic coupling constant.\\
\verb?ALPHA_T?   & The $\alpha_T$ variable\cite{Randall:2008rw}.\\
\verb?MEFF?      & Effective mass being defined as the sum of the transverse
                   momentum of all final-state objects and the missing
                   transverse energy.\\
\verb?MET?       & Missing transverse energy.\\
\verb?MHT?       & Missing transverse energy defined from the jet activity
                   only.\\
\verb?NPID?      & Particle content (PDG code distribution).\\
\verb?NAPID?     & Particle content (PDG code distribution in
                   absolute value).\\
\verb?SQRTS?     & Partonic center-of-mass energy.\\
\verb?SCALE?     & Energy scale of the event.\\
\verb?TET?       & Scalar sum of the transverse energy of all final-state
                   objects.\\
\verb?THT?       & Scalar sum of the transverse energy of all final-state
                   jets.\\
\verb?WEIGHTS?   & Event weights.\\
\hline
\end{tabular}
\end{center}
The set of observables provided in the following table can be used to study the
properties of a given object or particle, and thus requires to provide one
four-momentum or one combination of four-momenta as an argument of the
observable function.
\renewcommand{\arraystretch}{1.2}%
\begin{center}\begin{tabular}{l p{10.2cm}}
\hline
Symbol& Description\\
\hline
\verb?ABSETA? & Absolute value of the pseudorapidity.\\
\verb?BETA?   & Velocity $\beta=v/c$ (relatively to the speed of light).\\
\verb?E?      & Energy.\\
\verb?EE_HE?  & Ratio of the electromagnetic energy to the hadronic energy (for
   a jet).\\
\verb?ET?     & Transverse energy.\\
\verb?ETA?    & Pseudorapidity.\\
\verb?GAMMA?  & Lorentz-factor.\\
\verb?HE_EE?  & Ratio of the hadronic energy to the electromagnetic energy (for
   a jet).\\
\verb?M?      & Invariant mass.\\
\verb?MT?     & Transverse mass.\\
\verb?MT_MET? & Transverse mass of the system comprised of the object and the
   missing momentum.\\
\verb?NTRACKS?& Number of tracks (inside a jet).\\
\verb?P?      & Magnitude of the three-momentum.\\
\verb?PHI?    & Azimuthal angle.\\
\verb?PT?     & Transverse momentum.\\
\verb?PX?     & $x$-component of the momentum.\\
\verb?PY?     & $y$-component of the momentum.\\
\verb?PZ?     & $z$-component of the momentum.\\
\verb?R?      & Position in the $(\eta, \phi)$ plane.\\
\verb?Y?      & Rapidity.\\
\hline
\end{tabular}
\end{center}
Three additional observables involving two objects are also available, the
methods given in the table below taking thus two arguments separated by a comma.
\renewcommand{\arraystretch}{1.2}%
\begin{center}\begin{tabular}{l p{10.0cm}}
\hline
Symbol& Description\\
\hline
\verb?DELTAR?     & Angular distance, in the transverse plane, between the
  objects.\\
\verb?DPHI_0_PI?  & Angular distance in azimuth between the objects. The
  bounds for the angle are $[0,\pi]$.\\
\verb?DPHI_0_2PI? & Angular distance in azimuth between the objects. The
  bounds for the angle are $[0,2\pi]$.\\
\hline
\end{tabular}
\end{center}
For all arguments of any of the above observable, any sequence of momenta
separated with spaces will lead to a sum of these momenta before computing the
observable. For instance,
\begin{verbatim}
  plot M(e+ e-)
\end{verbatim}
allows for the computation of the invariant of an electron-positron system.

\subsection*{Options for histograms}
The command \verb+plot+ accept varied options \verb+[opts]+,
\begin{verbatim}
plot [obs]([part1] [part2] ...) nbins min max [ [opts] ] { [regs] }
\end{verbatim}
to be provided between squared brackets. The list of all available choices is
given in the table below.
\renewcommand{\arraystretch}{1.2}%
\begin{center}\begin{tabular}{l p{9.4cm}}
\hline
Symbol& Description\\
\hline
\verb?Eordering?     & Sorts the objects in increasing energy.\\
\verb?ETordering?    & Sorts the objects in increasing transverse energy.\\
\verb?ETAordering?   & Sorts the objects in increasing pseudorapidity.\\
\verb?Pordering?     & Sorts the objects in increasing three-momentum magnitude.
  \\
\verb?PTordering?    & Sorts the objects in increasing transverse momentum.\\
\verb?PXordering?    & Sorts the objects in increasing momentum $x$-component.\\
\verb?PYordering?    & Sorts the objects in increasing momentum $y$-component.\\
\verb?PZordering?    & Sorts the objects in increasing momentum $z$-component.\\
\verb?allstate?      & Considers all (initial-state, final-state and
  intermediate-state) objects in the events.\\
\verb?finalstate?    & Considers only final-state objects (default).\\
\verb?initialstate?  & Considers only initial-state objects.\\
\verb?interstate?    & Considers only objects that are neither initial-state nor
  final-state objects.\\
\verb?logX?          & Logarithmic scale for the $x$-axis.\\
\verb?logY?          & Logarithmic scale for the $y$-axis.\\
\verb?normalize2one? & Normalizes the histogram to 1.\\
\verb?stack?         & Stacks the contributions of different datasets in the
  histogram.\\
\verb?superimpose?   & Superimposes the contributions of different datasets in
  the histogram.\\
\hline
\end{tabular}
\end{center}

\subsection*{Options for datasets}

Once one or more samples have been imported as a dataset, properties that will
impact the display of the subsequent contribution in histograms can be modified
by using the \verb+set+ command. For instance, a dataset named \verb?defaultset?
can be imported as
\begin{verbatim}
  import <path-to-sample> as defaultset
\end{verbatim}
and its properties can be modified by typing in
\begin{verbatim}
  set defaulset.<property> = <value>
\end{verbatim}
The list of available properties is given in the table below, together with the
allowed values.
\renewcommand{\arraystretch}{1.2}%
\begin{center}\begin{tabular}{l p{9.1cm}}
\hline
Symbol& Description\\
\hline
\verb?backcolor? & Background color in a histogram. The available colors are
   \verb?auto?, \verb?black?, \verb?blue?, \verb?cyan?, \verb?green?,
   \verb?grey?, \verb?none? (transparent), \verb?orange?, \verb?purple?,
   \verb?red?, \verb?white? and \verb?yellow?. The color can be made lighter or
    darker by adding an explicit $\pm 1$, $\pm 2$ or $\pm 3$.\\
\verb?backstyle? & Background texture in a histogram. The available values are
   \verb?dline? (diagonal-lines), \verb?dotted? (dots), \verb?hline? (horizontal
   lines), \verb?solid? (uniform color) and \verb?vline? (vertical lines).\\
\verb?linecolor? & Color of the histogram lines. The available colors are the
   same as for the \verb?backcolor? attribute.\\
\verb?linestyle? & Style of the histogram lines. The available values are
   \verb?dash-dotted?, \verb?dashed?, \verb?dotted? and \verb?solid?.\\
\verb?linewidth? & Width of the histogram lines, given as an integer smaller
   than 10.\\
\verb?title?     & Name of the dataset (for histogram legends).\\
\verb?type?      & \verb?background? or \verb?signal? nature of a given sample
  (for figure-of-merit calculations).\\
\verb?weight?    & Reweights each histogram entry with a constant factor. The
  value has to be a floating-point number.\\
\verb?weighted_events? & Allows \MA\ to ignore the weights of the events
  (property to be set to \verb?true? or \verb?false?).\\
\verb?xsection?  & This overwrites the event sample cross section. The value has
  to be given in pb.\\
\hline
\end{tabular}
\end{center}

\subsection*{Using \FJnorm\ throuh \MAnorm}
In order to activate the usage of \FJ\ through \MA, the program has to be
started in the reconstructed mode and the first command to be typed in the
interpreter has to be
\begin{verbatim}
  set main.fastsim.package = fastjet
\end{verbatim}
This allows for various option of the \verb+main.fastsim+ object, tuning the
properties of the jet algorithm that has to be employed. Those options are set
by typing in
\begin{verbatim}
  set main.fastsim.<property> = <value>
\end{verbatim}
the list of all available properties being presented, together with the allowed
values, in the following table.
\renewcommand{\arraystretch}{1.2}%
\begin{center}\begin{tabular}{l p{9.4cm}}
\hline
Symbol& Description\\
\hline
\verb?algorithm? & Sets up the jet algorithm to use. The allowed
  values are \verb?antikt?\cite{Cacciari:2008gp}, \verb?Cambridge?\cite{%
   Dokshitzer:1997in,Wobisch:1998wt}, \verb?cdfjetclu?\cite{Abe:1991ui},
   \verb?cdfmidpoint?\cite{Blazey:2000qt}, \verb?genkt?\cite{Cacciari:2011ma},
   \verb?gridjet?\cite{Cacciari:2011ma}, \verb?kt?\cite{Catani:1993hr,%
   Ellis:1993tq}, \verb?none? and \verb?siscone?\cite{Salam:2007xv}.\\
\verb?areafraction? & Controls the size of the cones in the CDF midpoint
  algorithm.\\
\verb?exclusive_id? & Exclusive mode for jet reconstruction. If set
  to \verb+false+, electrons muons and photons issued from hadron decays are
  included into the electron, muon and photon collections respectively.\\
\verb?input_ptmin? & Soft protojet threshold in the siscone
  algorithm.\\
\verb?iratch? & Switching on ratcheting for the CDF jet clustering algorithm.\\
\verb?npassmax? & Number of iterations in the siscone algorithm.\\
\verb?overlap? & Fraction of overlapping momentum used to combine
  protojets in the siscone and CDF reconstruction algorithms.\\
\verb?p? & $p$ parameter of the generalized $k_T$ algorithm.\\
\verb?ptmin? & Threshold for the transverse momentum of the
  reconstructed jets.\\
\verb?radius? & Radius parameter relevant for most jet
  clustering algorithms.\\
\verb?seed? & Threshold parameter used in the constituent merging
  procedure of the CDF reconstruction algorithms.\\
\verb?spacing? & Grid spacing in the grid jet algorithm.\\
\verb?ymax? & Maximum rapidity value in the grid jet algorithm.\\
\hline
\end{tabular}
\end{center}
In addition, basic detector simulation effects can be mimicked through setting
up the parameters included in the table below.
\renewcommand{\arraystretch}{1.2}%
\begin{center}\begin{tabular}{l p{8.4cm}}
\hline
Symbol& Description\\
\hline
\verb?bjet_id.efficiency? & $b$-tagging efficiency, as a float.\\
\verb?bjet_id.exclusive? & Allows several $b$-jets to be issued from a single
  $B$-hadron.\\
\verb?bjet_id.matching_dr? & Angular distance parameter matching a $b$-jet
  with a $B$-hadron.\\
\verb?bjet_id.misid_cjet? & Mistagging rate of a $c$-jet as a $b$-jet, as
  a float.\\
\verb?bjet_id.misid_ljet? & Mistagging rate of a light jet as a $b$-jet,
  as a float.\\
\verb?tau_id.efficiency? & Tau-tagging efficiency, as a float.\\
\verb?tau_id.misid_ljet? &  Mistagging rate of a light jet as a hadronic tau,
   as a float.\\
\hline
\end{tabular}
\end{center}

\subsection*{Using \DELnorm\ through \MAnorm}
In order to activate the usage of \DEL\ through \MA, the program has to be
started in the reconstructed mode and the first command to be typed in the
interpreter has to be
\begin{verbatim}
  set main.fastsim.package = delphes
\end{verbatim}
The properties of the simulation of the detector can then be adjusted through
\begin{verbatim}
  set main.fastsim.<property> = <value>
\end{verbatim}
the allowed choices being given in the table below.
\renewcommand{\arraystretch}{1.2}%
\begin{center}\begin{tabular}{l p{8.4cm}}
\hline
Symbol& Description\\
\hline
\verb?detector? & Determines which detector card to use. The card can be further
  modified on run time).\\
\verb?output?   & Saves the output \ROOT\ file (\verb+true+ or \verb+false+).\\
\verb?rootfile? & Name of the output file.\\
\verb?pileup? & Specifies the path to the input pile-up event file.\\
\verb?skim_genparticles? & If set to \verb?true?, the generator-level
  particles are not stored in the output file.\\
\verb?skim_tracks? & If set to \verb?true?, the track collection is not stored
  in the output file.\\
\verb?skim_towers? & If set to \verb?true?, the collection of calorimetric
  towers is not stored in the output file.\\
\verb?skim_eflow? & If set to \verb?true?, the collection of particle-flow
  towers is not stored in the output file.\\
\hline
\end{tabular}
\end{center}

\subsection*{Multipartonic matrix element merging}
\MA\ can be used to double check the merging procedure of event samples related
to a given hard process but with matrix elements featuring a different
final-state jet multiplicity. \MA\ has to be started in the hadronic mode, and
the check is then performed by typing in
\begin{verbatim}
 set main.merging.check = true
\end{verbatim}
Two extra options are available,
\begin{verbatim}
  set main.merging.ma5_mode = <true or false>
  set main.merging.njets = <integer>
\end{verbatim}
the first one indicating to extract the number of extra hard jets from the
process identifier, and the second one setting up the maximum number of extra
jets to consider.

\newpage
\section{Reference card for the expert mode}
\label{app:expert}

\subsection*{Creating an analysis template in the expert mode}
For sophisticated analysis going beyond the capabilities of the normal mode of
running of \MA, the user has to rely on the so-called expert mode, where
analyses are implemented directly in the C++ framework of the platform. A blank
template analysis can be generated by typing, in a shell,
\begin{verbatim}
  bin/ma5 --expert      bin/ma5 -e      bin/ma5 -E
\end{verbatim}
\MA\ then asks information about the name of the working directory to create, as
well as information on the name of the analysis class that will have to be
designed. Those two pieces of information can be provided as arguments when
calling \MA\ from the shell,
\begin{verbatim}
  bin/ma5 -E <dirname> <analysis>
\end{verbatim}
where \verb+<dirname>+ consists in the working directory name, and
\verb+analysis+ the name of the analysis class.

For a proper use of the program, the user has to setup some environment
variables accordingly. This can be done by entering the \verb+Build+ subfolder
of the working directory and typing in a shell,
\begin{verbatim}
  source setup.sh         source setup.csh
\end{verbatim}
depending on the shell nature. The \verb+Build+ folder also contains a makefile
allowing for standard \verb+make+ commands,
\begin{verbatim}
  make clean       make proper       make
\end{verbatim}
The first command allows one to remove all intermediate object and backup
files whilst the second command yields the removal of all files that have been
created by the make action. Finally, the last command allows to build the code.

The code can then be run from the \verb+Build+
directory by typing,
\begin{verbatim}
  MadAnalysis5Job [options] [inputfile]
\end{verbatim}
The file \verb?[inputfile]? consists in a text file with a list of paths
pointing to the event samples to analyze, with one filename per line. Several
options are available, as summarized in the table below.
\renewcommand{\arraystretch}{1.2}%
\begin{center}\begin{tabular}{l p{8.4cm}}
\hline
Option & Description\\
\hline
\verb?--check_event?      & Sanity check of the input file.\\
\verb?--no_event_weight?  & Ignores the event weights.\\
\verb?--ma5_version=XXXX? & Allows to specify which version of the \MA\ console
  to use.\\
\hline
\end{tabular}
\end{center}

\subsection*{Portable datatypes}
The \spla\ data format includes several portable data types, that we
show in the table below together with the corresponding bit width.
\renewcommand{\arraystretch}{1.1}%
\begin{center}\begin{tabular}{l l p{8.0cm}}
\hline
Name & Bit width & Description\\
\hline
\verb?MAbool?    & 1  & Boolean.\\
\hline
\verb?MAint8?    & 8  & Byte integer.\\
\verb?MAint16?   & 16 & Short integer.\\
\verb?MAint32?   & 32 & Integer.\\
\verb?MAint64?   & 64 & Long integer.\\
\hline
\verb?MAuint8?   & 8  & Unsigned byte integer.\\
\verb?MAuint16?  & 16 & Unsigned short integer.\\
\verb?MAuint32?  & 32 & Unsigned integer.\\
\verb?MAuint64?  & 64 & Unsigned long integer.\\
\hline
\verb?MAfloat32? & 32 & Single-precision floating-point number.\\
\verb?MAfloat64? & \multirow{2}{*}{64} & \multirow{2}{*}{Double-precision
   floating-point number.}\\
\verb?MAdouble64?& &\\
\hline
\end{tabular}
\end{center}
Moreover, four-vectors can be implemented as instances of the
\verb+MALorentzVector+ class that contains the same methods as the \ROOT\
\verb+TLorentzVector+ class~\cite{Brun:1997pa}.

\subsection*{Data format for an event sample}
The analysis class contains an \verb+Initialize+, an \verb+Execute+ and a
\verb+Finalize+ method that are respectively executed before starting to read an
event sample, on each event and after having read all events. The \verb+Execute+
method requires two arguments, an instance of the \verb+SampleFormat+ class and
an event passed as an \verb+EventFormat+ instance (see the next subsection).
Monte Carlo event samples are generally accompanied with global information on
the sample, such as the identifier of the parton density set that has been used
or the total cross section associated with the described process. Those pieces
of information are stored as attributes of the above-mentioned
\verb+SampleFormat+ object, and can be retrieved through the methods given in
the table below (in particular through attributes of the \verb+mc()+ and
\verb+rec()+ objects) on run time.
\verbdef{\expa}{const std::vector<std::string>& header() const}
\renewcommand{\arraystretch}{1.2}%
\begin{center}\begin{tabular}{l p{6.2cm}}
\hline
Method name and type & Description\\
\hline
\verb+MCSampleFormat* mc()+            & Monte Carlo information (see below).\\
\verb+RecSampleFormat* rec()+          & Reconstruction information (see below).
    \\
\verb+const std::string name()+        & Sample name.\\
\verb+const MAuint64& nevents() const+ & Number of events in the sample.\\
\multicolumn{2}{l}{\expa}\\    &  Sample header.\\
\hline
\end{tabular}
\end{center}
The Monte Carlo information is available through the \verb+mc()+ object, a
pointer to an instance of the \verb+MCSampleFormat+ class whose attributes are
given in the table below.
\verbdef{\expb}{const std::pair<MAint32,MAint32>& beamPDGID() const}
\verbdef{\expc}{const std::pair<MAuint32,MAuint32>& beamPDFauthor() const}
\verbdef{\expd}{const MAint32& weightMode() const}
\verbdef{\expe}{const MAfloat64& xsection() const}
\verbdef{\expf}{const MAfloat64& xsection_error() const}
\verbdef{\expg}{const MAfloat64& sumweight_positive() const}
\verbdef{\exph}{const MAfloat64& sumweight_negative() const}
\verbdef{\expi}{const std::vector<ProcessFormat>& processes()}
\verbdef{\expj}{const WeightDefinition& weight_definition()}
\verbdef{\expk}{const std::pair<MAfloat64,MAfloat64>& beamE() const}
\verbdef{\expl}{const std::pair<MAuint32,MAuint32>& beamPDFID() const}
\renewcommand{\arraystretch}{1.2}%
\begin{center}\begin{tabular}{l p{8.2cm}}
\hline
Method name and type & Description\\
\hline
\multicolumn{2}{l}{\expb}\\    &
  PDG codes of the intial partons.\\
\multicolumn{2}{l}{\expk}\\    &
  Beam energy.\\
\multicolumn{2}{l}{\expc}\\    &
  Group associated with the used parton density set.\\
\multicolumn{2}{l}{\expl}\\    &
  Identifier of the used parton density set.\\
\multicolumn{2}{l}{\expd}\\    &
  Information on the event weights.\\
\multicolumn{2}{l}{\expe}\\    &
  Cross section associated with the sample.\\
\multicolumn{2}{l}{\expf}\\    &
  Uncertainty on the sample cross section.\\
\multicolumn{2}{l}{\expg}\\    &
  Sum of the weights of all positively-weighted events.\\
\multicolumn{2}{l}{\exph}\\    &
  Sum of the weights of all negatively-weighted events.\\
\multicolumn{2}{l}{\expi}\\    &
  List of all described processes (see below).\\
\multicolumn{2}{l}{\expj}\\    &
  Definitions of the different weights for events featuring multiple weights.\\
\hline
\end{tabular}
\end{center}
Those attributes rely on two classes, the \verb+ProcessFormat+ one allowing
for the description of a physical process and the \verb+WeightDefinition+ one
connected to the potential assignment of multiple weights to a given
event\cite{Andersen:2014efa}. Whilst the
following methods have been implemented for the former class,
\renewcommand{\arraystretch}{1.2}%
\begin{center}\begin{tabular}{l p{4.8cm}}
\hline
Method name and type & Description\\
\hline
\verb+const MAfloat64& xsection() const+ & Associated cross section.\\
\verb+const MAfloat64& xsectionError() const+ & Error on the cross
  section.\\
\verb+const MAfloat64& weightMax() const+ & Maximum weight for an event.\\
\verb+const MAuint32& processId() const+ & Process identification number.\\
\hline
\end{tabular}
\end{center}
the latter class only allows for listing the names of the set of weights
associated with each event,
\begin{verbatim}
 void Print() const
\end{verbatim}
The \verb+rec()+ method of the \verb+MCSampleFormat+ class consists in an
instance of the \verb?RecSampleFormat? class that does not come with any
built-in method. This is left for future developments.

\subsection*{Data format for an event}

An event object possesses two attributes \verb+mc()+ and \verb+rec()+ that are
this time instances of the \verb+MCEventFormat+ and \verb+RecEventFormat+
classes respectively. These two classes are respectively connected to Monte
Carlo events (as simulated by a Monte Carlo event generator) and reconstructed
events as obtained after gathering all final-state objects into reconstructed
physical objects to be used for specific analyses. The \verb+MCEventFormat+
class comes with the methods summarized in the following table.
\verbdef{\expzza}{const MCParticleFormat& MET() const}
\verbdef{\expzzb}{const MCParticleFormat& MHT() const}
\verbdef{\expzzc}{MCParticleFormat& MET() const}
\verbdef{\expzzd}{MCParticleFormat& MHT() const}
\verbdef{\expzze}{const std::vector<MCParticleFormat>& particles() const}
\verbdef{\expzzf}{const WeightCollection& multiweights() const}
\renewcommand{\arraystretch}{1.2}%
\begin{center}\begin{tabular}{l p{5.5cm}}
\hline
\multicolumn{2}{l}{\expzza}\\
\multicolumn{2}{l}{\expzzc}\\    & Missing transverse energy\\
\multicolumn{2}{l}{\expzzb}\\
\multicolumn{2}{l}{\expzzd}\\    & Missing transverse hadronic energy. \\
\verb+const MAfloat64& TET() const+ &
   \multirow{2}{*}{Total (visible) transverse energy.}\\
\verb+MAfloat64& TET() const+ &\\
\verb+const MAfloat64& THT() const+ &
  \multirow{2}{*}{Total hadronic transverse energy.}\\
\verb+const MAfloat64& THT() const+ &\\
\verb+const MAfloat64& Meff() const+ & \multirow{2}{*}{Effective mass
  $\bigg(\sum_{\rm jets} p_T + \slashed{E}_T\bigg)$.}\\
\verb+MAfloat64& Meff() const+ &\\
\verb+const MAfloat64& alphaQED()  const+ & Used value for the electromagnetic
  coupling.\\
\verb+const MAfloat64& alphaQCD()  const+ & Used value for the strong
  coupling.\\
 \multicolumn{2}{l}{\expzzf}\\  & Container for all event weights.\\
 \multicolumn{2}{l}{\expzze}\\  & All particles of the event.\\
\verb+const MAuint32& processId() const+ & Identifier of the physical process
  related to the event.\\
\verb+const MAfloat64& scale() const+ & Factorization scale choice. \\
\verb+const MAfloat64& weight() const+ & Event weight.\\
\hline
\end{tabular}
\end{center}
Weights (as returned by the \verb+multiweights()+ method) are stored as an
instance of the \verb+WeightCollection+ class, which comes with the methods
given in the table below.
\verbdef{\expaaa}{const MAfloat64& Get(MAuint32 id) const}
\verbdef{\expaab}{const std::map<MAuint32,MAfloat64>& GetWeights() const}
\renewcommand{\arraystretch}{1.2}%
\begin{center}\begin{tabular}{p{2.7cm} p{9.0cm}}
\hline
\multicolumn{2}{l}{\expaaa}\\    & Weight value corresponding to
  the weight identifier \verb+id+.\\
\multicolumn{2}{l}{\expaab}\\    &  The full list of weights as identifier-value
  pairs.\\
\hline
\end{tabular}
\end{center}
As indicated above, the event particle content can be obtained via the method
\verb+particles()+ of the \verb+MCEventFormat+ class. This returns a vector of
\verb+MCParticleFormat+ objects, each element corresponding to an
initial-state, a final-state or an intermediate-state particle. This class
inherits all methods available from the \verb+ParticleBaseFormat+ class, that
are collected in the table below.
\verbdef{\expaac}{const MAfloat32 mt_met(const MALorentzVector& MET) const}
\verbdef{\expaad}{const MALorentzVector& momentum() const}
\verbdef{\expaae}{MALorentzVector& momentum()}
\verbdef{\expaaf}{const MAfloat32 dphi_0_pi(const ParticleBaseFormat* p) const}
\verbdef{\expaag}{const MAfloat32 dphi_0_pi(const ParticleBaseFormat& p) const}
\verbdef{\expaah}{const MAfloat32 dphi_0_2pi(const ParticleBaseFormat* p) const}
\verbdef{\expaai}{const MAfloat32 dphi_0_2pi(const ParticleBaseFormat& p) const}
\verbdef{\expaaj}{const MAfloat32 dr(const ParticleBaseFormat& p) const}
\verbdef{\expaak}{const MAfloat32 dr(const ParticleBaseFormat* p) const}
\verbdef{\expaal}{const MAfloat32 angle(const ParticleBaseFormat& p) const}
\verbdef{\expaam}{const MAfloat32 angle(const ParticleBaseFormat* p) const}
\renewcommand{\arraystretch}{1.2}%
\begin{center}\begin{tabular}{l p{6.3cm}}
\hline
\multicolumn{2}{l}{\expaae}\\
\multicolumn{2}{l}{\expaad}\\ & {Four-momentum.}\\
\verb+const MAfloat32 beta() const+ & Velocity (in $c$ units).\\
\verb+const MAfloat32 e() const+    & Energy.\\
\verb+const MAfloat32 et() const+   & Transverse energy.\\
\verb+const MAfloat32 eta() const+  & Pseudorapidity.\\
\verb+const MAfloat32 abseta() const+ & Pseudorapidity in absolute
  value.\\
\verb+const MAfloat32 gamma() const+ & Lorentz factor.\\
\verb+const MAfloat32 m() const+   & Invariant mass.\\
\verb+const MAfloat32 mt() const+  & Transverse mass.\\
\verb+const MAfloat32 phi() const+ & Azimuthal angle.\\
\verb+const MAfloat32 p() const+   & Magnitude of the momentum.\\
\verb+const MAfloat32 pt() const+  & Transverse momentum.\\
\verb+const MAfloat32 px() const+  & $x$-component of the momentum.\\
\verb+const MAfloat32 py() const+  & $y$-component of the momentum.\\
\verb+const MAfloat32 pz() const+  & $z$-component of the momentum.\\
\verb+const MAfloat32 r() const+   & Position in the $\eta-\phi$ plane.\\
\verb+const MAfloat32 theta() const+  & Polar angle.\\
\verb+const MAfloat32 y() const+ & Rapidity.\\
\hline
\end{tabular}
\end{center}
The \verb+ParticleBaseFormat+ class also includes a set of methods that
involve the particle itself as well as another object. These are given in the
following table.
\renewcommand{\arraystretch}{1.1}%
\begin{center}\begin{tabular}{p{2.7cm} p{9.0cm}}
\hline
\multicolumn{2}{l}{\expaal}\\
\multicolumn{2}{l}{\expaam}\\ & Angular separation between the
  particle momentum and the momentum of another particle \verb+p+.\\
\multicolumn{2}{l}{\expaac}\\      & Transverse mass of the system made of
    the particle and the missing momentum.\\
\multicolumn{2}{l}{\expaaf}\\
\multicolumn{2}{l}{\expaag}\\ & Azimuthal separation between the particle
momentum and the momentum of another particle \verb+p+, normalized in $[0,\pi]$.
\\
\multicolumn{2}{l}{\expaah}\\
\multicolumn{2}{l}{\expaai}\\ & Azimuthal separation between the particle
  momentum and the momentum of another particle \verb+p+, normalized in $[0,
  2\pi]$.\\
\multicolumn{2}{l}{\expaaj}\\
\multicolumn{2}{l}{\expaak}\\ & Angular distance, in the $\eta-\phi$ plane,
  between the particle momentum and the momentum of another particle \verb+p+.\\
\hline
\end{tabular}
\end{center}
Moreover, the \verb+MCParticleFormat+ class includes the extra methods
listed below.
\verbdef{\expbba}{const std::vector<MCParticleFormat*>& daughters() const}
\verbdef{\expbbb}{const std::vector<MCParticleFormat*>& mothers() const}
\verbdef{\expbbc}{const MALorentzVector& decay_vertex() const}
\renewcommand{\arraystretch}{1.2}%
\begin{center}\begin{tabular}{l p{6.1cm}}
\hline
\verb+const MAfloat64& ctau() const+ & Particle decay length.\\
\verb+const MAbool& isPU()  const+ & Tests whether the particle belongs to the
  pile-up event.\\
\verb+const MAint32& pdgid()  const+ & PDG identifier of the particle.\\
\verb+const MAfloat32& spin() const+ & Cosine of the angle between the particle
  momentum and its spin vector, computed in the laboratory frame.\\
\verb+const MAint16& statuscode() const+& Code indicating the particle initial-,
intermediate- or final-state nature.\\
\multicolumn{2}{l}{\expbba}\\ & Particles in which the current particle
  decays into.\\
\multicolumn{2}{l}{\expbbb}\\ & Particles from which the current
  particle originates from.\\
\multicolumn{2}{l}{\expbbc}\\ & Spacetime position of the particle decay.\\
\hline
\end{tabular}
\end{center}

A slightly different format is available for reconstructed events. An event is
here provided as an instance of the \verb+RecEventFormat+ class, which comes
with the following methods.
\verbdef{\expcca}{const std::vector<RecPhotonFormat>& photons() const}
\verbdef{\expccb}{const std::vector<RecLeptonFormat>& electrons() const}
\verbdef{\expccc}{const std::vector<RecLeptonFormat>& muons() const}
\verbdef{\expccd}{const std::vector<RecTauFormat>& taus() const}
\verbdef{\expcce}{const std::vector<RecJetFormat>& fatjets() const}
\verbdef{\expccf}{const std::vector<RecJetFormat>& jets() const}
\verbdef{\expccg}{const std::vector<RecJetFormat>& genjets() const}
\verbdef{\expcch}{const std::vector<RecTrackFormat>& tracks() const}
\verbdef{\expcci}{const std::vector<RecTowerFormat>& towers() const}
\verbdef{\expccj}{const std::vector<RecTrackFormat>& EFlowTracks() const}
\verbdef{\expcck}{const std::vector<RecParticleFormat>& EFlowPhotons() const}
\verbdef{\expccl}{const std::vector<RecParticleFormat>& EFlowNeutralHadrons() const}
\verbdef{\expccm}{const RecParticleFormat& MET() const}
\verbdef{\expccn}{const RecParticleFormat& MHT() const}
\verbdef{\expcco}{const MAfloat64& TET() const}
\verbdef{\expccp}{const MAfloat64& THT() const}
\verbdef{\expccq}{const MAfloat64& Meff() const}
\verbdef{\expccr}{const std::vector<const MCParticleFormat*>& MCHadronicTaus() const}
\verbdef{\expccs}{const std::vector<const MCParticleFormat*>& MCElectronicTaus() const}
\verbdef{\expcct}{const std::vector<const MCParticleFormat*>& MCMuonicTaus() const}
\verbdef{\expccu}{const std::vector<const MCParticleFormat*>& MCBquarks() const}
\verbdef{\expccv}{const std::vector<const MCParticleFormat*>& MCCquarks() const}
\renewcommand{\arraystretch}{1.2}%
\begin{center}\begin{tabular}{p{2.7cm} p{9.0cm}}
\hline
\multicolumn{2}{l}{\expcca}\\  & Reconstructed photons.\\
\multicolumn{2}{l}{\expccb}\\  & Reconstructed electrons.\\
\multicolumn{2}{l}{\expccc}\\  & Reconstructed muons.\\
\multicolumn{2}{l}{\expccd}\\  & Reconstructed hadronic taus.\\
\multicolumn{2}{l}{\expcce}\\  & Reconstructed fat jets.\\
\multicolumn{2}{l}{\expccf}\\  & Reconstructed jets.\\
\multicolumn{2}{l}{\expccg}\\  & Parton-level jets.\\
\multicolumn{2}{l}{\expcch}\\  & Tracks left in a detector.\\
\multicolumn{2}{l}{\expcci}\\  & Calorimetric deposits left in a detector.\\
\multicolumn{2}{l}{\expccj}\\  & Tracks left in a detector, reconstructed from
  the particle flow information.\\
\multicolumn{2}{l}{\expcck}\\  & Photons, reconstructed from the particle flow
  information.\\
\multicolumn{2}{l}{\expccl}\\  & Neutral hadrons, reconstructed from the
  particle flow information.\\
\multicolumn{2}{l}{\expccm}\\  & Missing transverse energy.\\
\multicolumn{2}{l}{\expccn}\\  & Missing hadronic energy.\\
\multicolumn{2}{l}{\expcco}\\  & Visible transverse energy.\\
\hline
\end{tabular}
\end{center}
\renewcommand{\arraystretch}{1.2}%
\begin{center}\begin{tabular}{p{2.7cm} p{9.0cm}}
\hline
\multicolumn{2}{l}{\expccp}\\  & Hadronic transverse energy.\\
\multicolumn{2}{l}{\expccq}\\  & Effective mass $\Big(\sum_{\rm jets} p_T +
   \slashed{E}_T\Big)$.\\
\multicolumn{2}{l}{\expccr}\\  & Parton-level taus that decayed hadronically.\\
\multicolumn{2}{l}{\expccs}\\  & Parton-level taus that decayed into an electron
  and missing energy. \\
\multicolumn{2}{l}{\expcct}\\  & Parton-level taus that decayed into a muon and
  missing energy. \\
\multicolumn{2}{l}{\expccu}\\  & Parton-level $b$-quarks of the event.\\
\multicolumn{2}{l}{\expccv}\\  & Parton-level $c$-quarks of the event.\\
\hline
\end{tabular}
\end{center}
The above methods introduce the data format implemented for all reconstructed
objects. It relies on various types (\verb+RecLeptonFormat+,
\verb+RecJetFormat+,
\verb+RecPhotonFormat+, \verb+RecTauFormat+, \verb+RecTrackFormat+ and
\verb+RecTowerFormat+) that inherit from the \verb+RecParticleFormat+ class
based on the \verb+BaseParticleFormat+ class (see above).
All these new classes include the following set of methods.
\verbdef{\expdda}{const MAfloat32& HEoverEE() const}
\verbdef{\expddb}{const MAfloat32& EEoverHE() const}
\verbdef{\expddc}{const MAuint16 ntracks() const}
\verbdef{\expddd}{const int charge() const}
\verbdef{\expdde}{MAbool isElectron() const}
\verbdef{\expddf}{MAbool isMuon() const}
\verbdef{\expddg}{MAfloat32 d0() const}
\verbdef{\expddh}{MAfloat32 d0error() const}
\verbdef{\expddi}{const MAbool& btag() const}
\verbdef{\expddj}{const MAbool& ctag() const}
\verbdef{\expddk}{const MAbool& true_btag() const}
\verbdef{\expddl}{const MAbool& true_ctag() const}
\verbdef{\expddm}{const std::vector<MAint32>& constituents() const}
\verbdef{\expddn}{const MAint32 DecayMode() const}
\renewcommand{\arraystretch}{1.2}%
\begin{center}\begin{tabular}{p{2.7cm} p{9.0cm}}
\hline
\multicolumn{2}{l}{\expddb}\\  & Ratio of the electromagnetic to
  hadronic calorimetric energy associated with the object.\\
\multicolumn{2}{l}{\expdda}\\  & Ratio of the hadronic to
  electromagnetic calorimetric energy associated with the object.\\
\multicolumn{2}{l}{\expddn}\\  & Identifier of the decay mode of a
  \verb+RecTauFormat+ object. The available choices are 1 ($\tau\to e \nu \nu$),
  2 ($\tau \to \mu \nu \nu$),  3 ($\tau \to K \nu$), 4 ($\tau \to K^* \nu$),
  5 ($\tau \to \rho (\to \pi \pi^0) \nu$), 6 ($\tau \to a_1 (\to \pi\pi^0\pi^0)
  \nu$), 7 ($\tau \to a_1 (\to \pi\pi\pi) \nu$), 8 ($\tau \to \pi \nu$), 
  9 ($\tau \to \pi\pi\pi \pi^0 \nu$) and 0 (any other decay mode).\\
\multicolumn{2}{l}{\expddi}\\  & Indicates whether a \verb+RecJetFormat+ object
  has been $b$-tagged.\\
\hline
\end{tabular}
\end{center}
\renewcommand{\arraystretch}{1.2}%
\begin{center}\begin{tabular}{p{2.7cm} p{9.0cm}}
\hline
\multicolumn{2}{l}{\expddj}\\  & Indicates whether a \verb+RecJetFormat+ object
  has been $c$-tagged.\\
\multicolumn{2}{l}{\expddm}\\  & Returns the constituents of a
  \verb+RecJetFormat+ object.\\
\multicolumn{2}{l}{\expddd}\\  & The electric charge of the object. This method
  is available for the \verb+RecLeptonFormat+, \verb+RecTauFormat+ and
  \verb+RecTrackFormat+ classes.\\
\multicolumn{2}{l}{\expddg}\\ & Impact parameter of a \verb+RecLeptonFormat+
  object. An extension to the other classes of reconstructed objects is
  foreseen.\\
\multicolumn{2}{l}{\expddh}\\ & Uncertainty on the impact parameter of a
  \verb+RecLeptonFormat+ object.\\
\multicolumn{2}{l}{\expddc}\\ & Number of charged tracks associated with a
  reconstructed object. This method is available for the \verb+RecJetFormat+ and
  and \verb+RecTauFormat+ classes.\\
\multicolumn{2}{l}{\expdde}\\ & Indicates if a \verb+RecLeptonFormat+ object is
  an electron.\\
\multicolumn{2}{l}{\expddf}\\ & Indicates if a \verb+RecLeptonFormat+ object is
  a muon.\\
\multicolumn{2}{l}{\expddk}\\  & Indicates whether a \verb+RecJetFormat+ object
  is a true $b$-jet.\\
\multicolumn{2}{l}{\expddl}\\  & Indicates whether a \verb+RecJetFormat+ object
  is a true $c$-jet.\\
\hline
\end{tabular}
\end{center}

\subsection*{Lepton and photon isolation}

In the context of reconstructed events, lepton and photon isolation can be
ensured by relying either on track information, on calorimetric information, on
the combination of both or on a reconstruction based on the energy flow.
Corresponding isolation methods are available, within the \MA\ data format,
through the four classes,
\begin{verbatim}
  PHYSICS->Isol->tracker           PHYSICS->Isol->calorimeter
  PHYSICS->Isol->combined          PHYSICS->Isol->eflow
\end{verbatim}
respectively connected to the four ways to enforce object isolation. All
those classes come with two methods summarized in the table below.
\verbdef{\expyya}{MAfloat64 relIsolation(const <x>& prt,const RecEventFormat* evt,}
\verbdef{\expyyb}{       const double& DR, double PTmin=0.5) const}
\verbdef{\expyyc}{MAfloat64 sumIsolation(const <x>& prt,const RecEventFormat* evt,}
\renewcommand{\arraystretch}{1.2}%
\begin{center}\begin{tabular}{p{2.7cm} p{9.0cm}}
\hline
\multicolumn{2}{l}{\expyya}\\
\multicolumn{2}{l}{\expyyb}\\ & Sum of the transverse momenta of all objects
  lying in a cone of radius \verb+DR+ centered on the considered object
  \verb+prt+, and whose transverse momentum is larger than \verb+PTmin+. The sum
  is evaluated relatively to the transverse momentum of the considered object
  \verb+prt+ that can be either a \verb+RecLeptonFormat+ or a
  \verb+RecPhotonFormat+ object ({\it i.e.} the value of the \verb+<x>+ type).\\
\multicolumn{2}{l}{\expyyc}\\
\multicolumn{2}{l}{\expyyb}\\ & Sum of the transverse momenta of all objects
  lying in a cone of radius \verb+DR+ centered on the considered object
  \verb+prt+, and whose transverse momentum is larger than \verb+PTmin+. The
  object \verb+prt+ can be either a \verb+RecLeptonFormat+ or a
  \verb+RecPhotonFormat+ object ({\it i.e.} the value of the \verb+<x>+ type).\\
\hline
\end{tabular}
\end{center}
When isolation is imposed on the basis of the energy flow
(\verb?PHYSICS->Isol->eflow?), those methods take an extra argument,
\begin{verbatim}
MAfloat64 relIsolation(const <x>& prt, const RecEventFormat* evt,
    const double& DR, double PTmin=0.5, ComponentType type) const

MAfloat64 sumIsolation(const <x>& prt, const RecEventFormat* evt,
    const double& DR, double PTmin=0.5, ComponentType type) const
\end{verbatim}
where \verb+type+ can take one of the four values,
\begin{verbatim}
 TRACK_COMPONENT                    PHOTON_COMPONENT
 NEUTRAL_COMPONENT                  ALL_COMPONENTS
\end{verbatim}
In the first case, the activity around the considered object \verb+prt+ is
evaluated only from the charged track information, whilst in the
second case, only the photon information is considered. In the
third case, the neutral hadron activity is accounted for whilst the last option
consists in the sum of the three previous cases.

In addition, a series of \verb+JetCleaning+ functions are provided in the aim of
cleaning jet collections from objects present in a lepton collection or a photon
collection.
\verbdef{\expwwa}{std::vector<const RecJetFormat*> PHYSICS->Isol->JetCleaning(}
\verbdef{\expwwb}{   const std::vector<const RecJetFormat*>& uncleaned,}
\verbdef{\expwwc}{   const std::vector<const RecLeptonFormat*>& leptons,}
\verbdef{\expwwd}{   double DeltaRmax = 0.1, double PTmin = 0.5) const}
\verbdef{\expwwe}{   const std::vector<RecJetFormat>& uncleaned,}
\verbdef{\expwwf}{   const std::vector<const RecPhotonFormat*>& photons,}
\renewcommand{\arraystretch}{1.13}%
\begin{center}\begin{tabular}{p{2.7cm} p{9.0cm}}
\hline
\multicolumn{2}{l}{\expwwa}\\
\multicolumn{2}{l}{\expwwb}\\
\multicolumn{2}{l}{\expwwc}\\
\multicolumn{2}{l}{\expwwd}\\
\multicolumn{2}{l}{\expwwa}\\
\multicolumn{2}{l}{\expwwe}\\
\multicolumn{2}{l}{\expwwc}\\
\multicolumn{2}{l}{\expwwd}\\ & Removal from the \verb+uncleaned+ jet collection
  of all leptons included in the \verb+leptons+ collection lying at an angular
  distance of at most \verb+DeltaRmax+ of the jet, and whose transverse momentum
  is of at least \verb+PTmin+.\\
\multicolumn{2}{l}{\expwwa}\\
\multicolumn{2}{l}{\expwwb}\\
\multicolumn{2}{l}{\expwwf}\\
\multicolumn{2}{l}{\expwwd}\\
\multicolumn{2}{l}{\expwwa}\\
\multicolumn{2}{l}{\expwwe}\\
\multicolumn{2}{l}{\expwwf}\\
\multicolumn{2}{l}{\expwwd}\\ & Removal from the \verb+uncleaned+ jet collection
  of all photons included in the \verb+photons+ collection lying at an angular
  distance of at most \verb+DeltaRmax+ of the jet, and whose transverse momentum
  is of at least \verb+PTmin+.\\
\hline
\end{tabular}
\end{center}

\subsection*{Observables}
The \spla\ data format contains various methods to compute observables
connected to the entire event, which includes in particular a small set of
common transverse variables and a set of methods related to object
identification. They are available through a series of \verb+PHYSICS+ services,
that first contain  the two general methods below.
\verbdef{\expxxa}{MAint32 PHYSICS->GetTauDecayMode (const MCParticleFormat* part)}
\verbdef{\expxxb}{double PHYSICS->SqrtS(const MCEventFormat* event) const}
\renewcommand{\arraystretch}{1.2}%
\begin{center}\begin{tabular}{p{2.7cm} p{9.0cm}}
\hline
\multicolumn{2}{l}{\expxxa}\\ & Returns the identifier of the decay mode of a
  tau particle. The available values are 1 ($e \nu \nu$),
  2 ($\mu \nu \nu$),  3 ($K \nu$), 4 ($K^* \nu$),
  5 ($\rho (\to \pi \pi^0) \nu$), 6 ($a_1 (\to \pi\pi^0\pi^0)
  \nu$), 7 ($a_1 (\to \pi\pi\pi) \nu$), 8 ($\pi \nu$),
  9 ($\pi\pi\pi \pi^0 \nu$) and 0 (any other decay mode).\\
\multicolumn{2}{l}{\expxxb}\\ & Returns the partonic center-of-mass energy.\\
\hline
\end{tabular}
\end{center}
Identification functions are collected as methods attached to the
\verb+PHYSICS->Id+ object. The list of available methods is given in the table
below.
\verbdef{\expxxc}{MAbool IsInitialState(const MCParticleFormat& part) const}
\verbdef{\expxxd}{MAbool IsFinalState(const MCParticleFormat& part) const}
\verbdef{\expxxe}{MAbool IsInterState(const MCParticleFormat& part) const}
\verbdef{\expxxf}{MAbool IsInitialState(const MCParticleFormat* part) const}
\verbdef{\expxxg}{MAbool IsFinalState(const MCParticleFormat* part) const}
\verbdef{\expxxh}{MAbool IsInterState(const MCParticleFormat* part) const}
\verbdef{\expxxj}{bool IsHadronic(const RecParticleFormat* part) const}
\verbdef{\expxxk}{bool IsHadronic(const MCParticleFormat* part) const}
\verbdef{\expxxl}{bool IsHadronic(MAint32 pdgid) const}
\verbdef{\expxxm}{bool IsInvisible(const RecParticleFormat* part) const}
\verbdef{\expxxn}{bool IsInvisible(const MCParticleFormat* part) const}
\verbdef{\expxxo}{MAbool IsBHadron(MAint32 pdg)}
\verbdef{\expxxp}{MAbool IsBHadron(const MCParticleFormat& part)}
\verbdef{\expxxq}{MAbool IsBHadron(const MCParticleFormat* part)}
\verbdef{\expxxs}{MAbool IsCHadron(MAint32 pdg)}
\verbdef{\expxxt}{MAbool IsCHadron(const MCParticleFormat& part)}
\verbdef{\expxxu}{MAbool IsCHadron(const MCParticleFormat* part)}
\renewcommand{\arraystretch}{1.2}%
\begin{center}\begin{tabular}{p{2.7cm} p{9.0cm}}
\hline
\multicolumn{2}{l}{\expxxc}\\
\multicolumn{2}{l}{\expxxf}\\ & Tests the initial-state nature of an object.\\
\multicolumn{2}{l}{\expxxh}\\
\multicolumn{2}{l}{\expxxe}\\ & Tests the intermediate-state nature of an
  object.\\
\multicolumn{2}{l}{\expxxd}\\
\multicolumn{2}{l}{\expxxg}\\ & Tests the final-state nature of an object.\\
\multicolumn{2}{l}{\expxxj}\\
\multicolumn{2}{l}{\expxxk}\\
\multicolumn{2}{l}{\expxxl}\\ & Tests the hadronic nature of an object.\\
\multicolumn{2}{l}{\expxxm}\\
\multicolumn{2}{l}{\expxxn}\\ & Tests the invisible nature of an object.\\
\multicolumn{2}{l}{\expxxo}\\
\multicolumn{2}{l}{\expxxp}\\
\multicolumn{2}{l}{\expxxq}\\ & Tests whether the object is a $B$-hadron.\\
\multicolumn{2}{l}{\expxxs}\\
\multicolumn{2}{l}{\expxxt}\\
\multicolumn{2}{l}{\expxxu}\\ & Tests whether the object is a $C$-hadron.\\
\hline
\end{tabular}
\end{center}
Finally, a set of transverse variables can be evaluated by relying on the
\verb+PHYSICS->Transverse+ object. The following methods are available within
the \spla\ data format.
\verbdef{\expqqa}{double EventTET(const MCEventFormat* event) const}
\verbdef{\expqqb}{double EventTET(const RecEventFormat* event) const}
\verbdef{\expqqc}{double EventMET(const MCEventFormat* event) const}
\verbdef{\expqqd}{double EventMET(const RecEventFormat* event) const}
\verbdef{\expqqe}{double EventTHT(const MCEventFormat* event) const}
\verbdef{\expqqf}{double EventTHT(const RecEventFormat* event) const}
\verbdef{\expqqg}{double EventMEFF(const MCEventFormat* event) const}
\verbdef{\expqqh}{double EventMEFF(const RecEventFormat* event) const}
\verbdef{\expqqi}{double EventMHT(const MCEventFormat* event) const}
\verbdef{\expqqj}{double EventMHT(const RecEventFormat* event) const}
\verbdef{\expqqka}{double MT2(const MALorentzVector* p1, const MALorentzVector* p2,}
\verbdef{\expqqkb}{   const MALorentzVector& met, const double &mass)}
\verbdef{\expqqla}{double MT2W(std::vector<const RecJetFormat*> jets,}
\verbdef{\expqqlb}{   const RecLeptonFormat* lep, const ParticleBaseFormat& met)}
\verbdef{\expqqma}{double MT2W(std::vector<const MCParticleFormat*> jets}
\verbdef{\expqqmb}{  const MCParticleFormat* lep,const ParticleBaseFormat& met)}
\verbdef{\expqqn}{double AlphaT(const MCEventFormat*)}
\verbdef{\expqqo}{double AlphaT(const RecEventFormat*)}
\renewcommand{\arraystretch}{1.2}%
\begin{center}\begin{tabular}{p{2.7cm} p{9.0cm}}
\hline
\multicolumn{2}{l}{\expqqn}\\
\multicolumn{2}{l}{\expqqo}\\ & The $\alpha_T$ variable\cite{Randall:2008rw}.\\
\multicolumn{2}{l}{\expqqg}\\
\multicolumn{2}{l}{\expqqh}\\ & The effective mass of the event.\\
\hline
\end{tabular}
\end{center}
\renewcommand{\arraystretch}{1.2}%
\begin{center}\begin{tabular}{p{2.7cm} p{9.0cm}}
\hline
\multicolumn{2}{l}{\expqqc}\\
\multicolumn{2}{l}{\expqqd}\\ & The event missing transverse energy.\\
\multicolumn{2}{l}{\expqqi}\\
\multicolumn{2}{l}{\expqqj}\\ & The event missing transverse hadronic energy.\\
\multicolumn{2}{l}{\expqqa}\\
\multicolumn{2}{l}{\expqqb}\\ & The event total transverse energy.\\
\multicolumn{2}{l}{\expqqe}\\
\multicolumn{2}{l}{\expqqf}\\ & The event total transverse hadronic energy.\\
\multicolumn{2}{l}{\expqqka}\\
\multicolumn{2}{l}{\expqqkb}\\ & The event $m_{T2}$ variable computed from a system
  of two visible objects \verb+p1+ and \verb+p2+, the event missing momentum
  \verb+met+ and a test mass \verb+mass+\cite{Lester:1999tx,Cheng:2008hk}.\\
\multicolumn{2}{l}{\expqqla}\\
\multicolumn{2}{l}{\expqqlb}\\
\multicolumn{2}{l}{\expqqma}\\
\multicolumn{2}{l}{\expqqmb}\\ & The event $m_{T2}^W$ variable computed from a
  system of jets \verb+jets+, a lepton \verb+lep+ and the missing
  momentum \verb+met+~\cite{Bai:2012gs}.\\
\hline
\end{tabular}
\end{center}

\subsection*{Signal regions, histograms and cuts}
The implementation of an analysis in \MA\ requires to deal with
signal regions, selection cuts and histograms. Each analysis comes with an
instance of the analysis manager class \verb+RegionSelectionManager+, named
\verb+Manager()+, which allows the user to use the methods presented in the
table below.
\verbdef{\SRa}{void AddCut(const std::string&name, const std::string &RSname)}
\verbdef{\SRba}{template<int NRS> void AddCut(const std::string&name,}
\verbdef{\SRbb}{  std::string const(& RSnames)[NRS])}
\verbdef{\SRbc}{void AddCut(const std::string &name)}
\verbdef{\SRca}{void AddHisto(const std::string&name,unsigned int nb,}
\verbdef{\SRcb}{  double xmin, double xmax)}
\verbdef{\SRcd}{void AddHistoLogX(const std::string&name,unsigned int nb,}
\verbdef{\SRce}{  double xmin, double xmax, const std::string &RSname)}
\verbdef{\SRcf}{template <int NRS> void AddHisto(const std::string&name,}
\verbdef{\SRcg}{  unsigned int nb, double xmin, double xmax,}
\verbdef{\SRch}{template <int NRS> void AddHistoLogX(const std::string&name,}
\verbdef{\SRd}{void AddRegionSelection(const std::string& name)}
\verbdef{\SRe}{bool ApplyCut(bool cond, std::string const &name)}
\verbdef{\SRf}{void FillHisto(std::string const&name, double val)}
\verbdef{\SRg}{void InitializeForNewEvent(double EventWeight)}
\verbdef{\SRh}{bool IsSurviving(const std::string &RSname)}
\verbdef{\SRi}{void SetCurrentEventWeight(double weight)}
\renewcommand{\arraystretch}{1.1}%
\begin{center}\begin{tabular}{p{2.7cm} p{9.0cm}}
\hline
\multicolumn{2}{l}{\SRa}\\\multicolumn{2}{l}{\SRba}\\
  \multicolumn{2}{l}{\SRbb}\\
  \multicolumn{2}{l}{\SRbc}\\&
  Declares a cut named \verb+name+ and associates it with one region (the second
  argument is a string), with a set of regions (the second argument is an
  array of strings) or with all regions (the second argument is omitted).\\
\hline
\end{tabular}
\end{center}
\renewcommand{\arraystretch}{1.2}%
\begin{center}\begin{tabular}{p{2.7cm} p{9.0cm}}
\hline
\multicolumn{2}{l}{\SRca}\\ \multicolumn{2}{l}{\SRcb}\\
  \multicolumn{2}{l}{\SRcd}\\ \multicolumn{2}{l}{\SRcb}\\&
  Declares a histogram named \verb+name+ of \verb+nb+ bins ranging from
  \verb+xmin+ to \verb+xmax+. The histogram is associated with all regions and
  the $x$-axis can rely on a logarithmic scale (the second method).\\
\multicolumn{2}{l}{\SRca}\\ \multicolumn{2}{l}{\SRce}\\
  \multicolumn{2}{l}{\SRcd}\\ \multicolumn{2}{l}{\SRce}\\&
  Same as above but the histogram is associated with a single region
  \verb+RSname+.\\
\multicolumn{2}{l}{\SRcf}\\ \multicolumn{2}{l}{\SRcg}\\
  \multicolumn{2}{l}{\SRbb}\\
  \multicolumn{2}{l}{\SRch}\\ \multicolumn{2}{l}{\SRcg}\\
  \multicolumn{2}{l}{\SRbb}\\ &
  Same as above but the histogram is associated with an array of regions.\\
\multicolumn{2}{l}{\SRd}\\ &
  Declares a new region named \verb+name+.\\
\multicolumn{2}{l}{\SRe}\\ &
  Applies the cut \verb+name+, an event passing this cut if the condition
  \verb+cond+ is realized. The method returns \texttt{true} if at least one
  region is passing all cuts applied so far, or \texttt{false} otherwise.\\
\multicolumn{2}{l}{\SRf}\\ &
  Fills the histogram named \verb+name+, the bin choice being driven by the
  value \verb+value+.\\
\multicolumn{2}{l}{\SRg}\\ &
  To be called at the beginning of the analysis of an event in order to tag all
  regions as surviving the cuts and initialize the event weight to the value
  \verb+EventWeight+.\\
\hline
\end{tabular}
\end{center}
\renewcommand{\arraystretch}{1.2}%
\begin{center}\begin{tabular}{p{2.7cm} p{9.0cm}}
\hline
\multicolumn{2}{l}{\SRh}\\ &
  Verifies whether the region \verb+RSname+ survives all cut applied so far.\\
\multicolumn{2}{l}{\SRi}\\ &
  Modifies the weight of the current event to the value \verb+weight+.\\
\hline
\end{tabular}
\end{center}

\subsection*{Message services}
\spla\ handles four levels of streamers, that can be cast within any analysis
code by typing one of the following lines,
\begin{verbatim}
 INFO    << "..." << endmsg;
 WARNING << "..." << endmsg;
 ERROR   << "..." << endmsg;
 DEBUG   << "..." << endmsg;
\end{verbatim}
This allows the user to print informative, warning, error and debugging
messages. Additionally, warning and error messages return information on the
line number responsible for printing the message.
The effect of a given message service can be modified by
means of the methods presented in the table below.
\verbdef{\ina}{void SetStream(std::ostream* stream)}
\renewcommand{\arraystretch}{1.2}%
\begin{center}\begin{tabular}{l p{8.3cm}}
\hline
\verb+void DisableColor()+ & Switches off the colored display of messages (that
  is on by default).\\
\verb+void EnableColor()+  & Switches on the colored display of messages.\\
\verb+void SetMute()+      & Switches entirely off a given message service (that
  is on by default).\\
\multicolumn{2}{l}{\ina}\\ & Redirects the output of a given service to a file.
  \\
\verb+void SetUnMute()+   & Switches on a specific message service.\\
\hline
\end{tabular}
\end{center}

\subsection*{Sorting particles and objects}
It is usally important to order particle as a function of one of their
properties, like their transverse momentum or their energy. For this reason,
\spla\ contains a series of routines allowing to sort a set of objects, which
can be called by implementing
\begin{verbatim}
  SORTER->sort(objects, criterion)
\end{verbatim}
where \verb+objects+ is the vector of objects that needs to be sorted and
\verb+criterion+ is the ordering variable. The latter can be \verb+ETAordering+
(pseudorapidity), \verb+ETordering+ (transverse energy), \verb+Eordering+
(energy), \verb+Pordering+ (the magnitude of the three-momentum),
\verb+PTordering+ (the transverse momentum), \verb+PXordering+ (the
$x$-component of the momentum), \verb+PYordering+ (the $y$-component of the
momentum) and \verb+PZordering+ (the $z$-component of the momentum). As a
result, the vector of objects is sorted by decreasing values of the ordering
variable.

\newpage

\bibliographystyle{ws-ijmpa}
\bibliography{recasting}

\end{document}